\documentclass[longauth]{aa}  

\usepackage{graphicx}
\usepackage{amsmath,amssymb,amsfonts}
\usepackage{txfonts}
\usepackage{color, colortbl}
\usepackage[T1]{fontenc} % if needed
\usepackage{braket}
\usepackage{euclid}
\usepackage{placeins}
\usepackage{tabularx}
\usepackage{subcaption}
\usepackage{afterpage}
\usepackage[normalem]{ulem}
\usepackage{fontawesome}
\usepackage{lineno}
\usepackage[pdfencoding=auto,psdextra]{hyperref}

\hypersetup{
    colorlinks=true,
    linkcolor=blue,
    filecolor=purple,      
    urlcolor=blue,
    citecolor=blue
}

\defcitealias{Winther:2015wla}{W15}

\newcommand{\Om}[1]{\Omega_\mathrm{#1}}

\newcommand{\de}[1]{\delta_\mathrm{#1}}
\newcommand{\rh}[1]{\rho_\mathrm{#1}}

\newcommand{\pysco}{\texttt{PySCo}\xspace}

\usepackage{orcidlink}
\newcommand{\orcid}[1]{\orcidlink{#1}}

\begin{document} 

%   \title{\Euclid preparation: Simulations and nonlinearities beyond $\mathsf{\Lambda}$CDM}
\title{\Euclid preparation}
\subtitle{Simulations and nonlinearities beyond $\mathsf{\Lambda}$CDM. 1. Numerical methods and validation}    
   \titlerunning{Simulations and nonlinearities beyond $\Lambda$CDM -- numerical methods and validation}

%   \subtitle{1. Numerical methods and validation}

%%%% please do not edit the author list -- contact ECEB Bureau for changes
%\newcommand{\orcid}[1]{} %% if already defined in aa.cls: comment, or use renewcommand			   
\author{Euclid Collaboration: J.~Adamek\orcid{0000-0002-0723-6740}\thanks{\email{julian.adamek@uzh.ch}}\inst{\ref{aff1}}
\and B.~Fiorini\orcid{0000-0002-0092-4321}\inst{\ref{aff2}}
\and M.~Baldi\orcid{0000-0003-4145-1943}\inst{\ref{aff3},\ref{aff4},\ref{aff5}}
\and G.~Brando\orcid{0000-0003-0805-1905}\inst{\ref{aff6}}
\and M.-A.~Breton\inst{\ref{aff7},\ref{aff8},\ref{aff9}}
\and F.~Hassani\orcid{0000-0003-2640-4460}\inst{\ref{aff10}}
\and K.~Koyama\orcid{0000-0001-6727-6915}\inst{\ref{aff2}}
\and A.~M.~C.~Le~Brun\orcid{0000-0002-0936-4594}\inst{\ref{aff9}}
\and G.~R\'acz\orcid{0000-0003-3906-5699}\inst{\ref{aff11}}
\and H.-A.~Winther\orcid{0000-0002-6325-2710}\inst{\ref{aff10}}
\and A.~Casalino\orcid{0000-0001-6709-5292}\inst{\ref{aff3}}
\and C.~Hern\'andez-Aguayo\orcid{0000-0001-9921-8832}\inst{\ref{aff12}}
\and B.~Li\orcid{0000-0002-1098-9188}\inst{\ref{aff13}}
\and D.~Potter\orcid{0000-0002-0757-5195}\inst{\ref{aff1}}
\and E.~Altamura\orcid{0000-0001-6973-1897}\inst{\ref{aff14}}
\and C.~Carbone\orcid{0000-0003-0125-3563}\inst{\ref{aff15}}
\and C.~Giocoli\orcid{0000-0002-9590-7961}\inst{\ref{aff4},\ref{aff16}}
\and D.~F.~Mota\orcid{0000-0003-3141-142X}\inst{\ref{aff10}}
\and A.~Pourtsidou\orcid{0000-0001-9110-5550}\inst{\ref{aff17},\ref{aff18}}
\and Z.~Sakr\orcid{0000-0002-4823-3757}\inst{\ref{aff19},\ref{aff20},\ref{aff21}}
\and F.~Vernizzi\orcid{0000-0003-3426-2802}\inst{\ref{aff22}}
\and A.~Amara\inst{\ref{aff23}}
\and S.~Andreon\orcid{0000-0002-2041-8784}\inst{\ref{aff24}}
\and N.~Auricchio\orcid{0000-0003-4444-8651}\inst{\ref{aff4}}
\and C.~Baccigalupi\orcid{0000-0002-8211-1630}\inst{\ref{aff25},\ref{aff26},\ref{aff27},\ref{aff28}}
\and S.~Bardelli\orcid{0000-0002-8900-0298}\inst{\ref{aff4}}
\and P.~Battaglia\orcid{0000-0002-7337-5909}\inst{\ref{aff4}}
\and D.~Bonino\orcid{0000-0002-3336-9977}\inst{\ref{aff29}}
\and E.~Branchini\orcid{0000-0002-0808-6908}\inst{\ref{aff30},\ref{aff31},\ref{aff24}}
\and M.~Brescia\orcid{0000-0001-9506-5680}\inst{\ref{aff32},\ref{aff33},\ref{aff34}}
\and J.~Brinchmann\orcid{0000-0003-4359-8797}\inst{\ref{aff35},\ref{aff36}}
\and A.~Caillat\inst{\ref{aff37}}
\and S.~Camera\orcid{0000-0003-3399-3574}\inst{\ref{aff38},\ref{aff39},\ref{aff29}}
\and V.~Capobianco\orcid{0000-0002-3309-7692}\inst{\ref{aff29}}
\and V.~F.~Cardone\inst{\ref{aff40},\ref{aff41}}
\and J.~Carretero\orcid{0000-0002-3130-0204}\inst{\ref{aff42},\ref{aff43}}
\and S.~Casas\orcid{0000-0002-4751-5138}\inst{\ref{aff44}}
\and F.~J.~Castander\orcid{0000-0001-7316-4573}\inst{\ref{aff7},\ref{aff45}}
\and M.~Castellano\orcid{0000-0001-9875-8263}\inst{\ref{aff40}}
\and G.~Castignani\orcid{0000-0001-6831-0687}\inst{\ref{aff4}}
\and S.~Cavuoti\orcid{0000-0002-3787-4196}\inst{\ref{aff33},\ref{aff34}}
\and A.~Cimatti\inst{\ref{aff46}}
\and C.~Colodro-Conde\inst{\ref{aff47}}
\and G.~Congedo\orcid{0000-0003-2508-0046}\inst{\ref{aff17}}
\and C.~J.~Conselice\orcid{0000-0003-1949-7638}\inst{\ref{aff14}}
\and L.~Conversi\orcid{0000-0002-6710-8476}\inst{\ref{aff48},\ref{aff49}}
\and Y.~Copin\orcid{0000-0002-5317-7518}\inst{\ref{aff50}}
\and F.~Courbin\orcid{0000-0003-0758-6510}\inst{\ref{aff51},\ref{aff52},\ref{aff53}}
\and H.~M.~Courtois\orcid{0000-0003-0509-1776}\inst{\ref{aff54}}
\and A.~Da~Silva\orcid{0000-0002-6385-1609}\inst{\ref{aff55},\ref{aff56}}
\and H.~Degaudenzi\orcid{0000-0002-5887-6799}\inst{\ref{aff57}}
\and G.~De~Lucia\orcid{0000-0002-6220-9104}\inst{\ref{aff26}}
\and M.~Douspis\orcid{0000-0003-4203-3954}\inst{\ref{aff58}}
\and F.~Dubath\orcid{0000-0002-6533-2810}\inst{\ref{aff57}}
\and X.~Dupac\inst{\ref{aff49}}
\and S.~Dusini\orcid{0000-0002-1128-0664}\inst{\ref{aff59}}
\and M.~Farina\orcid{0000-0002-3089-7846}\inst{\ref{aff60}}
\and S.~Farrens\orcid{0000-0002-9594-9387}\inst{\ref{aff61}}
\and S.~Ferriol\inst{\ref{aff50}}
\and P.~Fosalba\orcid{0000-0002-1510-5214}\inst{\ref{aff45},\ref{aff7}}
\and M.~Frailis\orcid{0000-0002-7400-2135}\inst{\ref{aff26}}
\and E.~Franceschi\orcid{0000-0002-0585-6591}\inst{\ref{aff4}}
\and M.~Fumana\orcid{0000-0001-6787-5950}\inst{\ref{aff15}}
\and S.~Galeotta\orcid{0000-0002-3748-5115}\inst{\ref{aff26}}
\and B.~Gillis\orcid{0000-0002-4478-1270}\inst{\ref{aff17}}
\and P.~G\'omez-Alvarez\orcid{0000-0002-8594-5358}\inst{\ref{aff62},\ref{aff49}}
\and A.~Grazian\orcid{0000-0002-5688-0663}\inst{\ref{aff63}}
\and F.~Grupp\inst{\ref{aff64},\ref{aff65}}
\and L.~Guzzo\orcid{0000-0001-8264-5192}\inst{\ref{aff66},\ref{aff24}}
\and S.~V.~H.~Haugan\orcid{0000-0001-9648-7260}\inst{\ref{aff10}}
\and W.~Holmes\inst{\ref{aff11}}
\and F.~Hormuth\inst{\ref{aff67}}
\and A.~Hornstrup\orcid{0000-0002-3363-0936}\inst{\ref{aff68},\ref{aff69}}
\and S.~Ili\'c\orcid{0000-0003-4285-9086}\inst{\ref{aff70},\ref{aff20}}
\and K.~Jahnke\orcid{0000-0003-3804-2137}\inst{\ref{aff71}}
\and M.~Jhabvala\inst{\ref{aff72}}
\and B.~Joachimi\orcid{0000-0001-7494-1303}\inst{\ref{aff73}}
\and E.~Keih\"anen\orcid{0000-0003-1804-7715}\inst{\ref{aff74}}
\and S.~Kermiche\orcid{0000-0002-0302-5735}\inst{\ref{aff75}}
\and A.~Kiessling\orcid{0000-0002-2590-1273}\inst{\ref{aff11}}
\and M.~Kilbinger\orcid{0000-0001-9513-7138}\inst{\ref{aff61}}
\and B.~Kubik\orcid{0009-0006-5823-4880}\inst{\ref{aff50}}
\and M.~K\"ummel\orcid{0000-0003-2791-2117}\inst{\ref{aff65}}
\and M.~Kunz\orcid{0000-0002-3052-7394}\inst{\ref{aff76}}
\and H.~Kurki-Suonio\orcid{0000-0002-4618-3063}\inst{\ref{aff77},\ref{aff78}}
\and S.~Ligori\orcid{0000-0003-4172-4606}\inst{\ref{aff29}}
\and P.~B.~Lilje\orcid{0000-0003-4324-7794}\inst{\ref{aff10}}
\and V.~Lindholm\orcid{0000-0003-2317-5471}\inst{\ref{aff77},\ref{aff78}}
\and I.~Lloro\inst{\ref{aff79}}
\and G.~Mainetti\orcid{0000-0003-2384-2377}\inst{\ref{aff80}}
\and E.~Maiorano\orcid{0000-0003-2593-4355}\inst{\ref{aff4}}
\and O.~Mansutti\orcid{0000-0001-5758-4658}\inst{\ref{aff26}}
\and O.~Marggraf\orcid{0000-0001-7242-3852}\inst{\ref{aff81}}
\and K.~Markovic\orcid{0000-0001-6764-073X}\inst{\ref{aff11}}
\and M.~Martinelli\orcid{0000-0002-6943-7732}\inst{\ref{aff40},\ref{aff41}}
\and N.~Martinet\orcid{0000-0003-2786-7790}\inst{\ref{aff37}}
\and F.~Marulli\orcid{0000-0002-8850-0303}\inst{\ref{aff82},\ref{aff4},\ref{aff5}}
\and R.~Massey\orcid{0000-0002-6085-3780}\inst{\ref{aff13}}
\and E.~Medinaceli\orcid{0000-0002-4040-7783}\inst{\ref{aff4}}
\and S.~Mei\orcid{0000-0002-2849-559X}\inst{\ref{aff83}}
\and M.~Melchior\inst{\ref{aff84}}
\and Y.~Mellier\inst{\ref{aff85},\ref{aff86}}
\and M.~Meneghetti\orcid{0000-0003-1225-7084}\inst{\ref{aff4},\ref{aff5}}
\and E.~Merlin\orcid{0000-0001-6870-8900}\inst{\ref{aff40}}
\and G.~Meylan\inst{\ref{aff51}}
\and M.~Moresco\orcid{0000-0002-7616-7136}\inst{\ref{aff82},\ref{aff4}}
\and L.~Moscardini\orcid{0000-0002-3473-6716}\inst{\ref{aff82},\ref{aff4},\ref{aff5}}
\and C.~Neissner\orcid{0000-0001-8524-4968}\inst{\ref{aff87},\ref{aff43}}
\and S.-M.~Niemi\inst{\ref{aff88}}
\and C.~Padilla\orcid{0000-0001-7951-0166}\inst{\ref{aff87}}
\and S.~Paltani\orcid{0000-0002-8108-9179}\inst{\ref{aff57}}
\and F.~Pasian\orcid{0000-0002-4869-3227}\inst{\ref{aff26}}
\and K.~Pedersen\inst{\ref{aff89}}
\and W.~J.~Percival\orcid{0000-0002-0644-5727}\inst{\ref{aff90},\ref{aff91},\ref{aff92}}
\and V.~Pettorino\inst{\ref{aff88}}
\and S.~Pires\orcid{0000-0002-0249-2104}\inst{\ref{aff61}}
\and G.~Polenta\orcid{0000-0003-4067-9196}\inst{\ref{aff93}}
\and M.~Poncet\inst{\ref{aff94}}
\and L.~A.~Popa\inst{\ref{aff95}}
\and L.~Pozzetti\orcid{0000-0001-7085-0412}\inst{\ref{aff4}}
\and F.~Raison\orcid{0000-0002-7819-6918}\inst{\ref{aff64}}
\and A.~Renzi\orcid{0000-0001-9856-1970}\inst{\ref{aff96},\ref{aff59}}
\and J.~Rhodes\orcid{0000-0002-4485-8549}\inst{\ref{aff11}}
\and G.~Riccio\inst{\ref{aff33}}
\and E.~Romelli\orcid{0000-0003-3069-9222}\inst{\ref{aff26}}
\and M.~Roncarelli\orcid{0000-0001-9587-7822}\inst{\ref{aff4}}
\and R.~Saglia\orcid{0000-0003-0378-7032}\inst{\ref{aff65},\ref{aff64}}
\and A.~G.~S\'anchez\orcid{0000-0003-1198-831X}\inst{\ref{aff64}}
\and D.~Sapone\orcid{0000-0001-7089-4503}\inst{\ref{aff97}}
\and B.~Sartoris\orcid{0000-0003-1337-5269}\inst{\ref{aff65},\ref{aff26}}
\and M.~Schirmer\orcid{0000-0003-2568-9994}\inst{\ref{aff71}}
\and T.~Schrabback\orcid{0000-0002-6987-7834}\inst{\ref{aff98}}
\and A.~Secroun\orcid{0000-0003-0505-3710}\inst{\ref{aff75}}
\and G.~Seidel\orcid{0000-0003-2907-353X}\inst{\ref{aff71}}
\and S.~Serrano\orcid{0000-0002-0211-2861}\inst{\ref{aff45},\ref{aff99},\ref{aff7}}
\and C.~Sirignano\orcid{0000-0002-0995-7146}\inst{\ref{aff96},\ref{aff59}}
\and G.~Sirri\orcid{0000-0003-2626-2853}\inst{\ref{aff5}}
\and L.~Stanco\orcid{0000-0002-9706-5104}\inst{\ref{aff59}}
\and J.~Steinwagner\orcid{0000-0001-7443-1047}\inst{\ref{aff64}}
\and P.~Tallada-Cresp\'{i}\orcid{0000-0002-1336-8328}\inst{\ref{aff42},\ref{aff43}}
\and D.~Tavagnacco\orcid{0000-0001-7475-9894}\inst{\ref{aff26}}
\and I.~Tereno\inst{\ref{aff55},\ref{aff100}}
\and R.~Toledo-Moreo\orcid{0000-0002-2997-4859}\inst{\ref{aff101}}
\and F.~Torradeflot\orcid{0000-0003-1160-1517}\inst{\ref{aff43},\ref{aff42}}
\and I.~Tutusaus\orcid{0000-0002-3199-0399}\inst{\ref{aff20}}
\and E.~A.~Valentijn\inst{\ref{aff102}}
\and L.~Valenziano\orcid{0000-0002-1170-0104}\inst{\ref{aff4},\ref{aff103}}
\and T.~Vassallo\orcid{0000-0001-6512-6358}\inst{\ref{aff65},\ref{aff26}}
\and G.~Verdoes~Kleijn\orcid{0000-0001-5803-2580}\inst{\ref{aff102}}
\and A.~Veropalumbo\orcid{0000-0003-2387-1194}\inst{\ref{aff24},\ref{aff31},\ref{aff104}}
\and Y.~Wang\orcid{0000-0002-4749-2984}\inst{\ref{aff105}}
\and J.~Weller\orcid{0000-0002-8282-2010}\inst{\ref{aff65},\ref{aff64}}
\and G.~Zamorani\orcid{0000-0002-2318-301X}\inst{\ref{aff4}}
\and E.~Zucca\orcid{0000-0002-5845-8132}\inst{\ref{aff4}}
\and A.~Biviano\orcid{0000-0002-0857-0732}\inst{\ref{aff26},\ref{aff25}}
\and C.~Burigana\orcid{0000-0002-3005-5796}\inst{\ref{aff106},\ref{aff103}}
\and M.~Calabrese\orcid{0000-0002-2637-2422}\inst{\ref{aff107},\ref{aff15}}
\and D.~Di~Ferdinando\inst{\ref{aff5}}
\and J.~A.~Escartin~Vigo\inst{\ref{aff64}}
\and G.~Fabbian\orcid{0000-0002-3255-4695}\inst{\ref{aff108},\ref{aff109}}
\and F.~Finelli\orcid{0000-0002-6694-3269}\inst{\ref{aff4},\ref{aff103}}
\and J.~Gracia-Carpio\inst{\ref{aff64}}
\and S.~Matthew\orcid{0000-0001-8448-1697}\inst{\ref{aff17}}
\and N.~Mauri\orcid{0000-0001-8196-1548}\inst{\ref{aff46},\ref{aff5}}
\and A.~Pezzotta\orcid{0000-0003-0726-2268}\inst{\ref{aff64}}
\and M.~P\"ontinen\orcid{0000-0001-5442-2530}\inst{\ref{aff77}}
\and V.~Scottez\inst{\ref{aff85},\ref{aff110}}
\and M.~Tenti\orcid{0000-0002-4254-5901}\inst{\ref{aff5}}
\and M.~Viel\orcid{0000-0002-2642-5707}\inst{\ref{aff25},\ref{aff26},\ref{aff28},\ref{aff27},\ref{aff111}}
\and M.~Wiesmann\orcid{0009-0000-8199-5860}\inst{\ref{aff10}}
\and Y.~Akrami\orcid{0000-0002-2407-7956}\inst{\ref{aff112},\ref{aff113}}
\and V.~Allevato\orcid{0000-0001-7232-5152}\inst{\ref{aff33}}
\and S.~Anselmi\orcid{0000-0002-3579-9583}\inst{\ref{aff59},\ref{aff96},\ref{aff9}}
\and M.~Archidiacono\orcid{0000-0003-4952-9012}\inst{\ref{aff66},\ref{aff114}}
\and F.~Atrio-Barandela\orcid{0000-0002-2130-2513}\inst{\ref{aff115}}
\and A.~Balaguera-Antolinez\orcid{0000-0001-5028-3035}\inst{\ref{aff47},\ref{aff116}}
\and M.~Ballardini\orcid{0000-0003-4481-3559}\inst{\ref{aff117},\ref{aff4},\ref{aff118}}
\and A.~Blanchard\orcid{0000-0001-8555-9003}\inst{\ref{aff20}}
\and L.~Blot\orcid{0000-0002-9622-7167}\inst{\ref{aff119},\ref{aff9}}
\and H.~B\"ohringer\orcid{0000-0001-8241-4204}\inst{\ref{aff64},\ref{aff120},\ref{aff121}}
\and S.~Borgani\orcid{0000-0001-6151-6439}\inst{\ref{aff122},\ref{aff25},\ref{aff26},\ref{aff27}}
\and S.~Bruton\orcid{0000-0002-6503-5218}\inst{\ref{aff123}}
\and R.~Cabanac\orcid{0000-0001-6679-2600}\inst{\ref{aff20}}
\and A.~Calabro\orcid{0000-0003-2536-1614}\inst{\ref{aff40}}
\and B.~Camacho~Quevedo\orcid{0000-0002-8789-4232}\inst{\ref{aff45},\ref{aff7}}
\and G.~Ca\~nas-Herrera\orcid{0000-0003-2796-2149}\inst{\ref{aff88},\ref{aff124}}
\and A.~Cappi\inst{\ref{aff4},\ref{aff125}}
\and F.~Caro\inst{\ref{aff40}}
\and C.~S.~Carvalho\inst{\ref{aff100}}
\and T.~Castro\orcid{0000-0002-6292-3228}\inst{\ref{aff26},\ref{aff27},\ref{aff25},\ref{aff111}}
\and K.~C.~Chambers\orcid{0000-0001-6965-7789}\inst{\ref{aff126}}
\and S.~Contarini\orcid{0000-0002-9843-723X}\inst{\ref{aff64}}
\and A.~R.~Cooray\orcid{0000-0002-3892-0190}\inst{\ref{aff127}}
\and G.~Desprez\orcid{0000-0001-8325-1742}\inst{\ref{aff128}}
\and A.~D\'iaz-S\'anchez\orcid{0000-0003-0748-4768}\inst{\ref{aff129}}
\and J.~J.~Diaz\inst{\ref{aff130}}
\and S.~Di~Domizio\orcid{0000-0003-2863-5895}\inst{\ref{aff30},\ref{aff31}}
\and H.~Dole\orcid{0000-0002-9767-3839}\inst{\ref{aff58}}
\and S.~Escoffier\orcid{0000-0002-2847-7498}\inst{\ref{aff75}}
\and A.~G.~Ferrari\orcid{0009-0005-5266-4110}\inst{\ref{aff46},\ref{aff5}}
\and P.~G.~Ferreira\orcid{0000-0002-3021-2851}\inst{\ref{aff131}}
\and I.~Ferrero\orcid{0000-0002-1295-1132}\inst{\ref{aff10}}
\and A.~Finoguenov\orcid{0000-0002-4606-5403}\inst{\ref{aff77}}
\and F.~Fornari\orcid{0000-0003-2979-6738}\inst{\ref{aff103}}
\and L.~Gabarra\orcid{0000-0002-8486-8856}\inst{\ref{aff131}}
\and K.~Ganga\orcid{0000-0001-8159-8208}\inst{\ref{aff83}}
\and J.~Garc\'ia-Bellido\orcid{0000-0002-9370-8360}\inst{\ref{aff112}}
\and T.~Gasparetto\orcid{0000-0002-7913-4866}\inst{\ref{aff26}}
\and V.~Gautard\inst{\ref{aff132}}
\and E.~Gaztanaga\orcid{0000-0001-9632-0815}\inst{\ref{aff7},\ref{aff45},\ref{aff2}}
\and F.~Giacomini\orcid{0000-0002-3129-2814}\inst{\ref{aff5}}
\and F.~Gianotti\orcid{0000-0003-4666-119X}\inst{\ref{aff4}}
\and G.~Gozaliasl\orcid{0000-0002-0236-919X}\inst{\ref{aff133}}
\and C.~M.~Gutierrez\orcid{0000-0001-7854-783X}\inst{\ref{aff134}}
\and A.~Hall\orcid{0000-0002-3139-8651}\inst{\ref{aff17}}
\and H.~Hildebrandt\orcid{0000-0002-9814-3338}\inst{\ref{aff135}}
\and J.~Hjorth\orcid{0000-0002-4571-2306}\inst{\ref{aff89}}
\and A.~Jimenez~Mu\~noz\orcid{0009-0004-5252-185X}\inst{\ref{aff136}}
\and S.~Joudaki\orcid{0000-0001-8820-673X}\inst{\ref{aff2}}
\and J.~J.~E.~Kajava\orcid{0000-0002-3010-8333}\inst{\ref{aff137},\ref{aff138}}
\and V.~Kansal\orcid{0000-0002-4008-6078}\inst{\ref{aff139},\ref{aff140}}
\and D.~Karagiannis\orcid{0000-0002-4927-0816}\inst{\ref{aff141},\ref{aff142}}
\and C.~C.~Kirkpatrick\inst{\ref{aff74}}
\and S.~Kruk\orcid{0000-0001-8010-8879}\inst{\ref{aff49}}
\and J.~Le~Graet\orcid{0000-0001-6523-7971}\inst{\ref{aff75}}
\and L.~Legrand\orcid{0000-0003-0610-5252}\inst{\ref{aff143}}
\and J.~Lesgourgues\orcid{0000-0001-7627-353X}\inst{\ref{aff44}}
\and T.~I.~Liaudat\orcid{0000-0002-9104-314X}\inst{\ref{aff144}}
\and A.~Loureiro\orcid{0000-0002-4371-0876}\inst{\ref{aff145},\ref{aff146}}
\and G.~Maggio\orcid{0000-0003-4020-4836}\inst{\ref{aff26}}
\and M.~Magliocchetti\orcid{0000-0001-9158-4838}\inst{\ref{aff60}}
\and F.~Mannucci\orcid{0000-0002-4803-2381}\inst{\ref{aff147}}
\and R.~Maoli\orcid{0000-0002-6065-3025}\inst{\ref{aff148},\ref{aff40}}
\and C.~J.~A.~P.~Martins\orcid{0000-0002-4886-9261}\inst{\ref{aff149},\ref{aff35}}
\and L.~Maurin\orcid{0000-0002-8406-0857}\inst{\ref{aff58}}
\and R.~B.~Metcalf\orcid{0000-0003-3167-2574}\inst{\ref{aff82},\ref{aff4}}
\and M.~Migliaccio\inst{\ref{aff150},\ref{aff151}}
\and M.~Miluzio\inst{\ref{aff49},\ref{aff152}}
\and P.~Monaco\orcid{0000-0003-2083-7564}\inst{\ref{aff122},\ref{aff26},\ref{aff27},\ref{aff25}}
\and A.~Montoro\orcid{0000-0003-4730-8590}\inst{\ref{aff7},\ref{aff45}}
\and A.~Mora\orcid{0000-0002-1922-8529}\inst{\ref{aff153}}
\and C.~Moretti\orcid{0000-0003-3314-8936}\inst{\ref{aff28},\ref{aff111},\ref{aff26},\ref{aff25},\ref{aff27}}
\and G.~Morgante\inst{\ref{aff4}}
\and S.~Nadathur\orcid{0000-0001-9070-3102}\inst{\ref{aff2}}
\and L.~Patrizii\inst{\ref{aff5}}
\and V.~Popa\orcid{0000-0002-9118-8330}\inst{\ref{aff95}}
\and P.~Reimberg\orcid{0000-0003-3410-0280}\inst{\ref{aff85}}
\and I.~Risso\orcid{0000-0003-2525-7761}\inst{\ref{aff154}}
\and P.-F.~Rocci\inst{\ref{aff58}}
\and M.~Sahl\'en\orcid{0000-0003-0973-4804}\inst{\ref{aff155}}
\and E.~Sarpa\orcid{0000-0002-1256-655X}\inst{\ref{aff28},\ref{aff111},\ref{aff27}}
\and A.~Schneider\orcid{0000-0001-7055-8104}\inst{\ref{aff1}}
\and M.~Sereno\orcid{0000-0003-0302-0325}\inst{\ref{aff4},\ref{aff5}}
\and A.~Silvestri\orcid{0000-0001-6904-5061}\inst{\ref{aff124}}
\and A.~Spurio~Mancini\orcid{0000-0001-5698-0990}\inst{\ref{aff156},\ref{aff157}}
\and K.~Tanidis\inst{\ref{aff131}}
\and C.~Tao\orcid{0000-0001-7961-8177}\inst{\ref{aff75}}
\and N.~Tessore\orcid{0000-0002-9696-7931}\inst{\ref{aff73}}
\and G.~Testera\inst{\ref{aff31}}
\and R.~Teyssier\orcid{0000-0001-7689-0933}\inst{\ref{aff158}}
\and S.~Toft\orcid{0000-0003-3631-7176}\inst{\ref{aff159},\ref{aff160}}
\and S.~Tosi\orcid{0000-0002-7275-9193}\inst{\ref{aff30},\ref{aff31}}
\and A.~Troja\orcid{0000-0003-0239-4595}\inst{\ref{aff96},\ref{aff59}}
\and M.~Tucci\inst{\ref{aff57}}
\and C.~Valieri\inst{\ref{aff5}}
\and J.~Valiviita\orcid{0000-0001-6225-3693}\inst{\ref{aff77},\ref{aff78}}
\and D.~Vergani\orcid{0000-0003-0898-2216}\inst{\ref{aff4}}
\and G.~Verza\orcid{0000-0002-1886-8348}\inst{\ref{aff161},\ref{aff162}}
\and P.~Vielzeuf\orcid{0000-0003-2035-9339}\inst{\ref{aff75}}
\and N.~A.~Walton\orcid{0000-0003-3983-8778}\inst{\ref{aff108}}}
										   
%%%% please do not edit the affiliation list -- contact ECEB Bureau for changes
\institute{Department of Astrophysics, University of Zurich, Winterthurerstrasse 190, 8057 Zurich, Switzerland\label{aff1}
\and
Institute of Cosmology and Gravitation, University of Portsmouth, Portsmouth PO1 3FX, UK\label{aff2}
\and
Dipartimento di Fisica e Astronomia, Universit\`a di Bologna, Via Gobetti 93/2, 40129 Bologna, Italy\label{aff3}
\and
INAF-Osservatorio di Astrofisica e Scienza dello Spazio di Bologna, Via Piero Gobetti 93/3, 40129 Bologna, Italy\label{aff4}
\and
INFN-Sezione di Bologna, Viale Berti Pichat 6/2, 40127 Bologna, Italy\label{aff5}
\and
Max Planck Institute for Gravitational Physics (Albert Einstein Institute), Am Muhlenberg 1, D-14476 Potsdam-Golm, Germany\label{aff6}
\and
Institute of Space Sciences (ICE, CSIC), Campus UAB, Carrer de Can Magrans, s/n, 08193 Barcelona, Spain\label{aff7}
\and
Institut de Ciencies de l'Espai (IEEC-CSIC), Campus UAB, Carrer de Can Magrans, s/n Cerdanyola del Vall\'es, 08193 Barcelona, Spain\label{aff8}
\and
Laboratoire Univers et Th\'eorie, Observatoire de Paris, Universit\'e PSL, Universit\'e Paris Cit\'e, CNRS, 92190 Meudon, France\label{aff9}
\and
Institute of Theoretical Astrophysics, University of Oslo, P.O. Box 1029 Blindern, 0315 Oslo, Norway\label{aff10}
\and
Jet Propulsion Laboratory, California Institute of Technology, 4800 Oak Grove Drive, Pasadena, CA, 91109, USA\label{aff11}
\and
Max-Planck-Institut f\"ur Astrophysik, Karl-Schwarzschild-Str.~1, 85748 Garching, Germany\label{aff12}
\and
Department of Physics, Institute for Computational Cosmology, Durham University, South Road, DH1 3LE, UK\label{aff13}
\and
Jodrell Bank Centre for Astrophysics, Department of Physics and Astronomy, University of Manchester, Oxford Road, Manchester M13 9PL, UK\label{aff14}
\and
INAF-IASF Milano, Via Alfonso Corti 12, 20133 Milano, Italy\label{aff15}
\and
Istituto Nazionale di Fisica Nucleare, Sezione di Bologna, Via Irnerio 46, 40126 Bologna, Italy\label{aff16}
\and
Institute for Astronomy, University of Edinburgh, Royal Observatory, Blackford Hill, Edinburgh EH9 3HJ, UK\label{aff17}
\and
Higgs Centre for Theoretical Physics, School of Physics and Astronomy, The University of Edinburgh, Edinburgh EH9 3FD, UK\label{aff18}
\and
Institut f\"ur Theoretische Physik, University of Heidelberg, Philosophenweg 16, 69120 Heidelberg, Germany\label{aff19}
\and
Institut de Recherche en Astrophysique et Plan\'etologie (IRAP), Universit\'e de Toulouse, CNRS, UPS, CNES, 14 Av. Edouard Belin, 31400 Toulouse, France\label{aff20}
\and
Universit\'e St Joseph; Faculty of Sciences, Beirut, Lebanon\label{aff21}
\and
Institut de Physique Th\'eorique, CEA, CNRS, Universit\'e Paris-Saclay 91191 Gif-sur-Yvette Cedex, France\label{aff22}
\and
School of Mathematics and Physics, University of Surrey, Guildford, Surrey, GU2 7XH, UK\label{aff23}
\and
INAF-Osservatorio Astronomico di Brera, Via Brera 28, 20122 Milano, Italy\label{aff24}
\and
IFPU, Institute for Fundamental Physics of the Universe, via Beirut 2, 34151 Trieste, Italy\label{aff25}
\and
INAF-Osservatorio Astronomico di Trieste, Via G. B. Tiepolo 11, 34143 Trieste, Italy\label{aff26}
\and
INFN, Sezione di Trieste, Via Valerio 2, 34127 Trieste TS, Italy\label{aff27}
\and
SISSA, International School for Advanced Studies, Via Bonomea 265, 34136 Trieste TS, Italy\label{aff28}
\and
INAF-Osservatorio Astrofisico di Torino, Via Osservatorio 20, 10025 Pino Torinese (TO), Italy\label{aff29}
\and
Dipartimento di Fisica, Universit\`a di Genova, Via Dodecaneso 33, 16146, Genova, Italy\label{aff30}
\and
INFN-Sezione di Genova, Via Dodecaneso 33, 16146, Genova, Italy\label{aff31}
\and
Department of Physics "E. Pancini", University Federico II, Via Cinthia 6, 80126, Napoli, Italy\label{aff32}
\and
INAF-Osservatorio Astronomico di Capodimonte, Via Moiariello 16, 80131 Napoli, Italy\label{aff33}
\and
INFN section of Naples, Via Cinthia 6, 80126, Napoli, Italy\label{aff34}
\and
Instituto de Astrof\'isica e Ci\^encias do Espa\c{c}o, Universidade do Porto, CAUP, Rua das Estrelas, PT4150-762 Porto, Portugal\label{aff35}
\and
Faculdade de Ci\^encias da Universidade do Porto, Rua do Campo de Alegre, 4150-007 Porto, Portugal\label{aff36}
\and
Aix-Marseille Universit\'e, CNRS, CNES, LAM, Marseille, France\label{aff37}
\and
Dipartimento di Fisica, Universit\`a degli Studi di Torino, Via P. Giuria 1, 10125 Torino, Italy\label{aff38}
\and
INFN-Sezione di Torino, Via P. Giuria 1, 10125 Torino, Italy\label{aff39}
\and
INAF-Osservatorio Astronomico di Roma, Via Frascati 33, 00078 Monteporzio Catone, Italy\label{aff40}
\and
INFN-Sezione di Roma, Piazzale Aldo Moro, 2 - c/o Dipartimento di Fisica, Edificio G. Marconi, 00185 Roma, Italy\label{aff41}
\and
Centro de Investigaciones Energ\'eticas, Medioambientales y Tecnol\'ogicas (CIEMAT), Avenida Complutense 40, 28040 Madrid, Spain\label{aff42}
\and
Port d'Informaci\'{o} Cient\'{i}fica, Campus UAB, C. Albareda s/n, 08193 Bellaterra (Barcelona), Spain\label{aff43}
\and
Institute for Theoretical Particle Physics and Cosmology (TTK), RWTH Aachen University, 52056 Aachen, Germany\label{aff44}
\and
Institut d'Estudis Espacials de Catalunya (IEEC),  Edifici RDIT, Campus UPC, 08860 Castelldefels, Barcelona, Spain\label{aff45}
\and
Dipartimento di Fisica e Astronomia "Augusto Righi" - Alma Mater Studiorum Universit\`a di Bologna, Viale Berti Pichat 6/2, 40127 Bologna, Italy\label{aff46}
\and
Instituto de Astrof\'isica de Canarias, Calle V\'ia L\'actea s/n, 38204, San Crist\'obal de La Laguna, Tenerife, Spain\label{aff47}
\and
European Space Agency/ESRIN, Largo Galileo Galilei 1, 00044 Frascati, Roma, Italy\label{aff48}
\and
ESAC/ESA, Camino Bajo del Castillo, s/n., Urb. Villafranca del Castillo, 28692 Villanueva de la Ca\~nada, Madrid, Spain\label{aff49}
\and
Universit\'e Claude Bernard Lyon 1, CNRS/IN2P3, IP2I Lyon, UMR 5822, Villeurbanne, F-69100, France\label{aff50}
\and
Institute of Physics, Laboratory of Astrophysics, Ecole Polytechnique F\'ed\'erale de Lausanne (EPFL), Observatoire de Sauverny, 1290 Versoix, Switzerland\label{aff51}
\and
Institut de Ci\`{e}ncies del Cosmos (ICCUB), Universitat de Barcelona (IEEC-UB), Mart\'{i} i Franqu\`{e}s 1, 08028 Barcelona, Spain\label{aff52}
\and
Instituci\'o Catalana de Recerca i Estudis Avan\c{c}ats (ICREA), Passeig de Llu\'{\i}s Companys 23, 08010 Barcelona, Spain\label{aff53}
\and
UCB Lyon 1, CNRS/IN2P3, IUF, IP2I Lyon, 4 rue Enrico Fermi, 69622 Villeurbanne, France\label{aff54}
\and
Departamento de F\'isica, Faculdade de Ci\^encias, Universidade de Lisboa, Edif\'icio C8, Campo Grande, PT1749-016 Lisboa, Portugal\label{aff55}
\and
Instituto de Astrof\'isica e Ci\^encias do Espa\c{c}o, Faculdade de Ci\^encias, Universidade de Lisboa, Campo Grande, 1749-016 Lisboa, Portugal\label{aff56}
\and
Department of Astronomy, University of Geneva, ch. d'Ecogia 16, 1290 Versoix, Switzerland\label{aff57}
\and
Universit\'e Paris-Saclay, CNRS, Institut d'astrophysique spatiale, 91405, Orsay, France\label{aff58}
\and
INFN-Padova, Via Marzolo 8, 35131 Padova, Italy\label{aff59}
\and
INAF-Istituto di Astrofisica e Planetologia Spaziali, via del Fosso del Cavaliere, 100, 00100 Roma, Italy\label{aff60}
\and
Universit\'e Paris-Saclay, Universit\'e Paris Cit\'e, CEA, CNRS, AIM, 91191, Gif-sur-Yvette, France\label{aff61}
\and
FRACTAL S.L.N.E., calle Tulip\'an 2, Portal 13 1A, 28231, Las Rozas de Madrid, Spain\label{aff62}
\and
INAF-Osservatorio Astronomico di Padova, Via dell'Osservatorio 5, 35122 Padova, Italy\label{aff63}
\and
Max Planck Institute for Extraterrestrial Physics, Giessenbachstr. 1, 85748 Garching, Germany\label{aff64}
\and
Universit\"ats-Sternwarte M\"unchen, Fakult\"at f\"ur Physik, Ludwig-Maximilians-Universit\"at M\"unchen, Scheinerstrasse 1, 81679 M\"unchen, Germany\label{aff65}
\and
Dipartimento di Fisica "Aldo Pontremoli", Universit\`a degli Studi di Milano, Via Celoria 16, 20133 Milano, Italy\label{aff66}
\and
Felix Hormuth Engineering, Goethestr. 17, 69181 Leimen, Germany\label{aff67}
\and
Technical University of Denmark, Elektrovej 327, 2800 Kgs. Lyngby, Denmark\label{aff68}
\and
Cosmic Dawn Center (DAWN), Denmark\label{aff69}
\and
Universit\'e Paris-Saclay, CNRS/IN2P3, IJCLab, 91405 Orsay, France\label{aff70}
\and
Max-Planck-Institut f\"ur Astronomie, K\"onigstuhl 17, 69117 Heidelberg, Germany\label{aff71}
\and
NASA Goddard Space Flight Center, Greenbelt, MD 20771, USA\label{aff72}
\and
Department of Physics and Astronomy, University College London, Gower Street, London WC1E 6BT, UK\label{aff73}
\and
Department of Physics and Helsinki Institute of Physics, Gustaf H\"allstr\"omin katu 2, 00014 University of Helsinki, Finland\label{aff74}
\and
Aix-Marseille Universit\'e, CNRS/IN2P3, CPPM, Marseille, France\label{aff75}
\and
Universit\'e de Gen\`eve, D\'epartement de Physique Th\'eorique and Centre for Astroparticle Physics, 24 quai Ernest-Ansermet, CH-1211 Gen\`eve 4, Switzerland\label{aff76}
\and
Department of Physics, P.O. Box 64, 00014 University of Helsinki, Finland\label{aff77}
\and
Helsinki Institute of Physics, Gustaf H{\"a}llstr{\"o}min katu 2, University of Helsinki, Helsinki, Finland\label{aff78}
\and
NOVA optical infrared instrumentation group at ASTRON, Oude Hoogeveensedijk 4, 7991PD, Dwingeloo, The Netherlands\label{aff79}
\and
Centre de Calcul de l'IN2P3/CNRS, 21 avenue Pierre de Coubertin 69627 Villeurbanne Cedex, France\label{aff80}
\and
Universit\"at Bonn, Argelander-Institut f\"ur Astronomie, Auf dem H\"ugel 71, 53121 Bonn, Germany\label{aff81}
\and
Dipartimento di Fisica e Astronomia "Augusto Righi" - Alma Mater Studiorum Universit\`a di Bologna, via Piero Gobetti 93/2, 40129 Bologna, Italy\label{aff82}
\and
Universit\'e Paris Cit\'e, CNRS, Astroparticule et Cosmologie, 75013 Paris, France\label{aff83}
\and
University of Applied Sciences and Arts of Northwestern Switzerland, School of Engineering, 5210 Windisch, Switzerland\label{aff84}
\and
Institut d'Astrophysique de Paris, 98bis Boulevard Arago, 75014, Paris, France\label{aff85}
\and
Institut d'Astrophysique de Paris, UMR 7095, CNRS, and Sorbonne Universit\'e, 98 bis boulevard Arago, 75014 Paris, France\label{aff86}
\and
Institut de F\'{i}sica d'Altes Energies (IFAE), The Barcelona Institute of Science and Technology, Campus UAB, 08193 Bellaterra (Barcelona), Spain\label{aff87}
\and
European Space Agency/ESTEC, Keplerlaan 1, 2201 AZ Noordwijk, The Netherlands\label{aff88}
\and
DARK, Niels Bohr Institute, University of Copenhagen, Jagtvej 155, 2200 Copenhagen, Denmark\label{aff89}
\and
Waterloo Centre for Astrophysics, University of Waterloo, Waterloo, Ontario N2L 3G1, Canada\label{aff90}
\and
Department of Physics and Astronomy, University of Waterloo, Waterloo, Ontario N2L 3G1, Canada\label{aff91}
\and
Perimeter Institute for Theoretical Physics, Waterloo, Ontario N2L 2Y5, Canada\label{aff92}
\and
Space Science Data Center, Italian Space Agency, via del Politecnico snc, 00133 Roma, Italy\label{aff93}
\and
Centre National d'Etudes Spatiales -- Centre spatial de Toulouse, 18 avenue Edouard Belin, 31401 Toulouse Cedex 9, France\label{aff94}
\and
Institute of Space Science, Str. Atomistilor, nr. 409 M\u{a}gurele, Ilfov, 077125, Romania\label{aff95}
\and
Dipartimento di Fisica e Astronomia "G. Galilei", Universit\`a di Padova, Via Marzolo 8, 35131 Padova, Italy\label{aff96}
\and
Departamento de F\'isica, FCFM, Universidad de Chile, Blanco Encalada 2008, Santiago, Chile\label{aff97}
\and
Universit\"at Innsbruck, Institut f\"ur Astro- und Teilchenphysik, Technikerstr. 25/8, 6020 Innsbruck, Austria\label{aff98}
\and
Satlantis, University Science Park, Sede Bld 48940, Leioa-Bilbao, Spain\label{aff99}
\and
Instituto de Astrof\'isica e Ci\^encias do Espa\c{c}o, Faculdade de Ci\^encias, Universidade de Lisboa, Tapada da Ajuda, 1349-018 Lisboa, Portugal\label{aff100}
\and
Universidad Polit\'ecnica de Cartagena, Departamento de Electr\'onica y Tecnolog\'ia de Computadoras,  Plaza del Hospital 1, 30202 Cartagena, Spain\label{aff101}
\and
Kapteyn Astronomical Institute, University of Groningen, PO Box 800, 9700 AV Groningen, The Netherlands\label{aff102}
\and
INFN-Bologna, Via Irnerio 46, 40126 Bologna, Italy\label{aff103}
\and
Dipartimento di Fisica, Universit\`a degli studi di Genova, and INFN-Sezione di Genova, via Dodecaneso 33, 16146, Genova, Italy\label{aff104}
\and
Infrared Processing and Analysis Center, California Institute of Technology, Pasadena, CA 91125, USA\label{aff105}
\and
INAF, Istituto di Radioastronomia, Via Piero Gobetti 101, 40129 Bologna, Italy\label{aff106}
\and
Astronomical Observatory of the Autonomous Region of the Aosta Valley (OAVdA), Loc. Lignan 39, I-11020, Nus (Aosta Valley), Italy\label{aff107}
\and
Institute of Astronomy, University of Cambridge, Madingley Road, Cambridge CB3 0HA, UK\label{aff108}
\and
School of Physics and Astronomy, Cardiff University, The Parade, Cardiff, CF24 3AA, UK\label{aff109}
\and
Junia, EPA department, 41 Bd Vauban, 59800 Lille, France\label{aff110}
\and
ICSC - Centro Nazionale di Ricerca in High Performance Computing, Big Data e Quantum Computing, Via Magnanelli 2, Bologna, Italy\label{aff111}
\and
Instituto de F\'isica Te\'orica UAM-CSIC, Campus de Cantoblanco, 28049 Madrid, Spain\label{aff112}
\and
CERCA/ISO, Department of Physics, Case Western Reserve University, 10900 Euclid Avenue, Cleveland, OH 44106, USA\label{aff113}
\and
INFN-Sezione di Milano, Via Celoria 16, 20133 Milano, Italy\label{aff114}
\and
Departamento de F{\'\i}sica Fundamental. Universidad de Salamanca. Plaza de la Merced s/n. 37008 Salamanca, Spain\label{aff115}
\and
Departamento de Astrof\'isica, Universidad de La Laguna, 38206, La Laguna, Tenerife, Spain\label{aff116}
\and
Dipartimento di Fisica e Scienze della Terra, Universit\`a degli Studi di Ferrara, Via Giuseppe Saragat 1, 44122 Ferrara, Italy\label{aff117}
\and
Istituto Nazionale di Fisica Nucleare, Sezione di Ferrara, Via Giuseppe Saragat 1, 44122 Ferrara, Italy\label{aff118}
\and
Center for Data-Driven Discovery, Kavli IPMU (WPI), UTIAS, The University of Tokyo, Kashiwa, Chiba 277-8583, Japan\label{aff119}
\and
Ludwig-Maximilians-University, Schellingstrasse 4, 80799 Munich, Germany\label{aff120}
\and
Max-Planck-Institut f\"ur Physik, Boltzmannstr. 8, 85748 Garching, Germany\label{aff121}
\and
Dipartimento di Fisica - Sezione di Astronomia, Universit\`a di Trieste, Via Tiepolo 11, 34131 Trieste, Italy\label{aff122}
\and
Minnesota Institute for Astrophysics, University of Minnesota, 116 Church St SE, Minneapolis, MN 55455, USA\label{aff123}
\and
Institute Lorentz, Leiden University, Niels Bohrweg 2, 2333 CA Leiden, The Netherlands\label{aff124}
\and
Universit\'e C\^{o}te d'Azur, Observatoire de la C\^{o}te d'Azur, CNRS, Laboratoire Lagrange, Bd de l'Observatoire, CS 34229, 06304 Nice cedex 4, France\label{aff125}
\and
Institute for Astronomy, University of Hawaii, 2680 Woodlawn Drive, Honolulu, HI 96822, USA\label{aff126}
\and
Department of Physics \& Astronomy, University of California Irvine, Irvine CA 92697, USA\label{aff127}
\and
Department of Astronomy \& Physics and Institute for Computational Astrophysics, Saint Mary's University, 923 Robie Street, Halifax, Nova Scotia, B3H 3C3, Canada\label{aff128}
\and
Departamento F\'isica Aplicada, Universidad Polit\'ecnica de Cartagena, Campus Muralla del Mar, 30202 Cartagena, Murcia, Spain\label{aff129}
\and
Instituto de Astrof\'isica de Canarias (IAC); Departamento de Astrof\'isica, Universidad de La Laguna (ULL), 38200, La Laguna, Tenerife, Spain\label{aff130}
\and
Department of Physics, Oxford University, Keble Road, Oxford OX1 3RH, UK\label{aff131}
\and
CEA Saclay, DFR/IRFU, Service d'Astrophysique, Bat. 709, 91191 Gif-sur-Yvette, France\label{aff132}
\and
Department of Computer Science, Aalto University, PO Box 15400, Espoo, FI-00 076, Finland\label{aff133}
\and
Instituto de Astrof\'\i sica de Canarias, c/ Via Lactea s/n, La Laguna E-38200, Spain. Departamento de Astrof\'\i sica de la Universidad de La Laguna, Avda. Francisco Sanchez, La Laguna, E-38200, Spain\label{aff134}
\and
Ruhr University Bochum, Faculty of Physics and Astronomy, Astronomical Institute (AIRUB), German Centre for Cosmological Lensing (GCCL), 44780 Bochum, Germany\label{aff135}
\and
Univ. Grenoble Alpes, CNRS, Grenoble INP, LPSC-IN2P3, 53, Avenue des Martyrs, 38000, Grenoble, France\label{aff136}
\and
Department of Physics and Astronomy, Vesilinnantie 5, 20014 University of Turku, Finland\label{aff137}
\and
Serco for European Space Agency (ESA), Camino bajo del Castillo, s/n, Urbanizacion Villafranca del Castillo, Villanueva de la Ca\~nada, 28692 Madrid, Spain\label{aff138}
\and
ARC Centre of Excellence for Dark Matter Particle Physics, Melbourne, Australia\label{aff139}
\and
Centre for Astrophysics \& Supercomputing, Swinburne University of Technology,  Hawthorn, Victoria 3122, Australia\label{aff140}
\and
School of Physics and Astronomy, Queen Mary University of London, Mile End Road, London E1 4NS, UK\label{aff141}
\and
Department of Physics and Astronomy, University of the Western Cape, Bellville, Cape Town, 7535, South Africa\label{aff142}
\and
ICTP South American Institute for Fundamental Research, Instituto de F\'{\i}sica Te\'orica, Universidade Estadual Paulista, S\~ao Paulo, Brazil\label{aff143}
\and
IRFU, CEA, Universit\'e Paris-Saclay 91191 Gif-sur-Yvette Cedex, France\label{aff144}
\and
Oskar Klein Centre for Cosmoparticle Physics, Department of Physics, Stockholm University, Stockholm, SE-106 91, Sweden\label{aff145}
\and
Astrophysics Group, Blackett Laboratory, Imperial College London, London SW7 2AZ, UK\label{aff146}
\and
INAF-Osservatorio Astrofisico di Arcetri, Largo E. Fermi 5, 50125, Firenze, Italy\label{aff147}
\and
Dipartimento di Fisica, Sapienza Universit\`a di Roma, Piazzale Aldo Moro 2, 00185 Roma, Italy\label{aff148}
\and
Centro de Astrof\'{\i}sica da Universidade do Porto, Rua das Estrelas, 4150-762 Porto, Portugal\label{aff149}
\and
Dipartimento di Fisica, Universit\`a di Roma Tor Vergata, Via della Ricerca Scientifica 1, Roma, Italy\label{aff150}
\and
INFN, Sezione di Roma 2, Via della Ricerca Scientifica 1, Roma, Italy\label{aff151}
\and
HE Space for European Space Agency (ESA), Camino bajo del Castillo, s/n, Urbanizacion Villafranca del Castillo, Villanueva de la Ca\~nada, 28692 Madrid, Spain\label{aff152}
\and
Aurora Technology for European Space Agency (ESA), Camino bajo del Castillo, s/n, Urbanizacion Villafranca del Castillo, Villanueva de la Ca\~nada, 28692 Madrid, Spain\label{aff153}
\and
INAF-Osservatorio Astronomico di Brera, Via Brera 28, 20122 Milano, Italy, and INFN-Sezione di Genova, Via Dodecaneso 33, 16146, Genova, Italy\label{aff154}
\and
Theoretical astrophysics, Department of Physics and Astronomy, Uppsala University, Box 515, 751 20 Uppsala, Sweden\label{aff155}
\and
Department of Physics, Royal Holloway, University of London, TW20 0EX, UK\label{aff156}
\and
Mullard Space Science Laboratory, University College London, Holmbury St Mary, Dorking, Surrey RH5 6NT, UK\label{aff157}
\and
Department of Astrophysical Sciences, Peyton Hall, Princeton University, Princeton, NJ 08544, USA\label{aff158}
\and
Cosmic Dawn Center (DAWN)\label{aff159}
\and
Niels Bohr Institute, University of Copenhagen, Jagtvej 128, 2200 Copenhagen, Denmark\label{aff160}
\and
Center for Cosmology and Particle Physics, Department of Physics, New York University, New York, NY 10003, USA\label{aff161}
\and
Center for Computational Astrophysics, Flatiron Institute, 162 5th Avenue, 10010, New York, NY, USA\label{aff162}}    
    
    \authorrunning{Euclid Collaboration}

   \date{Received XXX; accepted ZZZ}

% \abstract{}{}{}{}{} 
% 5 {} token are mandatory
 
  \abstract{To constrain models beyond $\Lambda$CDM, the development of the \Euclid analysis pipeline requires simulations that capture the nonlinear phenomenology of such models.
  We present an overview of numerical methods and $N$-body simulation codes developed to study the nonlinear regime of structure formation in alternative dark energy and modified gravity theories. We review a variety of numerical techniques and approximations employed in cosmological $N$-body simulations to model the complex phenomenology of scenarios beyond $\Lambda$CDM. This includes discussions on solving nonlinear field equations, accounting for fifth forces, and implementing screening mechanisms. Furthermore, we conduct a code comparison exercise to assess the reliability and convergence of different simulation codes across a range of models. Our analysis demonstrates a high degree of agreement among the outputs of different simulation codes, providing confidence in current numerical methods for modelling cosmic structure formation beyond $\Lambda$CDM. We highlight recent advances made in simulating the nonlinear scales of structure formation, which are essential for leveraging the full scientific potential of the forthcoming observational data from the \Euclid mission.
   }


   \maketitle
%
%-------------------------------------------------------------------

\section{Introduction}
Significant progress in cosmological observations is expected
in the upcoming years, in particular from the \Euclid 
survey \citep{2011arXiv1110.3193L,Euclid:2024yrr,Euclid:2021icp,Euclid:2021frk,Euclid:2021cfn,Euclid:2022dbc},  Vera Rubin Observatory’s Legacy Survey of Space and Time \citep[LSST,][]{LSST:2008ijt}, the Roman Space Telescope \citep{2015arXiv150303757S} and the Dark Energy Spectroscopic Instrument \citep[DESI,][]{DESI:2016fyo}. 
These surveys will offer precision observations to high redshifts, allowing
us to study the evolution of the Universe with unprecedented accuracy and potentially uncover the nature of dark matter and dark energy (DE). 
Gaining a deeper understanding of the nature of DE and addressing the long-standing question of whether the cosmological constant ($\Lambda$) is responsible for the late-time accelerated expansion of the Universe is indeed one of the primary goals of the \textit{Euclid} survey \citep{Amendola:2016saw}.

The \Euclid space telescope was launched on July 1, 2023, and is going to observe billions of galaxies out to redshift $z \approx 2$, covering more than a third of the sky in optical and near-infrared wavelengths. \Euclid will deliver precise measurements of the shapes and redshifts of galaxies \citep{2022A&A...657A..90E, 2023A&A...671A.102E, 2023A&A...671A.101E, Euclid:2020gbk, Euclid:2021upd}, from which we will measure weak gravitational lensing \citep{Euclid:2023uha} and galaxy clustering \citep{Euclid:2019bue}. 
These primary probes can be used to rigorously investigate different cosmological scenarios, in particular those related to DE that go beyond the $\Lambda$-Cold-Dark-Matter ($\Lambda$CDM) concordance model.

Although the $\Lambda$CDM model is generally very successful in matching observations, the true identities of CDM and the cosmological constant $\Lambda$ remain unknown. Additionally, some tensions have persisted in recent years, most notably the Hubble tension \citep[see][for a summary and references]{DiValentino:2020zio} where local measurements of the Hubble parameter today, $H_0$, appear to disagree with those inferred from high-redshift observations by around $5\,\sigma$. Further examples are the $S_8$ tension \citep[see][for a summary and references]{DiValentino:2020vvd} and some anomalies found in measurements of the cosmic microwave background \citep{Abdalla:2022yfr}.
The presence of these tensions
may hint at a breakdown of the $\Lambda$CDM model and further motivates the exploration of alternative scenarios.

Over the past few years, cosmologists have explored different possibilities to account for the late-time accelerating expansion of the Universe \citep{Tsujikawa:2010zza, Clifton:2011jh, Joyce:2016vqv} either by introducing a new field, referred to as the DE field or by proposing a modified theory of gravity (MG). A wide range of MG or DE models is equivalent to adding a new light scalar degree of freedom to the theory of General Relativity (GR).

In these theories, the scalar degree of freedom exhibits
time evolution, sometimes accompanied by
spatial fluctuations within the cosmic horizon. Even in the absence of such fluctuations,
the background evolution may be different from $\Lambda$CDM,
leading to modifications in structure formation. 
Significant spatial
fluctuations in these models may arise due to various factors, including a low characteristic speed of sound in the theory 
\citep{Gleyzes:2014rba, Hassani:2019lmy}, or as a result of the non-minimal coupling of the scalar field to matter or gravity
\citep[see][for an example]{Amendola:2003wa}. 
MG and DE theories featuring a coupling of the scalar field to matter
can further affect perturbations at sub-horizon scales by mediating a fifth force. 
If the coupling is universal and includes baryons, a screening mechanism is essential to 
evade the precise constraints of local experiments \citep{Will:2014kxa}. Screening mechanisms are typically achieved through nonlinear phenomena in such theories.
If, on the other hand, the coupling to matter is non-universal and is confined entirely to the dark sector, local experiments have no constraining power, and cosmological observations provide the main constraints.

Given the diversity of possible DE or MG scenarios, a large information gain is expected from nonlinear scales in the cosmological large-scale structure. These scales must be studied using $N$-body simulations that capture the essential aspects of the DE or MG models under consideration. This usually means that at least one additional equation needs to be solved for the extra degree of freedom. In many cases, this leads to a difficult nonlinear problem that could require special techniques or approximations that need to be developed. This makes $N$-body simulations for models of DE or MG a challenging task.

In this paper, we first review the main features of the different classes of DE and MG models that have been proposed over the past years \citep[see also][and Frusciante et al.\ in prep.\ for a more comprehensive and detailed overview]{Amendola:2016saw}. For each of them, we then discuss the
numerical methods implemented within a selection of existing $N$-body codes
(summarised in Table~\ref{tab:codes}). 
Focusing on MG models with a universal coupling, we then compare the results of different $N$-body implementations for two well-studied theories, namely the Hu--Sawicki $f(R)$ gravity \citep{Hu:2007nk} and the `normal branch' of the Dvali--Gabadadze--Porrati braneworld model \citep[nDGP,][]{Dvali:2000hr,Schmidt:2009sv}. 
We choose simulation parameters following the code comparison paper by \citet{Winther:2015wla} [W15 hereafter], allowing us to validate a number of new codes against existing results.

This article is part of a series that collectively explores simulations and nonlinearities beyond the $\Lambda$CDM model:
\begin{enumerate}
    \item Numerical methods and validation (this work).
    \item Results from non-standard simulations (R\'acz et al.\ in prep.).
    \item Cosmological constraints on non-standard cosmologies from simulated Euclid probes (D'Amico et al.\ in prep.).
    \item Constraints on $f(R)$ models from the photometric primary probes (Koyama et al.\ in prep.).
\end{enumerate}
For further details, see our companion papers. The purpose of this first article in the series is to serve as a reference for models beyond $\Lambda$CDM and their existing implementations in various codes. This
paper is structured as follows. In Sect.\,\ref{sec:Methods} we give a broad overview of different numerical approaches to treat the additional physics of models beyond $\Lambda$CDM. In Sect.\,\ref{sec:Validation} we discuss a number of different codes that implement those approaches and carry out a validation exercise, comparing several recently developed codes with the existing state of the art. We conclude in Sect.\,\ref{sec:Conclusion}. In an Appendix, we discuss some performance considerations.

%--------------------------------------------------------------------

\section{Methods \label{sec:Methods}}

\subsection{Non-standard background evolution}

A wide range of models beyond the simplest cosmological constant scenario are based on an additional scalar degree of freedom -- e.g.\ a classical scalar field $\phi$ -- that evolves dynamically in the expanding Universe and whose background energy density $\rho_\phi$ provides the source for the observed DE abundance. To induce cosmic acceleration and to match existing constraints on the background expansion history, the equation-of-state parameter $w$ of such an additional field must be sufficiently negative at recent epochs but is poorly constrained at earlier times, which allows for models where $w$ also evolves dynamically as long as it converges to values close to $w \approx -1$ in the late Universe. For these models, the DE component modifies the background expansion history of the Universe, which is encoded by the general expression of the Hubble function,
\begin{equation}
    \frac{H^{2}(a)}{H^{2}_{0}} = \Om{m} a^{-3} + \Om{r} a^{-4} + \Om{k} a^{-2} + \Om{DE} \mathrm{e}^{ -3\!\int _{1}^{a}\frac{1+w(a')}{a'}\mathrm{d}a' }\,,
\label{eq:hubble}
\end{equation}
where the equation-of-state parameter of DE can be obtained by solving the background field equations -- including the evolution of the additional scalar degree of freedom $\phi$ -- or can be parameterised. A common parameterisation suggested by Chevallier, Polarski \& Linder \citep[CPL,][]{Chevallier:2000qy,Linder:2002et} is based on the desired evolution of $w$ at low redshifts,
\begin{equation}
    w(a) = w_{0} + w_{a} (1 - a)\,.
\end{equation}
Alternatively, one can set the desired relative abundance of DE at late ($\Om{DE} = 1-\Om{m}$) and early ($\Omega _{\rm EDE}$) epochs as in the Early Dark Energy \citep[EDE,][]{Wetterich:2004pv} parameterisation,
\begin{equation}
    w(a) = \frac{w_{0}}{1 + b\ln (1/a)}\,, \qquad
    b = \frac{3w_{0}}{\ln \frac{1 - \Omega _{\rm EDE}}{\Omega _{\rm EDE}} + \ln \frac{1 - \Om{m}}{\Om{m}}}\,.
\end{equation}
The modified expansion history expressed by Eq.~\eqref{eq:hubble} will indirectly affect the evolution of matter density fluctuations and modify the formation process of collapsed structures by changing the Hubble friction term in the equation for linear matter perturbations, which in Newtonian gauge and in Fourier space for sub-horizon scales reads:
\begin{equation}
    \ddot{\delta}_\mathrm{m} + 2H\dot{\delta}_\mathrm{m} = 4\pi G \left( \rh{m}\de{m} + \rh{DE}\de{DE}\right),
\label{eq:linear_growth}
\end{equation}
where $\de{m}$ and $\de{DE}$ are the density contrasts of matter and DE perturbations, respectively, $G$ is  Newton's constant, and a dot represents a derivative with respect to cosmic time.

Besides the richer background dynamics that is endowed by an evolving field, whenever DE is promoted from a cosmological constant to a dynamical degree of freedom, the model also acquires an additional layer of complexity: the presence and evolution of DE fluctuations around the mean-field configuration. This corresponds to the situation where $\delta_\mathrm{DE}$ in Eq.~\eqref{eq:linear_growth} is non-negligible, whereas in $\Lambda $CDM it would vanish identically
at all scales. Like any other density perturbations, inhomogeneities in the DE would then contribute to the peculiar gravitational potential that governs
the evolution of matter perturbations and thus the
formation of cosmic structures as shown by Eq.~\eqref{eq:linear_growth}.

However, in many of the simplest scalar-field scenarios, such perturbations are negligible at sub-horizon scales because the speed of sound $c_\mathrm{s}$ of the scalar field is naturally close to the speed of light. Ignoring them for the purpose of numerical simulations, the only modification of $N$-body algorithms required to simulate these DE models is given by an appropriate calculation of the cosmic expansion rate. The most common approach amounts to tabulating the specific expansion rate of the universe for the model to be simulated according to Eq.~\eqref{eq:hubble} and replacing the standard analytical calculation of the Hubble function within the {\it N}-body algorithm with an interpolated value from the tabulated solution that is provided to the code as an input.
This approach is implemented by most of the simulation codes employed within the Euclid Collaboration to perform cosmological simulations in homogeneous DE models beyond $\Lambda$CDM.

\subsection{Linearised DE perturbations}

Although a wide range of DE models are characterised by negligible DE fluctuations as discussed above, some specific scenarios may not fulfil such a condition at all scales and/or at all times, either because they feature a lower value of the DE speed of sound, allowing DE perturbations to grow on scales above the associated Jeans length that then falls inside the cosmological horizon, or because additional interactions -- besides gravity -- can induce the growth of such perturbations. The former case corresponds to the class of clustering DE models, while the latter is known as coupled DE.

\subsubsection{Clustering DE} 
The clustering DE models are characterised by two time-dependent variables: the speed of sound $c_\mathrm{s}$ and the equation of state parameter $w$. In these theories, the DE component clusters on scales larger than the associated
sound-horizon/Jeans scale, $\lambda_\mathrm{s} = H/c_\mathrm{s}$ and DE perturbations decay quickly below the sound-horizon scale. For a sufficiently small speed of sound, we may even
expect nonlinear DE structures to form. At a fundamental level, clustering DE models are analogous to the k-essence type of theories, so that the action reads \citep{Armendariz-Picon:2000ulo}
\begin{equation}
S = \int \mathrm{d}^4x \sqrt{-g}\left[\frac{ c^4 R}{ 16\pi G}  + P(X, \phi)  + \mathcal{L}_\mathrm{m} \right],
\end{equation}
where $P$ is a general function of the kinetic term $X \equiv -\frac{1}{2} \nabla_{\mu}\phi\nabla^{\mu} \phi$ and the scalar field $\phi$, and $\mathcal{L}_\mathrm{m}$ is the matter Lagrangian. For a given $P(X, \phi)$, the speed of sound and the equation of state are given by \citep{Armendariz-Picon:2000nqq}
\begin{equation}
w=\frac{P}{P-2X P_{,X}}\,, \qquad c_\mathrm{s}^2 = \frac{P_{,X}}{2XP_{,XX}+P_{,X}}\,,
\end{equation}
where the subscript ``$,{X}$'' denotes the partial derivative with respect to $X$. We therefore need to specify the function $P(X, \phi)$ to derive the equations of motion for the k-essence scalar field. However, since there are
many possible choices, we can instead employ the effective field theory (EFT) approach to model the dynamics of k-essence DE. The EFT framework, although
not a fundamental theory, offers several advantages \citep{Gleyzes:2014rba,Cheung:2007st}, such as being a description of a wide range of theories within some scales. The EFT is a perturbative approach based on the assumption that the scalar field perturbations remain small over the scales of interest. It is worth noting that the regime of nonlinear matter clustering is accessible to the EFT framework as long as the scalar field perturbations remain small.
The k-essence theories or clustering DE models are implemented in several $N$-body and Einstein--Boltzmann codes. In \texttt{CLASS} \citep{Lesgourgues:2011re} and \texttt{CAMB} \citep{Lewis:1999bs}, these theories are implemented using the fluid picture.
In \texttt{hi\_class} \citep{Zumalacarregui:2016pph}, the EFT equations are implemented and can be controlled using the EFT parameter $\alpha_\mathrm{K} \equiv 3(1+w)c_\mathrm{s}^{-2}$ within the code. On the other hand, in \texttt{k-evolution} \citep{Hassani:2019lmy, Hassani:2019wed}, which is an $N$-body code based on \texttt{gevolution} \citep{Adamek:2015eda}, nonlinear equations for clustering DE are implemented as an independent component, and the k-essence field for small $c_\mathrm{s}$ can form nonlinear structures. In some $N$-body codes, for example in \texttt{gevolution} \citep{Adamek:2016zes}, clustering DE is implemented through a linear solution from an Einstein--Boltzmann solver. This is a good assumption for large speeds of sound, but for small ones, this method does not allow for the response of DE to the nonlinear matter structures.

\subsubsection{Coupled quintessence}

Moving to the case of coupled DE models, the interaction can be formulated at a fundamental level by introducing a direct coupling between the scalar field and the spatial curvature $R$ in the so-called Jordan frame \citep[see e.g.][]{Pettorino:2008ez}, so that the action reads
\begin{equation}
S = \int \mathrm{d}^4x \sqrt{-g}\left[\frac{c^4 f(\phi, R)}{16\pi G}-\frac{1}{2}Z(\phi)\nabla^{\mu}\phi\nabla_\mu\phi - V(\phi) + \mathcal{L}_\mathrm{m}\right]\,, 
\end{equation}
where $f(\phi, R)$ is a function that couples the scalar field to the curvature, $Z(\phi)$ is a function that allows for non-standard kinetic terms, $V(\phi )$ is the scalar field self-interaction potential, and the matter Lagrangian contains at least one cold species characterised by some rest mass $m_0$.

Alternatively, the interaction can be formulated by including source terms in the covariant conservation equations of the interacting species in the so-called Einstein frame,
\begin{eqnarray}
\label{eq:dm_continuity}
\nabla^\mu T^{(\mathrm{c})}_{\mu\nu} &=& -\frac{\beta_{(\mathrm{c})}(\phi)}{M_{\rm Pl}} T^{(\mathrm{c})}\nabla_{\nu}\phi\,,\\ 
\label{eq:baryon_continuity}
\nabla^\mu T^{(\mathrm{b})}_{\mu\nu} &=& -\frac{\beta_{(\mathrm{b})}(\phi)}{M_{\rm Pl}}T^{(\mathrm{b})}\nabla_{\nu}\phi\,, \\
\label{eq:scalar_continuity}
\nabla^\mu T^{(\phi)}_{\mu\nu} &=& \frac{1}{M_{\rm Pl}}\left[\beta_{(\mathrm{c})}(\phi)T^{(\mathrm{c})} + \beta_{(\mathrm{b})}(\phi)T^{(\mathrm{b})} \right]\nabla_{\nu }\phi\,,
\end{eqnarray}
where $T_{\mu\nu}^{(\mathrm{Y})}$ is the stress-energy tensor of a given species Y, $T^{(\mathrm{Y})}$ is its trace, $\beta_{(\mathrm{Y})}(\phi)$ is the coupling function of species Y, the labels $\mathrm{c},\,\mathrm{b},\,\phi$ refer to the dark matter, baryon, and scalar field species, respectively, and $M_{\rm Pl}\equiv (\hbar c)^{1/2} (8\pi G)^{-1/2}$ is the reduced Planck mass.

While in the former case the interaction will be universal (i.e.\ involving all matter species with the same strength), which goes under the name of Extended Quintessence,  the latter approach allows for non-universal couplings that may selectively involve individual species, for example by separately choosing the coupling functions for baryons and dark matter. 

In the case of a universal coupling (that is, if $\beta_{(\mathrm{b})} = \beta_{(\mathrm{c})}$), the two approaches can be related to one another through a Weyl transformation of the metric \citep[see again][]{Pettorino:2008ez}, and are therefore equivalent.
On the other hand, the possibility to leave the baryonic component of the Universe only minimally coupled evades Solar System constraints \citep[see e.g.][]{Will:2014kxa} on the
deviations from standard gravity thereby avoiding the need for screening mechanisms. This is the case of Coupled Quintessence models \citep[][]{Wetterich:1994bg,Amendola:1999er}, where the direct coupling between the scalar field and massive (non-baryonic) particles can support stable perturbations of the DE field at sub-horizon scales \citep[see e.g.][]{Amendola:2003wa}. In general, such perturbations may even become nonlinear in the presence of a sufficiently strong coupling \citep[as in the case of Growing Neutrino Quintessence models, see e.g.][]{Amendola:2007yx, Mota:2008nj, Baldi:2011mt, Ayaita:2014una}. Nonetheless, a large class of widely studied coupled DE models is known to feature scalar perturbations of the order of the standard Newtonian gravitational potential \citep[$\delta_{\phi} \sim \Phi_\mathrm{N}$, see again][for an extended derivation]{Amendola:2003wa}, thereby remaining in the linear regime at all times and scales of cosmological interest. This allows us to linearise the corresponding field equations and derive modified equations of motions for massive particles, including the contribution of the additional force arising from the direct coupling with the scalar field \citep[see e.g.][]{Baldi:2010vv}. 

In fact, a general feature of coupled DE models is the existence of a `fifth force' mediated by the
scalar field. The new force can be expressed as an additional acceleration
experienced by a massive coupled particle, which in comoving coordinates will be given by
\begin{equation}
\label{eq:fifth-force}
    \vec{a}_{\mathrm{Y},5th} = -\beta_{(\mathrm{Y})}(\phi )\vec{\nabla}\delta \phi\,,
\end{equation}
where $\mathrm{Y}$ identifies a coupled matter species, and $\delta \phi $ is the scalar field fluctuation.
This extra acceleration term is added to the standard Newtonian acceleration acting on all massive particles,
\begin{equation}
    \vec{a}_{\mathrm{N}}= -H\vec{v} -\vec{\nabla }\Phi_\mathrm{N}\,, 
    \label{eq:Newtonian_acceleration}
\end{equation}
where $\vec{v}$ is the peculiar particle velocity in comoving coordinates, and $\Phi_\mathrm{N}$ is the peculiar Newtonian potential obeying the standard Poisson equation 
\begin{equation}
    \nabla ^{2}\Phi_\mathrm{N} = 4\pi G a^2 \sum_\mathrm{Y} \rho_\mathrm{Y} \delta_\mathrm{Y}\,,
\end{equation}
where the sum runs over all clustering species in the Universe.

Therefore, solving for the dynamical evolution of massive coupled particles requires solving for the scalar field perturbation $\delta \phi$ entering in Eq.~\eqref{eq:fifth-force}, which in the most general case follows a 
nonlinear elliptic equation,
\begin{equation}
    \nabla ^{2}\delta \phi = F(\delta \phi ) + \sum_\mathrm{Y} 8\pi Ga^2 \beta_{(\mathrm{Y})}(\phi)\delta_\mathrm{Y}\,,
    \label{eq:field_equation}
\end{equation}
with $F$ a function of the scalar field fluctuation $\delta \phi$, and where the sum runs over all the coupled matter species with their respective couplings $\beta _\mathrm{(Y)}(\phi )$. 

For the particular case of a coupled DE model with a non-universal interaction \citep[][]{Damour:1990tw} involving only dark matter and leaving baryons uncoupled (i.e. $\beta_{(\mathrm{b})} = 0$) the function $F(\delta \phi )$ in Eq.~\eqref{eq:field_equation} is negligible compared to the term associated with matter density perturbations \citep[see][for a derivation]{Amendola:2003wa} and can be safely discarded. As a result, the scalar-field equation reduces to
\begin{equation}
    \nabla ^{2} \delta \phi \approx 8\pi Ga^2 \beta_{(\mathrm{c})}(\phi)\delta_\mathrm{c} = 2\beta_{(\mathrm{c})}(\phi)\nabla ^{2}\Phi_{\mathrm{c}} \,,
\end{equation}
where $\Phi_{\mathrm{c}}$ is the Newtonian potential generated by the distribution of the coupled dark matter particles, that is
\begin{equation}
    \nabla ^{2}\Phi_\mathrm{c} = 4\pi G a^2  \rho_{\mathrm{c}} \delta_{\mathrm{c}}\,.
\end{equation}
Therefore, the solution for the scalar field perturbations will be directly proportional to the potential $\Phi_{\mathrm{c}}$ according to the relation
\begin{equation}
    \delta \phi \approx 2\beta_{(\mathrm{c})} \Phi_{\mathrm{c}}\,.
\end{equation}
The acceleration equation for a coupled particle can then be rewritten as 
\begin{equation}
    \vec{a}_{\mathrm{c}} = -H\vec{v}_{\mathrm{c}} - \vec{\nabla}\Phi_\mathrm{b} - \vec{\nabla} \left(1 + 2\beta_{(\mathrm{c})}^2(\phi)\right)\Phi_\mathrm{c}\,,
    \label{eq:cq_acceleration}
\end{equation}
assuming here for simplicity that other clustering species (such as massive neutrinos) give a negligible contribution to the total Newtonian potential such that $\Phi_\mathrm{N} = \Phi_{\mathrm{c}} + \Phi_{\mathrm{b}}$.
This modified acceleration equation introduces a further modification to be implemented in $N$-body simulation codes for Coupled Quintessence cosmologies besides the specific expansion history of each particular model. This often requires substantial modifications in the gravity solvers of conventional $N$-body codes, as the algorithms need to evolve coupled and uncoupled massive particles (typically dark matter and baryons, respectively) with different equations and should therefore treat these components separately. Even under the approximation of a purely collisionless treatment (i.e., ignoring the hydrodynamical and astrophysical processes that affect standard baryonic matter leading to the formation of stars and galaxies) that is often employed for large-volume simulations targeted at galaxy surveys such as \textit{Euclid}, both coupled and uncoupled matter species must be included in the simulation to provide a consistent representation of the dynamics at all scales: as baryons and dark matter evolve differently, assuming that all matter is dark would lead to an overestimation of the effects of the coupling and, thus, biased results. This approach is implemented in the \texttt{C-Gadget} code \citep[][]{Baldi:2008ay,Baldi:2011qi} which is employed for
Coupled DE simulations performed within the Euclid Collaboration.

Distinguishing between coupled and uncoupled particle types in simulations of Coupled Quintessence is also crucial for proper treatment of two other effects that characterise these cosmological models beyond the fifth force described by Eq.~\eqref{eq:fifth-force}.
The first is the mass variation of coupled particles due to the exchange of rest-frame energy with the DE scalar field, which arises as a direct consequence of the modified continuity equations \eqref{eq:dm_continuity} and of the assumption of particle number conservation. More specifically, the mass of coupled particles evolves as a result of the evolution of the background scalar field according to
\begin{equation}
    m_{\mathrm{Y}}(a) =  m_{0}\exp \left(- \int_{\phi _{0}}^{\phi (a)} \beta_{\mathrm{(Y)}} (\phi )\frac{\mathrm{d}\phi}{M_{\rm Pl}} \right)\,.
\label{eq:mass_variation}
\end{equation}
Such a mass variation, which involves only particle species with a non-vanishing coupling to the scalar field, must be taken into account in $N$-body algorithms by changing the mass of individual simulation particles at every time step. This is normally done by tabulating the mass as a function of scale factor $a$ by numerically integrating Eq.~\eqref{eq:mass_variation} along with the background dynamics of the scalar field $\phi$, and interpolating from that table as the simulation progresses.

The second effect is an additional force (on top of the fifth force) acting on coupled particles as a consequence of momentum conservation due to the particles'  mass variation described by Eq.~\eqref{eq:mass_variation}, which takes the form of a velocity-dependent extra acceleration behaving either as a friction or as a drag, depending on the relative signs of the coupling function $\beta (\phi )$ and of the scalar field velocity $\dot{\phi}$ \citep[see e.g. ][]{Baldi:2008ay},
\begin{equation}
    \vec{a}_{\mathrm{Y},v} = -\frac{\beta _{\mathrm{(Y)}}(\phi)}{M_{\rm Pl}}\dot{\phi }\vec{v}\,.
    \label{eq:friction-term}
\end{equation}
Such a velocity-dependent acceleration is responsible for a very rich phenomenology characterising Coupled Quintessence models, especially on highly nonlinear scales \citep[see e.g.][]{Baldi:2008ay,Baldi:2011qi,Li:2010re,Baldi:2022uwb}, and must be included in $N$-body simulations as well for a fully consistent treatment of these scenarios. This is done by adding the extra acceleration described in Eq.~\eqref{eq:friction-term} to the total acceleration (i.e.\ Newtonian plus fifth force) of all coupled particles in each time step,
\begin{equation}
    \vec{a}_{\mathrm{Y}} = \vec{a}_{\mathrm{N}} + \vec{a}_{\mathrm{Y},5th} + \vec{a}_{\mathrm{Y},v}\,.
\end{equation}
The relevant quantities $\beta (\phi)$ and $\dot{\phi}$ can again be interpolated from a table obtained by integrating the background dynamics of the system. This is the approach implemented in the \texttt{C-Gadget} code that has been used to run Coupled Quintessence simulations within the Euclid Collaboration.

\subsubsection{Momentum exchange and dark scattering}

A further example of interacting DE cosmologies characterised by scalar-field perturbations that always remain linear is given by models of pure momentum exchange \citep[see e.g.][]{Pourtsidou:2013nha,Skordis:2015yra} between the DE field and massive particles like dark matter or baryons. A limiting case is given by the Dark Scattering scenario \citep[][]{Simpson:2010vh} where the momentum transfer between the two components is modelled as the elastic scattering of massive particles moving through a homogeneous DE fluid with equation of state $w$. This results in an extra force acting on the moving massive particles which is proportional to their comoving velocity, similar to the velocity-dependent force described by Eq.~\eqref{eq:friction-term} for Coupled Quintessence models. However, the origin of this force is completely different in this case, as it does not originate from the mass variation of particles but rather from the momentum transfer with the DE field. As a result, the scattering acceleration can be expressed as
\begin{equation}
    \vec{a}_\mathrm{s} = -(1+w)\frac{3H^{2}\Omega_\mathrm{DE}}{8\pi G}\xi \vec{v}\,,
    \label{eq:scattering_force}
\end{equation}
where the parameter $\xi$ is defined as
\begin{equation}
    \xi \equiv \frac{\sigma}{m}\,,
\end{equation}
with $\sigma$ denoting the scattering cross section and $m$ the typical mass of the scattering particle species.

This type of interaction can be implemented in $N$-body algorithms \citep[see e.g.][]{Baldi:2014ica,Baldi:2016zom} in a very similar way as the velocity-dependent acceleration in Coupled Quintessence scenarios, as the factors entering Eq.~\eqref{eq:scattering_force} are all either constants or background quantities that can be interpolated at every timestep from tabulated data. This is the approach implemented in the \texttt{C-Gadget} code that has been used to run the \texttt{DAKAR} and \texttt{DAKAR2} simulations \citep[][R\'acz et al.\ in prep.]{Baldi:2016zom}.

Although Dark Scattering represents a limiting case of the more general class of pure momentum-exchange models between matter and DE \citep[also known as `Type 3' models in the classification of ][]{Skordis:2015yra}, for which further modifications to the standard particle dynamics are expected besides the drag force of Eq.~\eqref{eq:scattering_force}, recent works \citep[][]{Palma:2023ggq} have shown that such additional modifications are generally subleading with respect to the drag force so that their effect on structure formation can be neglected. This ensures that the current implementation of Dark Scattering within the simulations used in the Euclid Collaboration can be considered representative of the general class of momentum-exchange cosmologies.

\subsection{Nonlinear scalar field perturbations}
\label{sec:nonlinear_perts}

In models where a scalar field couples to matter universally, or at least to baryons in a relevant way, some mechanism to suppress the coupling is required to satisfy the stringent local tests of gravity. This is commonly referred to as `screening'. Screening mechanisms are achieved by nonlinearity in the scalar-field equation coupled to matter. 
The equation determining the evolution of the scalar field is typically a wave equation of the form 
\begin{equation}
\square \phi = S(\phi, \nabla_{\mu} \phi, \nabla_{\mu} \nabla_{\nu} \phi, \rh{m})\,.
\end{equation}
Here, $\square \equiv \nabla_{\mu} \nabla^{\mu}$ represents the d'Alembertian operator and $S$ is a nonlinear function that depends on the matter density, the scalar field, and its derivatives. 
Various methods have been developed to solve this nonlinear scalar field equation in $N$-body simulations, where the nonlinear density $\rh{m}$ is modelled by collisionless particles \citepalias[see][for more details]{Winther:2015wla}.

Several approximations are often used to solve these nonlinear equations. The most common one is the quasi-static approximation. The scalar field can be split into a background part, $\bar{\phi}$, and a perturbation, $\delta\phi$, as
$\phi = \bar{\phi} + \delta\phi$.
The quasi-static approximation amounts to ignoring the time dependence of the scalar field perturbation, i.e.\ assuming $\dot{\phi} \simeq \dot{\bar{\phi}}$.
The partial differential equation (PDE) of the field perturbation, which in its original form may have been of the hyperbolic or parabolic type, is therefore cast into an elliptic form so that the scalar field solution at any given time depends solely on the matter configuration at that time. This is a good approximation whenever the speed of sound of the scalar field is small
\citep{Sawicki:2015zya}, which is the case for the MG models considered here.
Non-quasistatic cosmological simulations have been conducted for several MG models using different techniques, such as the explicit leap-frog method and the implicit Newton--Gauss--Seidel method \citep{Llinares:2013qbh,Bose:2014zba,Winther:2015pta}.

The scalar-field solution is required for the computation of the total gravitational potential $\Phi$ that acts on the matter particles,
\begin{equation}
    \nabla^2 \Phi = 4 \pi G a^2 \Bigl(\delta \rh{m} + \delta \rh{eff} (\phi) \Bigr)\,, \label{Eq_m_poisson}
\end{equation}
where the effective density depends on the scalar field. There are two common ways of solving for this total gravitational force. The first option is to solve first for $\phi$ and then use this solution to compute the source term in Eq.~\eqref{Eq_m_poisson} and solve for $\Phi$ using a standard Poisson solver to get the total force $\nabla\Phi$. The other option is to 
apply the total force $\nabla\Phi_\mathrm{N} + \nabla\phi$ to the particles.

Under the quasi-static approximation, the scalar-field equation assumes the same form as the usual Poisson equation, 
\begin{equation}
    \nabla^2 \phi = S(\phi, \nabla_i \phi, \nabla_i \nabla_j \phi, \delta\rh{m})\,.
\end{equation}
The main difference is that the scalar-field equation is generally nonlinear. This nonlinear behaviour implies that conventional techniques, such as using Fourier analysis, cannot be used to solve the equation. Numerous approaches have been developed to address this challenge, and we refer the reader to \citetalias{Winther:2015wla} for details. For computational methods that aim to accurately solve a nonlinear equation on refined grids, the approach typically involves discretising the equation in a suitable way and employing an iterative algorithm, such as the Newton--Raphson method, to successively refine solutions based on an initial guess. To speed up convergence, many of these methods incorporate so-called `multigrid' acceleration techniques which we quickly review here.

\subsubsection{Nonlinear multigrid algorithm}
\label{sec:nonlinear-multigrid}

A generic way to solve nonlinear elliptic PDEs is to couple the multigrid algorithm to the Newton--Raphson method,
\begin{equation}\label{eq:newtonsphson}
    u^{\rm new} = u^{\rm old} - \frac{\mathcal{L}(u^{\rm old})}{\partial\mathcal{L}/\partial u^{\rm old}}\,,
\end{equation}
where $u$ is the discretised field, $\mathcal{L}$ is the differential operator (which is a Laplacian for Newtonian gravity) and the superscripts refer to the new or old estimate of the solution in one Newton--Raphson iteration. The Newton--Raphson method produces linear equations for the correction terms, which are solved by the \emph{Full-Approximation-Storage Multigrid} algorithm \citep{Brandt_1977,wesseling2004,GUILLET}. For a review of these methods applied to MG simulations, see e.g. \cite{Li_book}, \cite{Llinares_2018} and \citetalias{Winther:2015wla}.

A simple sketch of the algorithm goes as follows. One starts with a guess for the solution on a grid; this could be anything from a constant value across the grid to using the solution from the previous timestep in the simulation. One then loops a few times over all cells in the grid, updating the solution using Eq.~\eqref{eq:newtonsphson}. This solution is then restricted to a grid with half the resolution, the solution is updated again, and this process is repeated recursively up to the coarsest grid (one with only $2^3$ cells). The solution is then interpolated to the finer grid, updated once more, and this is done recursively until one reaches the finest grid we started with. One such cycle is called a V-cycle, and one repeats such V-cycles until convergence is achieved. The advantage of having this stack of coarser grids is that it helps to accelerate the convergence of the largest modes in the solution.

\subsubsection{Screening with nonlinearity in potentials}\label{sec:scr_pot} 
In models where screening is achieved by nonlinearity in a potential or coupling function, the equation for the scalar field becomes 
\begin{equation}
    \nabla^2 \phi = 4 \pi G a^2 \beta(\phi) \delta \rh{m} + V(\phi)\,,
\end{equation}
where $\beta(\phi)$ and $V(\phi)$ are nonlinear functions of $\phi$. A typical example is $f(R)$ gravity. In this class of models, the value of the scalar field changes by orders of magnitude. To enhance numerical stability, a common technique involves redefining the scalar field in terms of a new variable. The redefinition to choose depends on the specific model under consideration. It is typically chosen to prevent the occurrence of unphysical values of the scalar field during Newton--Raphson iterations. For example, for $f(R)$ models the scalar field $\phi = f_{,R}$ will be driven towards zero in high-density regions, but at the same time $f_{,R}$ cannot cross zero, as the potential becomes singular in this scenario. To avoid this issue, a commonly used field redefinition is $u \equiv \ln[f_{,R}/\bar{f}_{,R}(a)]$ \citep{Oyaizu:2008sr}. However, this transformation introduces additional nonlinearity and in some models, such as Hu--Sawicki $f(R)$ models, this transformation is not necessary and might even lead to considerable performance losses in a simulation.

In some cases, this can be avoided. For example, \citet{Bose:2016wms} noticed that for Hu--Sawicki $f(R)$ gravity with $n = 1$ \citep{Hu:2007nk}, when making the change of variable $u = \sqrt{-f_{,R}}$, the field equation could be recast as a depressed cubic equation,
\begin{equation}
    u^3 + pu + q = 0\,,
    \label{eq:cubic_equation}
\end{equation}
which possesses analytical solutions \citep{Ruan:2021wup}. Although the Gauss-Seidel smoothing procedure is still needed (because $p$ depends on the values of the field $u$ in neighbouring cells), this removes the Newton--Raphson part and expensive exponential/logarithmic operations from the method of \citet{Oyaizu:2008sr}, therefore leading to significant performance gains. \cite{Ruan:2021wup} also generalised this improved relaxation approach to the cases of $n=0$ (which strictly speaking is not a variation of the Hu--Sawicki model) and $n=2$.

For other models like the symmetron, which has a Higgs-like potential, the scalar field is free to cross zero, and no field redefinition is needed (apart from a simple rescaling).

\subsubsection{Screening with nonlinearity in kinetic terms}\label{sec:scr_kin}
In another class of models, nonlinearity emerges within the kinetic term.
For example, in models with the Vainshtein screening mechanism~\citep{Vainshtein:1972sx}, the equation exhibits nonlinearity in the second derivatives of the scalar field, 
\begin{equation}
    \nabla^2 \phi = 4 \pi G a^2\beta(a) \delta \rh{m} + g(\nabla_i \phi, \nabla_i \nabla_j \phi)\,,
\end{equation}
where $\beta(a)$ is a time-dependent coupling function. The simplest example here is the DGP model which was first simulated by \cite{Schmidt:2009sg}. In such cases, the operator-splitting trick \citep{Chan:2009ew} can be employed.  This approach can simplify the equations, avoiding potential issues associated with imaginary square roots, and improving code performance. This trick is particularly useful for the DGP braneworld models and other Vainshtein screening models, such as Cubic and Quartic Galileons \citepalias[see Sect.\,4.2.2 of][for more information]{Winther:2015wla}.

\subsubsection{Approximate treatments of screening}
\label{sec:approximate_methods}
Some models allow linearisation of the nonlinear equation using some approximation.
One approach (\citeauthor{Khoury:2009tk} \citeyear{Khoury:2009tk}; \citeauthor{Winther:2014cia} \citeyear{Winther:2014cia}; see also the appendix of \citeauthor{Schmidt:2009sv} \citeyear{Schmidt:2009sv}) is to introduce the screening factor for the matter density perturbation 
\begin{equation}
    \nabla^2 \phi = \frac{c^2m^2a^2}{\hbar^2} \phi  + 4\pi Ga^2 \delta \rh{m} \epsilon_{\rm screen}(\Phi_\mathrm{N}, |\nabla \Phi_\mathrm{N}|, \nabla^2 \Phi_\mathrm{N})\,,
\end{equation}
where the screening function depends on the Newtonian potential $\Phi_\mathrm{N}$. This type of parameterised modified gravity is referred to as `type 1' in Table~\ref{tab:codes}.
One specific method, developed in \citet{Brando:2023fzu}, starts from linearising Klein--Gordon's equation. In this formalism, one solves the Poisson equation in Fourier space,
\begin{equation}
    - k^2 \Phi= 4 \pi G_{\rm eff}(a,k) a^2 \delta \rh{m},,  
\end{equation}
where the function $G_{\rm eff}(a,k)$ approximates the effective Newton's constant introduced by the screening effect of the scalar field on small scales. This function is given by 
\begin{equation}\label{eq:Geff-ak}
    G_\mathrm{eff}(a,k) = G + \Delta G_\mathrm{eff}(a,k) = G + \left(G_\mathrm{eff}^\mathrm{lin}(a) - G\right)M(a,k)\,,
\end{equation}
with $G_{\rm eff}^{\rm lin}(a)$ being the asymptotic linear effective Newton's constant that depends only on time, and $M(a,k)$ is a function that approximately captures the nonlinear corrections introduced by the scalar field on small scales. This function allows Eq.~\eqref{eq:Geff-ak} to transition from $G_{\rm eff}(a,k)\to G_{\rm eff}^{\rm lin}(a)$ on large scales to $G_{\rm eff}(a,k)\to G$ on small scales. This type of parameterised modified gravity is referred to as `type 2' in Table~\ref{tab:codes}. A procedure to fix $M(a,k)$ is described in \citet{Brando:2023fzu}, which has the advantage of avoiding additional parameters to tune the screening efficiency.

One can also choose to parameterise the nonlinear contribution using an effective Newton's constant at both small and large scales. If the modifications of gravity are encoded in a scale-dependent function $\Delta G_\text{eff}(a,k)$ as in Eq.~\eqref{eq:Geff-ak},
then we can propose a similar equation in real space,
\begin{equation}\label{eq:G-eff-ar}
    \tilde{G}_\mathrm{eff}(a, r) = G +\Delta \tilde{G}_\mathrm{eff}(a, r)\,,
\end{equation}
where $\Delta \tilde{G}_\mathrm{eff}(a,r)$ is the Fourier transform of $\Delta G_\mathrm{eff}(a,k)$. In practice, an additional approximation is made, namely $\Delta \tilde{G}_\mathrm{eff}(a,r) \approx \Delta G_\mathrm{eff}(a,k \rightarrow 1/r)$.

\begingroup
\renewcommand{\arraystretch}{2.0}
\begin{table*}
    \caption{Summary table of the {\it N}-body codes implementing various extensions to the standard $\Lambda $CDM cosmology that have been used to produce simulations employed in \Euclid pre-launch analysis, validation, and forecasting.}
    \centering
    \begin{tabular}{p{80pt}|p{72pt}|p{72pt}|p{72pt}|p{72pt}|p{75pt}}
    \hline
    \hline
        Code (Reference) & Models & Solver\;type -- gravity & Solver\;type -- scalar field & Approximations & Treatment of massive neutrinos \\
        \hline
         \texttt{C-Gadget} \citep{Baldi:2008ay} & interacting\;DE, dark scattering & TreePM (FFT) & linear Poisson  & quasi-static approxi\-mation & $N$-body particles \\
         \texttt{ECOSMOG} \citep{Li:2011vk,Li:2013nua}  & $f(R)$,\;nDGP,\;cubic galileon & AMR\;+\;multi\-grid & NGS\;+\;multi\-grid & quasi-static approxi\-mation & -- \\
         \texttt{ISIS} \citep{Llinares:2013jza} & $f(R)$,\;nDGP, symmetron & AMR\;+\;multi\-grid & NGS\;+\;multi\-grid & quasi-static approxi\-mation &  -- \\
         \texttt{MG-Gadget} \citep{Puchwein:2013lza} & $f(R)$ & TreePM (FFT) & NGS\;+\;multi\-grid & quasi-static approxi\-mation & $N$-body particles \\
         \texttt{MG-Arepo} \citep{Arnold:2019vpg,Hernandez-Aguayo:2020kgq} & $f(R)$,\;nDGP +\;hydro & TreePM & NGS\;+\;multi\-grid & quasi-static approxi\-mation & $N$-body particles  \\
         \texttt{PANDA} (Casalino \& Baldi in prep.) & $f(R)$, nDGP & TreePM (FFT) & linear\;Poisson\;+ screening & quasi-static approxi\-mation, type\;2\;parame\-ter\-ised\;modified gravity & $N$-body particles  \\
         \pysco (Breton in prep.)  & $f(R)$ & PM + multigrid & cubic multigrid & quasi-static approxi\-mation & -- \\
         \texttt{MG-COLA} \citep{Winther:2017jof,Wright:2017dkw} & $f(R)$,\;nDGP;\,cubic galileon,\;sym\-metron & 2LPT\;+\;PM (FFT) & linear\;Poisson\;+ screening & quasi-static approx\-i\-mation, type\;1\;or\;type\;2 parameterised modified gravity & mesh  \\
         \hline
         \hline
    \end{tabular}
    \label{tab:codes}
\end{table*}
\endgroup

This approach allows the encoding of nonlinear contributions over the whole range of scales modelled by $N$-body algorithms through real-space equations, for instance, the Tree Particle-Mesh (TreePM) method implemented in codes like \texttt{Gadget4}.
Provided the parameterisation is effective, this is expected to increase the accuracy of the estimation of the nonlinear effects.

Several parameterisations have been proposed for this kind of approach, either with additional tuning parameters, such as in \citet{Lombriser:2016zfz}, or based on local small-scale environmental properties to avoid the need for any extra parameters, such as in \citet{Winther:2014cia}.

\section{Additional code validation}
\label{sec:Validation}
A code comparison for simulations that solve nonlinear scalar field perturbations is presented in \citetalias{Winther:2015wla}. Since then, various new codes have been developed.  In this section, we show a comparison of the predictions for the power spectrum and the halo mass function using the simulations of \citetalias{Winther:2015wla} as a reference and starting from the same initial conditions. These were generated using second-order Lagrangian perturbation theory in a $\Lambda$CDM cosmology with $\Om{m} = 0.269$, $\Omega_{\Lambda} = 0.731$, $h = 0.704$, $n_\mathrm{s} = 0.966$ and $\sigma_8 = 0.801$.
The simulations have $N_\mathrm{p} = 512^3$ particles of mass $M_{\rm p} \simeq 8.756 \times 10^9\,h^{-1}\,M_{\odot}$ in a box of size $B = 250\,h^{-1}\,\mathrm{Mpc}$ and start at redshift $z = 49$. As in \citetalias{Winther:2015wla}, we compare simulations for $f(R)$ and nDGP models. In these models \citep{Schmidt:2009sv}, the background expansion history is closely approximated by that of $\Lambda$CDM.
Furthermore, the effect of modified gravity can be ignored at $z = 49$, thus it is justified to use the initial conditions of $\Lambda$CDM. The measurements of the power spectrum and mass function are performed by the pipeline developed in the second article of this series, R\'acz et al.\ (in prep.). Based on two models only, our comparison does not encompass the full diversity of numerical methods discussed in the previous section. In many cases some validation of the various implementations can be found in the corresponding references.

\subsection{Summary of codes used in the validation}

Table~\ref{tab:codes} shows an overview of the simulation codes considered in this section and provides a quick reference of their capabilities and limitations. For each of them, a short summary is presented here. In the Appendix we comment on the trade-off between accuracy and computational cost of the implementations.

\subsubsection{\texttt{MG-Arepo}}

First presented in \cite{Arnold:2019vpg}, this code is based on the moving-mesh $N$-body and hydrodynamical simulation code \texttt{Arepo} \citep{Springel:2009aa,Weinberger:2020dt}, which uses a TreePM algorithm to calculate gravitational forces. The additional modified gravity force (fifth force) is calculated with a relaxation solver \citep{Bose:2016wms} that is accelerated by the multigrid method and uses adaptive mesh refinement (AMR). It currently also supports simulations for the nDGP model \citep{Hernandez-Aguayo:2020kgq}, as well as massive neutrinos implemented using the $\delta f$ method \citep{Elbers:2020lbn}.

To solve the modified gravity equations, the density field is projected onto the AMR grid, constructed in such a way that each cell on the highest refinement level contains at most one particle (except if a pre-set maximum refinement level is reached; the cell size at this level is of the order of the smoothing length of the standard gravity solver). Once the field equation is solved to obtain the scalar field configuration, the modified gravity force can be computed from its gradient using finite differencing. 

Since \texttt{MG-Arepo} computes the forces on all particles simultaneously, and the modified gravity field equations are generally highly nonlinear (with a poor convergence rate of the relaxation algorithm), this is computationally expensive compared to \texttt{Arepo}'s Newtonian gravity solver. However, the maximum acceleration of the modified gravity force is smaller than that of Newtonian gravity, mainly because the latter occurs in regions with high density where screening occurs. This allows the modified gravity solver to run using larger time steps \citep{Arnold:2019vpg}, resulting in significantly reduced computational cost. Together with \texttt{Arepo}'s efficient MPI parallelisation and lean memory footprint, these have made it possible to run the large number of $f(R)$ simulations used in various recent works, such as the \texttt{FORGE}-\texttt{BRIDGE} \citep{Arnold:2021xtm,Harnois-Deraps:2022bie,Ruan:2023mgq} simulation suite of 200 $f(R)$ and nDGP models. This has allowed accurate emulators of various physical quantities or observables to be constructed.

Another highlight of \texttt{MG-Arepo} is its capability to run realistic galaxy formation simulations in a cosmological box \citep{Arnold:2019vpg,Hernandez-Aguayo:2020kgq}, thanks to the use of the \textit{Illustris-TNG} subgrid physics model \citep{Weinberger:2017MNRAS,Pillepich:2017jle}. More recently, it has been used for larger-box hydrodynamical simulations with a realistic recalibrated \textit{Illustris-TNG} model \citep{Mitchell:2021ter}, enabling the study of galaxy clusters in modified gravity.

The \texttt{MG-Arepo} simulations used in this paper were run using a residual criterion of $\epsilon = 10^{-2}$ and a maximum refinement level (\texttt{MaxAMRLevel}) of 10 for the nDGP simulations and 18 for the $f(R)$ models with a gravitational softening of $0.01\,h^{-1}\,\mathrm{Mpc}$.

\subsubsection{\texttt{MG-Gadget}}

\texttt{MG-Gadget} \citep[][]{Puchwein:2013lza} is a modified version of the TreePM $N$-body code \texttt{Gadget-3} \citep[which in turn is based on the public code \texttt{Gadget-2}, see ][]{Springel:2005mi} implementing an AMR solver for the scalar degree of freedom $f_{,R}$ characterising the widely-studied Hu--Sawicki $f(R)$ gravity model.
In \texttt{MG-Gadget}, the same tree structure that is employed to solve for standard Newtonian gravity is also used as an adaptive grid to solve for the scalar field configuration through an iterative Newton--Gauss--Seidel (NGS) relaxation scheme (see Sect.\,\ref{sec:nonlinear_perts}) with the \textit{Full-Approximation-Storage Multigrid} method (see Sect.\,\ref{sec:nonlinear-multigrid}) and with the field redefinition $u \equiv \ln[f_{,R}/\bar{f}_{,R}(a)]$ (see Sect.\,\ref{sec:scr_pot}). \texttt{MG-Gadget} also allows MG simulations to be run with massive neutrinos \citep[see e.g.][]{Baldi:2013iza,Giocoli:2018gqh} using the neutrino particle method \citep[see][for a review on numerical methods for massive neutrino simulations]{Euclid:2022qde}.

For the simulations presented here, the relative tree opening criterion was used with an acceleration relative error threshold of 0.0025, and a uniform grid with $512^3$ cells was employed to compute long-range Newtonian forces. Concerning the MG field solver, a residual tolerance of $\epsilon = 10^{-2}$ was set for the V-cycle iteration and a maximum refinement level of $18$ was used for the AMR grid, corresponding to a spatial resolution of $1\,h^{-1}\,\mathrm{kpc}$ at the finest grid level, compared to a gravitational softening of $18\,h^{-1}\,\mathrm{kpc}$, following the setup adopted in \citetalias{Winther:2015wla}. 

\subsubsection{\texttt{ECOSMOG}}

\texttt{ECOSMOG} \citep{Li:2011vk} is a generic modified gravity simulation code based on the publicly-available $N$-body and hydrodynamical simulation code \texttt{RAMSES} \citep{Teyssier:2001cp}. Originally developed for $f(R)$ gravity, this code takes advantage of the adaptive mesh refinement of \texttt{RAMSES} to achieve the high resolution needed to solve the scalar field and hence the fifth force in high-density regions. The nonlinear $f(R)$ field equation is solved with the standard Gauss-Seidel approach as first applied by \cite{Oyaizu:2008sr}, but it was later replaced by the more efficient algorithm of \cite{Bose:2016wms}. The code has since been extended for simulations for the generalised chameleon \citep{Brax:2013mua}, symmetron and dilaton \citep{2012JCAP...10..002B}, nDGP \citep{Li:2013nua}, cubic Galileon \citep{Barreira:2013eea,Barreira:2015xvp}, quartic Galileon \citep{Li:2013tda}, vector Galileon \citep{Becker:2020azq} and nonlocal gravity \citep{Barreira:2014kra}.

For the \texttt{ECOSMOG} simulations used in this paper, we have used a domain grid (the uniform mesh that covers the whole simulation domain) with $2^9=512$ cells per dimension, and the cells are hierarchically refined if they contain $8$ or more effective\footnote{Since the code uses a cloud-in-cell mass assignment scheme, it is tricky to count particles in cells because a particle can contribute to the densities of 8 nearby cells. Here ``effective'' is used to mean that the density in a given cell, multiplied by the volume of the cell, is equivalent to the total mass of a specified number of particles.} particles. The highest refined levels have effectively $2^{16}$ cells, leading to a force resolution of about $0.0075\,h^{-1}\,\mathrm{Mpc}$. 

\subsubsection{ISIS}

The \texttt{ISIS} code \citep{Llinares:2013jza}, like the \texttt{ECOSMOG} code above, is based on \texttt{RAMSES} \citep{Teyssier:2001cp}. It contains a scalar field solver that can be used to simulate generic MG models with nonlinear equation of motion and has been used to simulate models such as $f(R)$ gravity, the symmetron model, nDGP and disformal coupled models \citep{Gronke:2013mea,Winther:2014cia,Winther:2015wla,Hagala:2015paa,Llinares:2019rbe}. It also allows for hydrodynamical simulations that have been used to study the interplay between baryonic physics and modified gravity \citep{Hammami:2015iwa}.  
Furthermore, the code has the capability to go beyond the quasistatic limit and study the full time dependence of the scalar field \citep{Llinares:2013qbh,Llinares:2013jua,Hagala:2016fks}.
The scalar field solver used in the code is a Gauss--Seidel relaxation method with multigrid acceleration, very similar to the one in \texttt{ECOSMOG} described above.

For the \texttt{ISIS} simulations presented in this paper, we have used a domain grid (the uniform mesh that covers the whole simulation domain) with $2^9 = 512$ cells per dimension, and the cells are hierarchically refined if they contain $8$ or more effective particles.

\subsubsection{\pysco}
\label{sec:pysco}
\pysco\footnote{\href{https://github.com/mianbreton/pysco}{\faicon{github}~https://github.com/mianbreton/pysco}} is a particle-mesh (PM) code written in \texttt{Python} and accelerated with \texttt{Numba} which currently supports Newtonian and $f(R)$ gravity \citep[parameterised as in][with $n = 1$ or $n = 2$]{Hu:2007nk}. While multiple flavours of solvers based on Fast Fourier Transforms (FFT) are available, for the present paper, we use a multigrid solver to propose something different from other codes in this comparison project (other codes that also use multigrid are AMR-based). \pysco uses a triangular-shaped cloud mass assignment scheme and solves the linear Poisson equation using multigrid V-cycles with a tolerance threshold of the residual of $10^{-3}$, and two F-cycles \citep[such cycles go through the mesh levels more often than V-cycles, resulting in a higher convergence rate of the residuals at the cost of increased runtime, for details on multigrid cycles see also][]{Ruan:2021wup} to solve the additional field in $f(R)$ gravity with the nonlinear multigrid method described in Sect.\,\ref{sec:nonlinear-multigrid} and Eq.~\eqref{eq:cubic_equation}. Our convergence threshold is very conservative since we do not intend to conduct a convergence study in this paper (a less conservative threshold could still give reasonable results at much lower computational cost). Furthermore, to resolve the small scales we use a coarse grid with $2048^3$ cells, resulting in roughly 500 time steps to complete the simulations.

\subsubsection{\texttt{MG-COLA}}

\texttt{MG-COLA} simulations were performed using the Fourier-Multigrid Library \texttt{FML}.\footnote{\href{https://github.com/HAWinther/FML}{\faicon{github}~https://github.com/HAWinther/FML}} These simulations are based on the COmoving Lagrangian Acceleration (COLA) method \citep{Tassev:2013pn}, which combines Lagrangian perturbation theory with the PM method to reduce the number of time steps that are required to recover clustering on large scales. The \texttt{MG-COLA} $N$-body solver in the \texttt{FML} library contains implementations of various DE and MG models like the DGP model, the symmetron model, $f(R)$ gravity and the Jordan--Brans--Dicke model \citep[see][for more details]{Winther:2017jof}.
For the PM part, we used $N_\mathrm{mesh}^{1/3} = 5 N_\mathrm{p}^{1/3}$, i.e.\ a mesh discretisation five times smaller than the mean particle separation, and the total of 150 time-steps linearly spaced along the scale factor to achieve a good agreement of the mass function with AMR simulations. We also ran low-resolution simulations with $N_\mathrm{mesh}^{1/3} = 3 N_\mathrm{p}^{1/3}$ with $100$ time steps, and 
checked that the nonlinear enhancement of the power spectrum from these low-resolution simulations agrees well with the one from the high-resolution ones. Screening is included using approximate treatments described in Sect.\,\ref{sec:approximate_methods}. 
For $f(R)$ gravity models, the \texttt{FML} code has one parameter to tune the strength of chameleon screening called \texttt{screening efficiency}. For the model with $\bar{f}_{,R} = 10^{-6}$, we used the default value \texttt{screening efficiency\,=\,1} while we used \texttt{screening efficiency\,=\,2} for $\bar{f}_{,R} = 10^{-5}$. For nDGP, we used Gaussian smoothing with a smoothing radius of $1\,h^{-1}\,\mathrm{Mpc}$ to compute the density field for screening and did not use an option to enforce the linear solution at small wavenumbers $k$. We also ran simulations based on the screening approximation using $G_\mathrm{eff}(a, k)$ given by Eq.~\eqref{eq:Geff-ak} for nDGP, and in these simulations, we have used the same \texttt{COLA} settings as in the other screening approximation implementation. The biggest advantage of this screening implementation is that it does not require any additional tuning parameter related to screening, that is, the function $G_\mathrm{eff}(a,k)$ is completely defined by the theoretical model one wants to simulate.

\subsubsection{\texttt{PANDA}}

\texttt{PANDA} is an extension of the TreePM code \texttt{Gadget4}. It introduces modifications at large scales in the PM part with an effective Newton's constant according to Eq.~\eqref{eq:Geff-ak}, while the force at small scales is modified in the tree part with Eq.~\eqref{eq:G-eff-ar}. 
In this respect, \texttt{PANDA} implements an approximate solver for the extra force induced by different possible MG theories. On the other hand, differently from other approximate methods, the dynamics of matter particles under the effect of the (dominant) standard gravity force is treated with the full TreePM solver of \texttt{Gadget4} of which it retains the accuracy in modeling the nonlinear density field.
For nDGP, the functional form of $\tilde{G}_\mathrm{eff}(a,r)$ is described in \citet{Lombriser:2016zfz}, while for $f(R)$ it is introduced as in \citet{Winther:2014cia}. In the latter case, no additional parameters are needed to describe the screening. For the first nDGP case instead, in addition to the theoretically defined parameters $N_0=1$, $B = 1/[3\beta(a)]$, $b=2$ and $a=3$, we consider the screening scale $k_\mathrm{th} = 0.4\,h\,\mathrm{Mpc}^{-1}$ defined by the parameterisation
\begin{equation}
      r_\text{V} \cong \frac{2}{3} \left(\frac{3 \Omega_\mathrm{m} H^2 r_\mathrm{c}^2}{\beta^2}\right)^{1/3} r_\text{th}\,.  
\end{equation}
where $r_{\rm V}$ is the Vainshtein radius of the model for a spherically symmetric density perturbation \citep[see][ for more details]{Lombriser:2016zfz}.

\subsection{Comparison of the power spectrum}

\begin{figure*}
    \centering
    \includegraphics[width=0.95\textwidth]{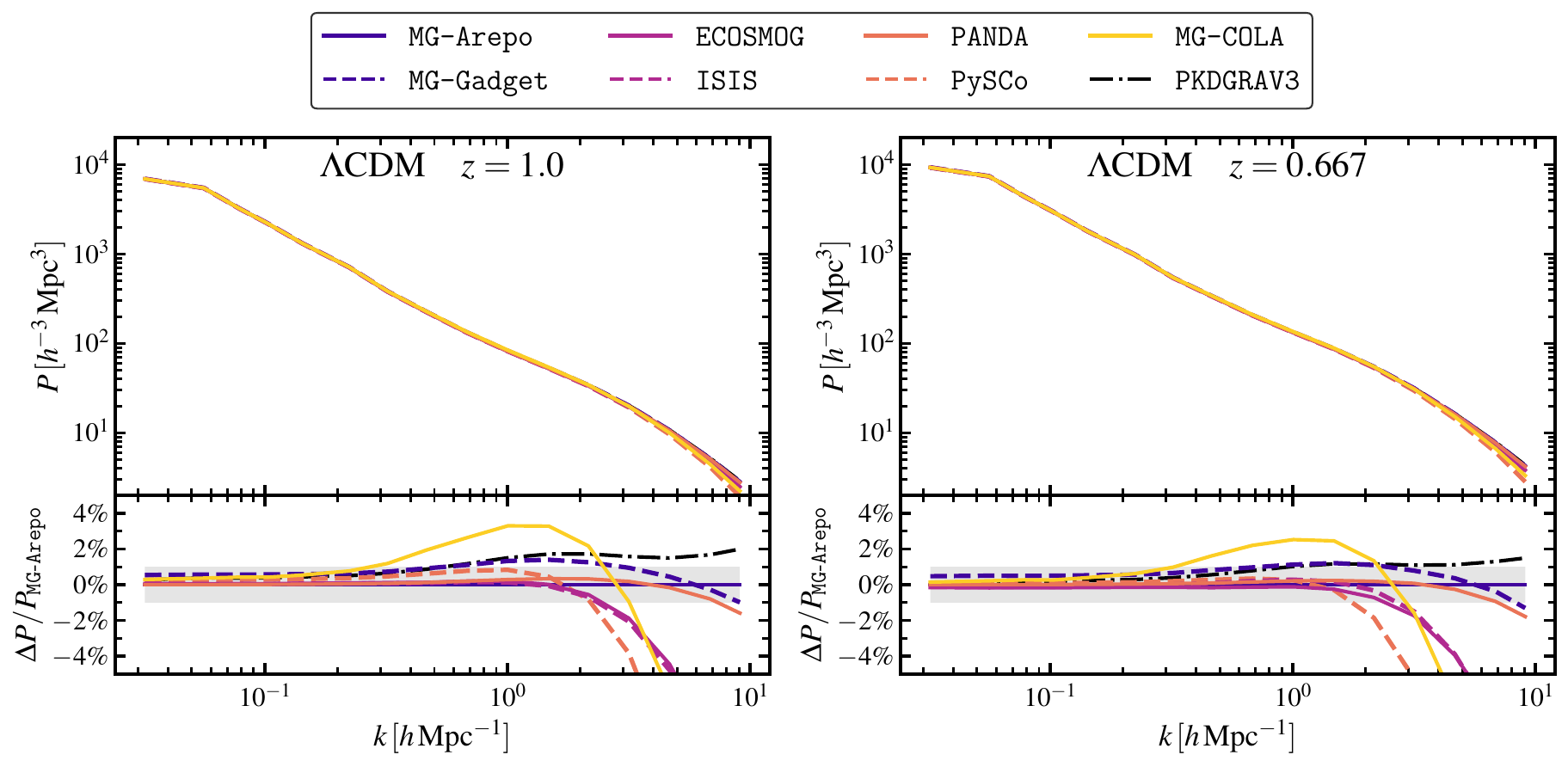}
    \caption{Matter power spectra from simulations carried out with different codes (different line styles) for a $\Lambda$CDM cosmology. The \emph{left panels} show the spectra at redshift $z=1$ whereas the \emph{right panels} show the spectra at redshift $z=0.667$. The \emph{bottom panels} show the relative difference with respect to the simulation carried out with \texttt{MG-Arepo}.} 
    \label{fig:Pk_LCDM}
\end{figure*}

Before discussing the effect of extra degrees of freedom, we compare the different codes within a $\Lambda$CDM cosmology. For reference and only for the case of $\Lambda$CDM, we include results from the code \texttt{PKDGRAV3} that was also used for the Flagship simulations in \Euclid \citep{Potter:2016ttn,Euclid:2024few}. Figure \ref{fig:Pk_LCDM} shows the matter power spectra at two different redshifts, $z=1$ (left panels) and $z=0.667$ (right panels). The lower panels show the relative difference to the simulation carried out with \texttt{MG-Arepo} that we use as a reference throughout. Since \texttt{ISIS} and \texttt{ECOSMOG} have both been developed from the \texttt{RAMSES} code, the agreement between these two codes is better than $1\%$. These AMR codes tend to underestimate the power on small scales when compared to tree-based codes, mainly due to the mesh refinement criterion used in our simulations. Fine-tuning some precision parameters, we expect that a better agreement can be achieved, cf.\ \citet{Schneider:2015yka}. As we can see, the codes that perform closely to our benchmark at small scales are the ones that also share the same tree-PM gravity solver, \texttt{PANDA} and \texttt{MG-Gadget}. \texttt{PKDGRAV3}, which uses a tree structure and the fast multipole method to compute gravitational forces, is in excellent agreement with \texttt{MG-Gadget} up to $k \sim 1\,h\,\mathrm{Mpc}^{-1}$, yielding slightly more power at higher $k$.

Also shown in the figure are the results from \texttt{COLA}, a pure PM code that uses a fixed grid to solve the Poisson equation. The fixed resolution of the PM grid explains why \texttt{COLA} consistently suffers from low force resolution at $k \gtrsim 1\,h\,\mathrm{Mpc}^{-1}$. 
All our simulations use the exact same initial particle data, such that our comparisons should not be contaminated by cosmic variance. However, for \texttt{COLA} simulations, we also reconstructed the initial density field from the initial particle distribution to estimate the displacement fields required for the 2LPT calculations in those simulations.
While \pysco incorporates a full $N$-body solver, its small-scale accuracy is constrained by resolution limitations that arise from the absence of AMR. Moreover, within the scope of this code comparison project, \pysco employs a multigrid algorithm, resulting in reduced small-scale clustering compared to FFT (on a regular grid), thereby explaining the observed deficiency in power at wavenumbers $k \gtrsim 1\,h\,\mathrm{Mpc}^{-1}$.

For modified gravity models, we compare the ratio between the power spectrum computed for that model and the one found in the $\Lambda$CDM reference cosmology. Taking such ratios largely cancels out the differences seen in $\Lambda$CDM between the codes, making the differences due to the treatments of the extra scalar field more apparent. Following \citetalias{Winther:2015wla}, we consider two MG models that exhibit two different types of growth dependence and screening mechanisms. The first one is the so-called Hu--Sawicki $f(R)$ which exhibits a scale-dependent growth factor at linear order, and a screening mechanism realised by nonlinearities in the potential, see Sect.\,\ref{sec:scr_pot}. The strength of the modification of gravity is characterised by the background value of the scalar degree of freedom, and we study the cases $\bar{f}_{,R} = 10^{-5}$ and $\bar{f}_{,R} = 10^{-6}$, labelled `F5' and `F6', respectively, the latter being closer to GR. The second one is nDGP, a braneworld theory defined in a five-dimensional spacetime. The growth factor in this theory is scale independent, and screening is realised through nonlinearities in the kinetic terms, as discussed in Sect.\,\ref{sec:scr_kin}. Here, the strength of the modification of gravity is characterised by the value of the cross-over scale, and we study the cases $r_\mathrm{c} = 1.2\,H_0^{-1}$ and $r_\mathrm{c} = 5.6\,H_0^{-1}$, labelled `N1.2' and `N5.6', respectively, the latter being closer to GR.

\begin{figure*}
    \centering
    \includegraphics[width=0.95\textwidth]{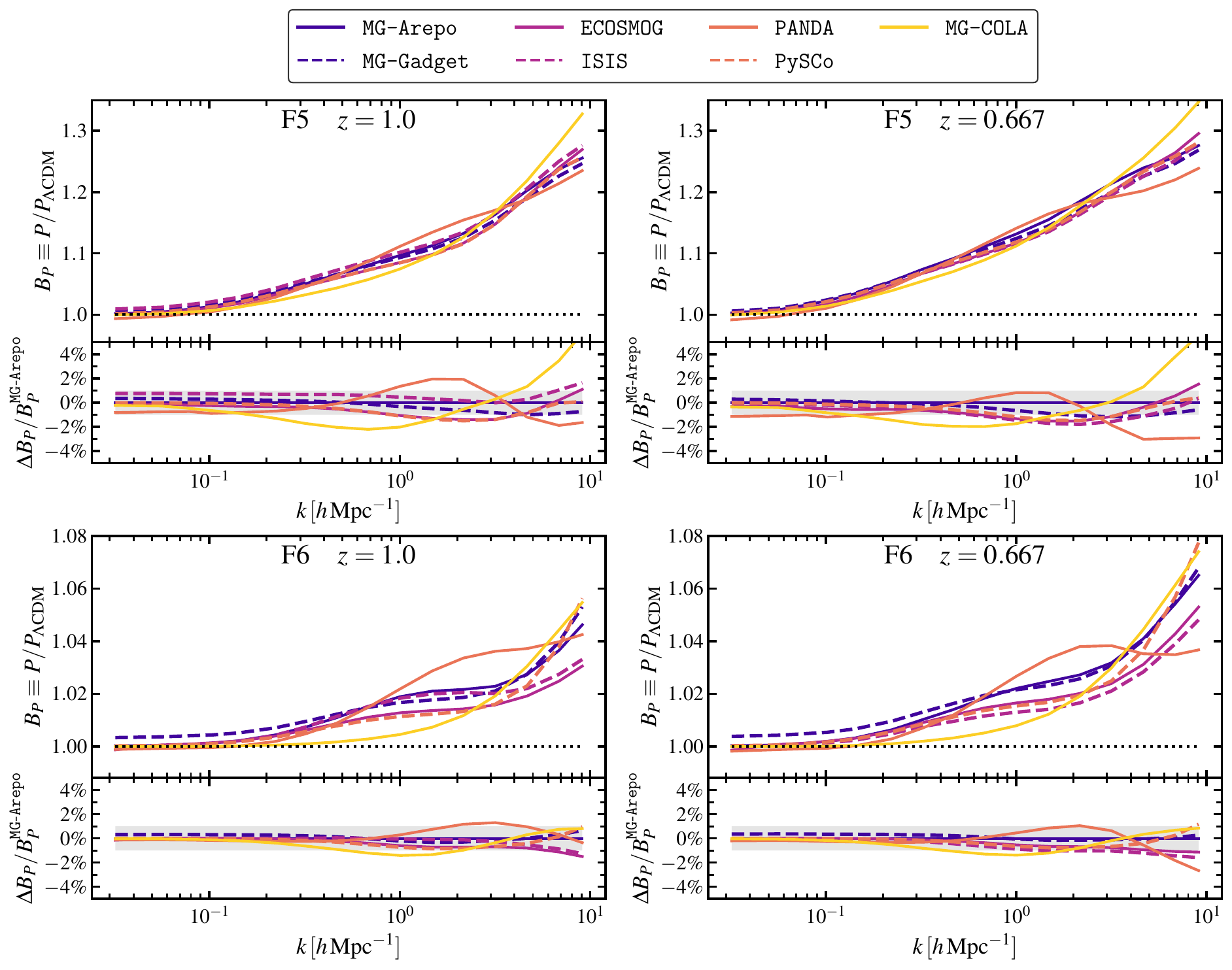}
    \caption{Amplification factor of the matter power spectra from simulations of two $f(R)$ scenarios, $\bar{f}_{,R} = 10^{-5}$ (F5, \emph{top row}) and $\bar{f}_{,R} = 10^{-6}$ (F6, \emph{bottom row}), relative to the reference $\Lambda$CDM cosmology. The amplification factor $B_P = P/P_{\Lambda\mathrm{CDM}}$ is measured for different codes (different line styles) at two different redshifts, $z=1$ (\emph{left panels}) and $z=0.667$ (\emph{right panels}). The \emph{bottom panel} of each plot shows the relative agreement of the individual measurements, using \texttt{MG-Arepo} as a common reference.} 
    \label{fig:Bk_fR}
\end{figure*}

\begin{figure*}
    \centering
    \includegraphics[width=0.95\textwidth]{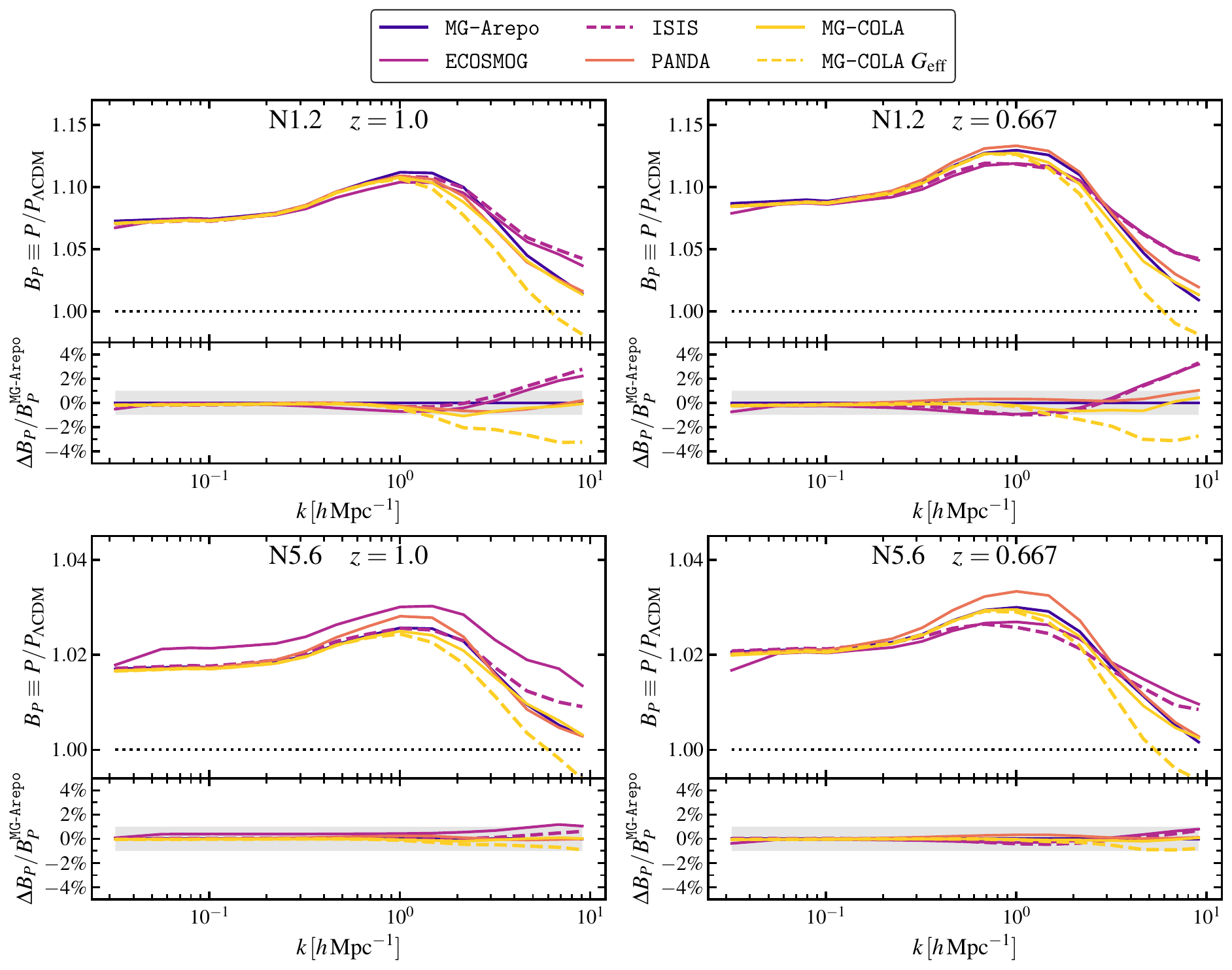}
    \caption{Amplification factor of the matter power spectra from simulations of two nDGP scenarios, $r_\mathrm{c} = 1.2\,H_0^{-1}$ (N1.2, \emph{top row}) and $r_\mathrm{c} = 5.6\,H_0^{-1}$ (N5.6, \emph{bottom row}), relative to the reference $\Lambda$CDM cosmology. The amplification factor $B_P = P/P_{\Lambda\mathrm{CDM}}$ is measured for different codes (different line styles) at two different redshifts, $z=1$ (\emph{left panels}) and $z=0.667$ (\emph{right panels}). The \emph{bottom panel} of each plot shows the relative agreement of the individual measurements, using \texttt{MG-Arepo} as a common reference.} 
    \label{fig:Bk_DGP}
\end{figure*}

Figure \ref{fig:Bk_fR} shows the relative change (with respect to $\Lambda$CDM) of the matter power spectrum due to modifications of gravity, sometimes called `boost', for the two $f(R)$ scenarios at redshifts $z=1$ (left panels) and $z=0.667$ (right panels). Since $f(R)$ already exhibits a scale-dependent growth factor at linear order, we can see that on scales of $k\approx 0.1\,h\,\mathrm{Mpc}^{-1}$ this effect is already present, and gets enhanced at large wavenumbers due to nonlinearities of the density field. As we can see the codes \texttt{MG-Arepo}, \texttt{MG-Gadget}, \texttt{ECOSMOG}, \texttt{ISIS}, and \pysco all roughly agree to better than one percent down to scales of $k\approx 10\,h\,\mathrm{Mpc}^{-1}$, as expected. These minor discrepancies can be caused by the refinement criterion used in the latter code or by slight variations in the redshift of the particle snapshot output. The present results are largely in agreement with a similar analysis performed in~\citetalias{Winther:2015wla}.

In the same plots, we can also see how approximation schemes to introduce screening from nonlinearities in the potential perform in contrast to the exact solutions. The results that use these methods are showcased by the examples of \texttt{COLA} and \texttt{PANDA}, where each uses a different scheme to approximate the effects of the dynamics of the scalar field in dense environments. As expected, the two codes do not exhibit the same level of agreement down to scales of $k\approx 10\,h\,\mathrm{Mpc}^{-1}$ as their counterparts using full solvers. However, in both cases, we can see that the deviations from the \texttt{MG-Arepo} reference results are limited to $2\%$ even in the most extreme regime of departure from GR. Through closer inspection, we can see small differences in the agreement between the approximate schemes and full solvers at different values of $k$. These deviations are caused by different implementations of the approximate MG solvers in the two codes. In fact, while \texttt{COLA} is an approximate method that uses a PM algorithm to solve the Poisson equation on a fixed grid with the use of the screening approximation to linearise the Poisson equation, \texttt{PANDA} is a new implementation that exploits the TreePM structure of the baseline \texttt{Gadget-4} code to solve for the small-scale particle dynamics by incorporating the MG effects (including the screening) through a scale-dependent Newton's constant in both real and Fourier space.

\begin{figure*}
    \centering
    \includegraphics[width=0.95\textwidth]{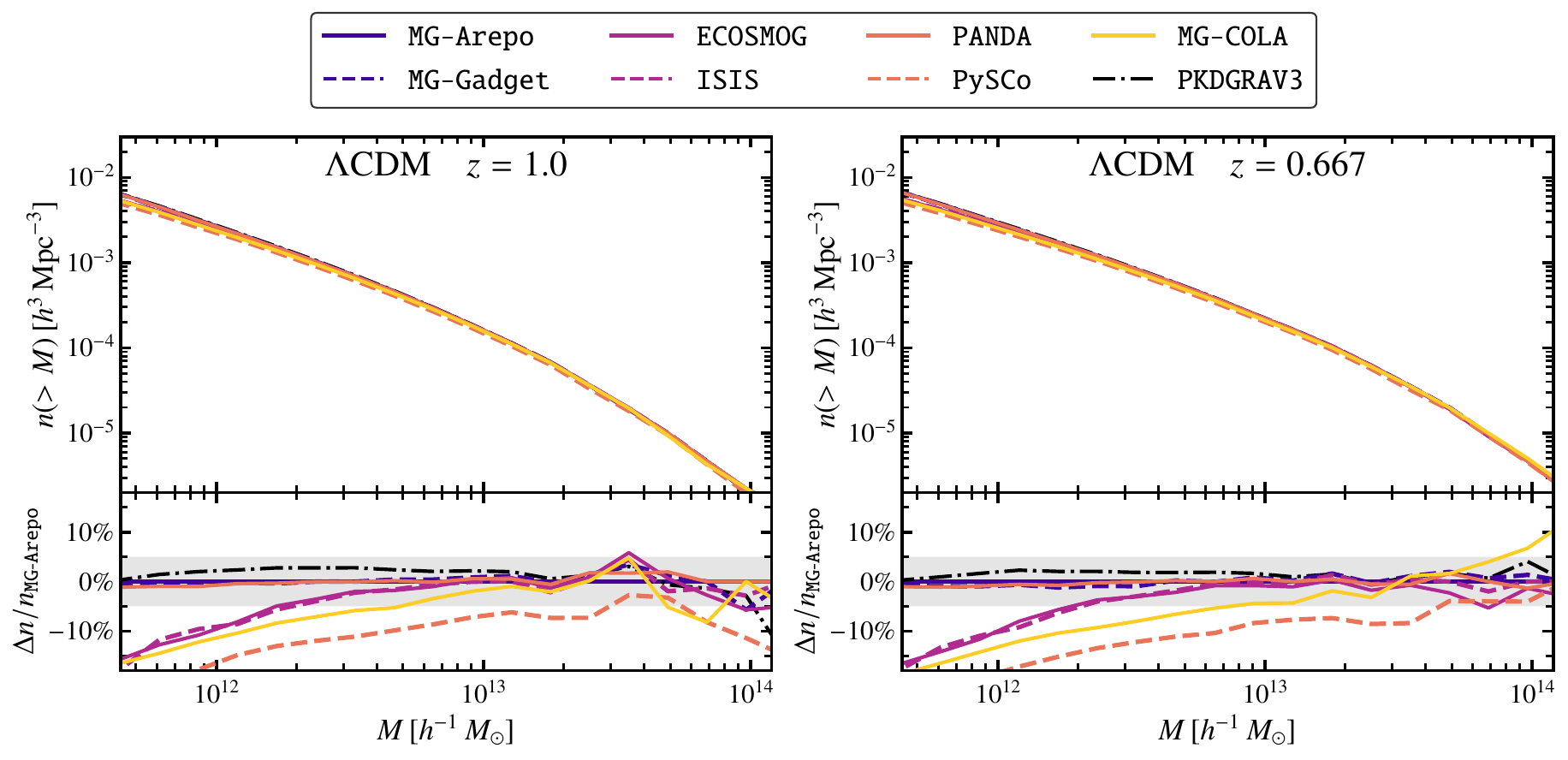}
    \caption{Cumulative halo mass function from simulations carried out with different codes (different line styles) for a $\Lambda$CDM cosmology. The \emph{left panels} show the halo mass function at redshift $z=1$ whereas the \emph{right panels} show the halo mass function at redshift $z=0.667$. The \emph{bottom panels} show the relative difference with respect to the simulation carried out with \texttt{MG-Arepo}.} 
    \label{fig:hmf_LCDM}
\end{figure*}

Figure~\ref{fig:Bk_DGP} shows results for the matter power spectra in the two nDGP scenarios we consider, characterised by the two values of the cross-over scale, $r_\mathrm{c} = 1.2\,H_0^{-1}$ (N1.2) and $r_\mathrm{c} = 5.6\,H_0^{-1}$ (N5.6). As before, we compare simulation results at two different redshifts, $z=1$ (left panels) and $z=0.667$ (right panels). Since the linear growth function has a scale-independent enhancement compared to $\Lambda$CDM, we see a constant amplification of power
at small values of $k$. The Vainshtein mechanism suppresses the deviation from $\Lambda$CDM on small scales, leading to a diminishing boost at large $k$. The agreement between different codes at $k < 1\,h\,\mathrm{Mpc}^{-1}$ is better than $1\%$ for all codes, including the approximate simulations with \texttt{COLA} and \texttt{PANDA}. At $k > 1\,h\,\mathrm{Mpc}^{-1}$, the AMR-based codes, \texttt{ECOSMOG} and \texttt{ISIS}, show larger deviations at the level of $2\%$, where the suppression of the deviation from $\Lambda$CDM is slightly underestimated compared with \texttt{MG-Arepo}. This could be caused by the difference in the $\Lambda$CDM power spectrum shown in Fig.\,\ref{fig:Pk_LCDM}, even though this difference in the baseline code is largely cancelled out in the boost. On the other hand, despite their approximations in the MG solvers, \texttt{COLA} and \texttt{PANDA} agree with \texttt{MG-Arepo} at the level of $1\%$ even on these scales.

\subsection{Comparison of the halo mass function}

To gain further insight into the nonlinear dynamics of the different models and their respective implementations we also compare the cumulative halo mass functions measured in our simulations. For this purpose, halo catalogues are obtained with the \texttt{Rockstar} halo finder by running a pipeline described in the second paper of this series (R\'acz et al.\ in prep.). To establish the baseline for the comparison, in the spirit of the previous section, Fig.\,\ref{fig:hmf_LCDM} shows a comparison of the cumulative halo mass function for the different codes in a $\Lambda$CDM cosmology. Since \texttt{MG-Arepo} and \texttt{MG-Gadget} are based on the same TreePM gravity solver, their agreement is excellent. They also agree very well with results from \texttt{PKDGRAV3} which, as we like to remind the reader, are only available for the case of $\Lambda$CDM and are shown for reference here. On the other hand, \texttt{ECOSMOG} and \texttt{ISIS} are based on the AMR method. They agree with each other, but these simulations underestimate the abundance of low-mass halos below $10^{13} h^{-1}\,M_{\odot}$. \texttt{COLA} uses a fixed-grid PM method, thus it also underestimates the abundance of low-mass halos. With relatively high resolution in these simulations ($N_\mathrm{mesh}^{1/3} = 5 N_\mathrm{p}^{1/3}$), \texttt{COLA} agrees well with \texttt{ECOSMOG} and \texttt{ISIS}. We note that \pysco exhibits a similar behaviour to \texttt{COLA}, \texttt{ECOSMOG} and \texttt{ISIS}, albeit with a lower amplitude. This discrepancy can be attributed to the fact that \pysco has the least small-scale clustering (see Fig.\,\ref{fig:Pk_LCDM}), resulting in a smaller number of halos within the simulation.

\begin{figure*}
    \centering
    \includegraphics[width=0.95\textwidth]{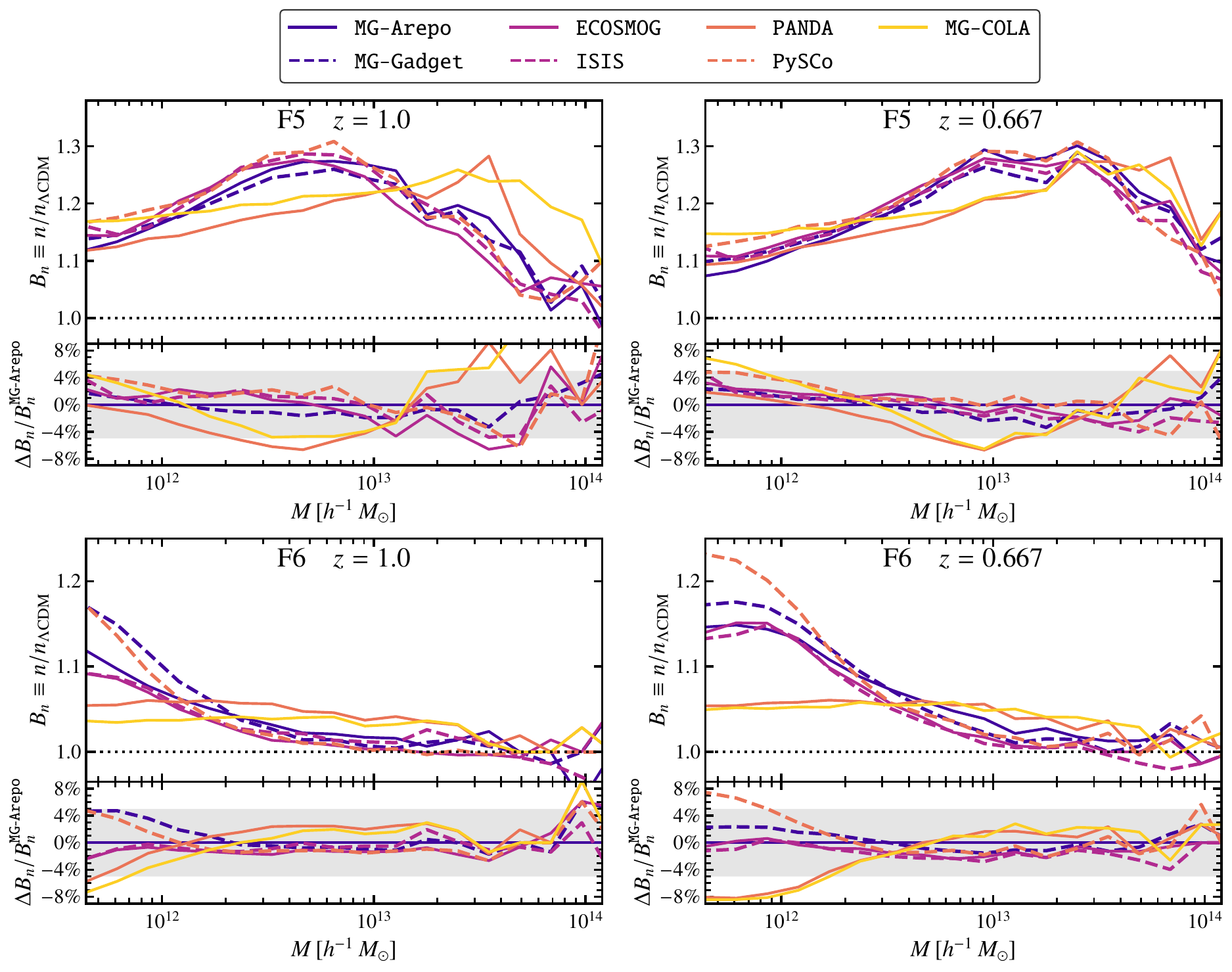}
    \caption{Increase of the cumulative halo mass function from simulations for the two $f(R)$ scenarios, $\bar{f}_{,R} = 10^{-5}$ (F5, \emph{top row}) and $\bar{f}_{,R} = 10^{-6}$ (F6, \emph{bottom row}), relative to the reference $\Lambda$CDM cosmology. The enhancement factor $B_n = n/n_{\Lambda\mathrm{CDM}}$ is measured for different codes (different line styles) at two different redshifts, $z=1$ (\emph{left panels}) and $z=0.667$ (\emph{right panels}). The \emph{bottom panel} of each plot shows the relative agreement of the individual measurements, using \texttt{MG-Arepo} as a common reference.} 
    \label{fig:Bhmf_fR}
\end{figure*}

The chameleon screening in $f(R)$ depends on the halo mass such that high-mass halos are typically screened. The critical mass above which screening is effective is determined by the parameter $\bar{f}_{R}$. 
As we can see from Fig.\,\ref{fig:Bhmf_fR}, the ratio of the halo mass function between $f(R)$ and $\Lambda$CDM is enhanced for low-mass halos but approaches unity at the high-mass end where all halos are effectively screened. The agreement between full simulations (\texttt{ECOSMOG} and \texttt{ISIS}, \texttt{MG-Arepo}, \texttt{MG-Gadget}) is around $4\%$ for all halo masses. \texttt{COLA} uses an approximation for screening based on the thin-shell condition for a spherically symmetric object. Although this captures an overall effect of screening, it fails for low-mass halos in F6 and high-mass halos in F5, leading to larger deviations.

\begin{figure*}
    \centering    \includegraphics[width=0.95\textwidth]{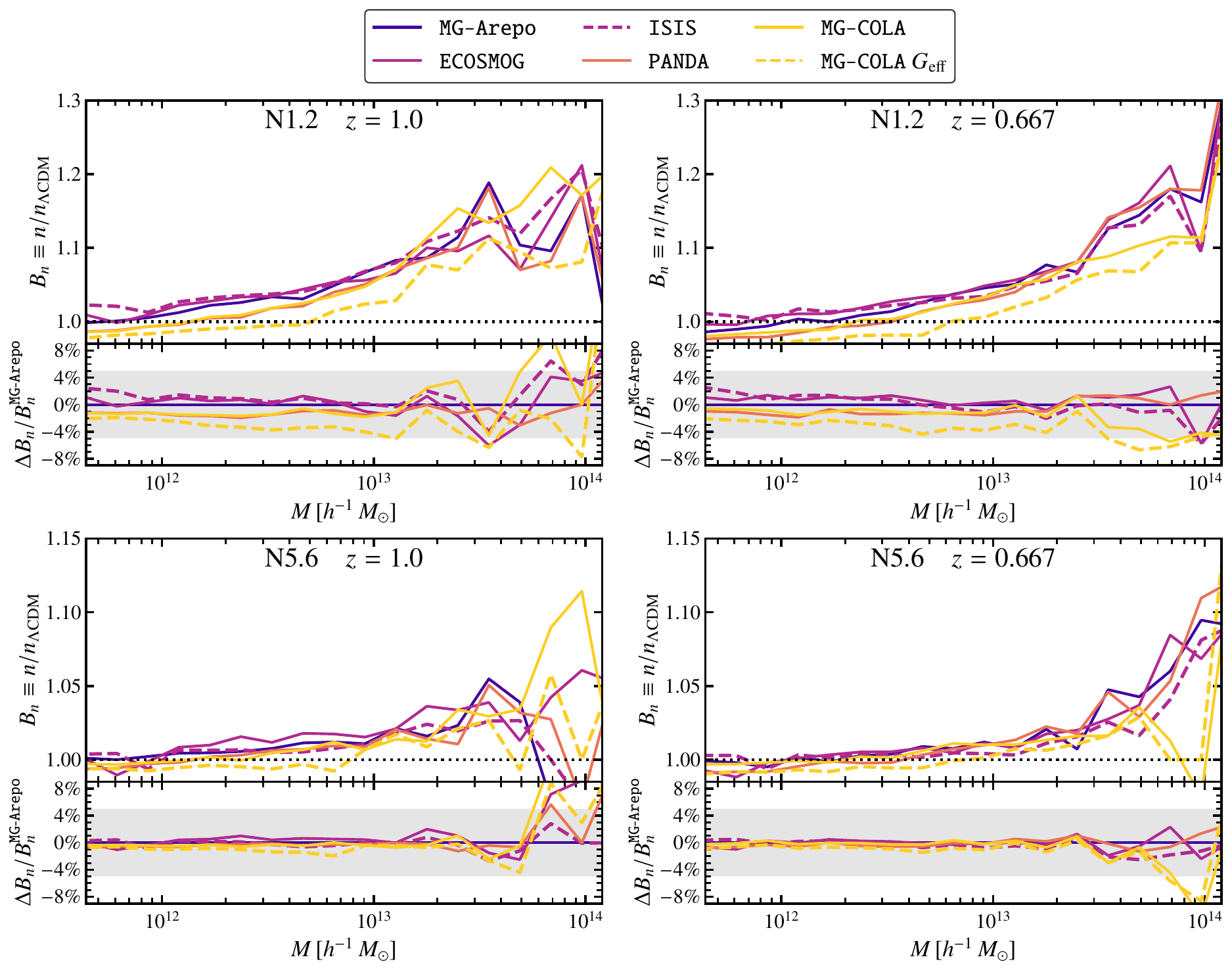}
    \caption{Increase of the cumulative halo mass function from simulations for the two nDGP scenarios, $r_\mathrm{c} = 1.2\,H_0^{-1}$ (N1.2, \emph{top row}) and $r_\mathrm{c} = 5.6\,H_0^{-1}$ (N5.6, \emph{bottom row}), relative to the reference $\Lambda$CDM cosmology. The enhancement factor $B_n = n/n_{\Lambda\mathrm{CDM}}$ is measured for different codes (different line styles) at two different redshifts, $z=1$ (\emph{left panels}) and $z=0.667$ (\emph{right panels}). The \emph{bottom panel} of each plot shows the relative agreement of the individual measurements, using \texttt{MG-Arepo} as a common reference.} 
    \label{fig:Bhmf_DGP}
\end{figure*}

In the case of the Vainshtein mechanism, there is no halo-mass dependence on the screening. Despite screening being effective inside dark matter halos, these halos still feel enhanced gravitational attraction. This increases the merger rate and ultimately leads to larger enhancements of the halo mass function for halos of larger masses. In nDGP, shown in Fig.\,\ref{fig:Bhmf_DGP}, all codes agree within $4\%$, with a sub-per cent agreement seen in the intermediate mass range 
$10^{12}\,h^{-1}\,M_{\odot} < M < 10^{13}\,h^{-1}\,M_{\odot}$. \texttt{ECOSMOG} shows a large deviation for $M \lesssim 5\,\times\,10^{11}\,h^{-1}\,M_{\odot}$. For these masses, there are less than 50 dark matter particles assigned to each halo, which indicates that the results could be affected by the refinement criteria of the simulations. The deviations are larger for the most massive halos, but the number of these halos is low and the halo mass function therefore becomes very noisy in this regime.

\section{Conclusions}
\label{sec:Conclusion}

In this work, we have presented a comprehensive review of numerical methods for cosmological $N$-body simulations in scenarios extending beyond the standard $\Lambda$CDM model. Our exploration spanned a variety of alternative DE and MG theories, highlighting the critical role of $N$-body simulations in connecting theoretical models with observational data. Through the detailed examination of numerical solvers and approximations tailored to these extended theories, we have showcased the state of the art of modelling the nonlinear scales of cosmic structure formation under a wide range of cosmological scenarios. Our code comparison exercise, based on the simulations from \citetalias{Winther:2015wla} and extended by incorporating new codes and approximation techniques, has demonstrated a fair consensus among different numerical implementations. This validation is particularly important for the \Euclid mission, as the forthcoming observational data will require precise nonlinear modelling to constrain cosmological parameters effectively.

This article is part of a series that explores simulations and nonlinearities in models beyond $\Lambda$CDM. The simulation codes that we have considered in this article are used to generate simulation products in the companion paper by R\'acz et al.\ (in preparation) and are crucial for generating the simulations needed to create the nonlinear modelling in the companion paper by Koyama et al.\ (in preparation), which forecasts constraints for $f(R)$ gravity from photometric primary probes of the \Euclid mission. The validation checks performed in this paper are therefore critical for being able to trust these results.

The main outcomes of this work can be summarised as follows:
\begin{itemize}
    \item[$\star $] $N$-body simulation codes have been developed for a wide range of extended cosmological scenarios, ranging from simple DE models to more complex interacting scalar field models and MG theories, to non-standard dark matter and initial conditions; the availability of such codes will be a crucial asset for the future developments of large galaxy surveys such as \Euclid; we have provided a concise yet comprehensive overview of several such codes, their main features, implementation methods, assumptions and approximations.
    \item[$\star $] As a matter of fact, among these simulation codes the ones involving algorithms for the solution of nonlinear differential equations of some additional degrees of freedom, as e.g.\ for the case of MG theories, are the most challenging in terms of implementation and numerical convergence; we have therefore performed a thorough validation effort of these methods through a code comparison study, extending the approach adopted in \citetalias{Winther:2015wla} to more recent and diverse algorithms.
    \item[$\star $] As a result of our validation effort, we found  agreement in the power spectrum boost at $\lesssim 1\%$ up to $k \lesssim 1\,h\,\mathrm{Mpc}^{-1}$ and at $\lesssim 3\%$ up to $k \lesssim 10\,h\,\mathrm{Mpc}^{-1}$ among all the codes implementing full field solvers (\texttt{MG-Arepo, MG-Gadget, ECOSMOG, ISIS, PySCo}), while approximate methods (\texttt{PANDA, MG-COLA}) display slightly larger deviations not exceeding $3\%$ up to $k \lesssim 7\,h\,\mathrm{Mpc}^{-1}$.
    \item[$\star $] The halo mass function shows larger deviations among the codes, also due to larger differences in the outcomes for the baseline $\Lambda $CDM simulations; nonetheless, all codes agree within less than $5\%$ on the relative change of the halo mass function except for the very largest and the smallest mass ranges where poor statistics and insufficient resolution, respectively, may impact the results.
    \item[$\star $] We also compared the computational requirements of the different codes by measuring the CPU time needed to complete a reference MG simulation starting from identical initial conditions; we found that while full field solvers generally imply a substantial increase -- up to a factor of ten -- of the total CPU time relative to a $\Lambda$CDM simulation, approximate solvers are not significantly more demanding for MG simulations compared to standard $\Lambda$CDM runs.
\end{itemize}

Looking forward, the continued evolution of simulation techniques will be paramount in leveraging the full potential of upcoming large-scale structure surveys such as \textit{Euclid}. $N$-body simulations therefore continue to set a solid foundation for the next generation of cosmological inquiries. By persistently pushing the boundaries of computational astrophysics, we are poised to uncover the underlying physics driving the accelerated expansion of the Universe, thereby opening new windows onto the fundamental nature of DE, dark matter, and gravity itself.

\begin{acknowledgements}
The lead authors thank A.\ Schneider and R.E.\ Smith for their diligent work as internal referees and P.\ Schneider for carefully proofreading the manuscript.
      The work of JA is supported by the Swiss National Science Foundation.
      During part of this work, AMCLB was supported by a fellowship of PSL University hosted by the Paris Observatory.
      This project was provided with computer and storage resources by GENCI at CINES thanks to the grant 2023-A0150402287 on the supercomputer Adastra's GENOA partition.
      GR’s research was supported by an appointment to the NASA Postdoctoral Program administered by Oak Ridge Associated Universities under contract with NASA. GR was supported by JPL, which is run under contract by the California Institute of Technology for NASA (80NM0018D0004).
      \AckEC
\end{acknowledgements}

\bibliographystyle{aa}
\bibliography{bibliography}

\begin{thebibliography}{126}
\expandafter\ifx\csname natexlab\endcsname\relax\def\natexlab#1{#1}\fi

\bibitem[{{Abdalla} {et~al.}(2022){Abdalla}, {Abell{\'a}n}, {Aboubrahim},
  {Agnello}, {Akarsu}, {Akrami}, {Alestas}, {Aloni}, {Amendola}, {Anchordoqui},
  {Anderson}, {Arendse}, {Asgari}, {Ballardini}, {Barger}, {Basilakos},
  {Batista}, {Battistelli}, {Battye}, {Benetti}, {Benisty}, {Berlin}, {de
  Bernardis}, {Berti}, {Bidenko}, {Birrer}, {Blakeslee}, {Boddy}, {Bom},
  {Bonilla}, {Borghi}, {Bouchet}, {Braglia}, {Buchert}, {Buckley-Geer},
  {Calabrese}, {Caldwell}, {Camarena}, {Capozziello}, {Casertano}, {Chen},
  {Chluba}, {Chen}, {Chen}, {Chudaykin}, {Cicoli}, {Copi}, {Courbin},
  {Cyr-Racine}, {Czerny}, {Dainotti}, {D'Amico}, {Davis}, {de Cruz P{\'e}rez},
  {de Haro}, {Delabrouille}, {Denton}, {Dhawan}, {Dienes}, {Di Valentino},
  {Du}, {Eckert}, {Escamilla-Rivera}, {Fert{\'e}}, {Finelli}, {Fosalba},
  {Freedman}, {Frusciante}, {Gazta{\~n}aga}, {Giar{\`e}}, {Giusarma},
  {G{\'o}mez-Valent}, {Handley}, {Harrison}, {Hart}, {Hazra}, {Heavens},
  {Heinesen}, {Hildebrandt}, {Hill}, {Hogg}, {Holz}, {Hooper}, {Hosseininejad},
  {Huterer}, {Ishak}, {Ivanov}, {Jaffe}, {Jang}, {Jedamzik}, {Jimenez},
  {Joseph}, {Joudaki}, {Kamionkowski}, {Karwal}, {Kazantzidis}, {Keeley},
  {Klasen}, {Komatsu}, {Koopmans}, {Kumar}, {Lamagna}, {Lazkoz}, {Lee},
  {Lesgourgues}, {Levi Said}, {Lewis}, {L'Huillier}, {Lucca}, {Maartens},
  {Macri}, {Marfatia}, {Marra}, {Martins}, {Masi}, {Matarrese}, {Mazumdar},
  {Melchiorri}, {Mena}, {Mersini-Houghton}, {Mertens}, {Milakovi{\'c}},
  {Minami}, {Miranda}, {Moreno-Pulido}, {Moresco}, {Mota}, {Mottola}, {Mozzon},
  {Muir}, {Mukherjee}, {Mukherjee}, {Naselsky}, {Nath}, {Nesseris},
  {Niedermann}, {Notari}, {Nunes}, {{\'O} Colg{\'a}in}, {Owens},
  {{\"O}z{\"u}lker}, {Pace}, {Paliathanasis}, {Palmese}, {Pan}, {Paoletti},
  {Perez Bergliaffa}, {Perivolaropoulos}, {Pesce}, {Pettorino}, {Philcox},
  {Pogosian}, {Poulin}, {Poulot}, {Raveri}, {Reid}, {Renzi}, {Riess}, {Sabla},
  {Salucci}, {Salzano}, {Saridakis}, {Sathyaprakash}, {Schmaltz},
  {Sch{\"o}neberg}, {Scolnic}, {Sen}, {Sehgal}, {Shafieloo}, {Sheikh-Jabbari},
  {Silk}, {Silvestri}, {Skara}, {Sloth}, {Soares-Santos}, {Sol{\`a} Peracaula},
  {Songsheng}, {Soriano}, {Staicova}, {Starkman}, {Szapudi}, {Teixeira},
  {Thomas}, {Treu}, {Trott}, {van de Bruck}, {Vazquez}, {Verde}, {Visinelli},
  {Wang}, {Wang}, {Wang}, {Watkins}, {Watson}, {Webb}, {Weiner}, {Weltman},
  {Witte}, {Wojtak}, {Yadav}, {Yang}, {Zhao}, \&
  {Zumalac{\'a}rregui}}]{Abdalla:2022yfr}
{Abdalla}, E., {Abell{\'a}n}, G.~F., {Aboubrahim}, A., {et~al.} 2022, Journal
  of High Energy Astrophysics, 34, 49

\bibitem[{{Adamek} {et~al.}(2023){Adamek}, {Angulo}, {Arnold}, {Baldi},
  {Biagetti}, {Bose}, {Carbone}, {Castro}, {Dakin}, {Dolag}, {Elbers},
  {Fidler}, {Giocoli}, {Hannestad}, {Hassani}, {Hern{\'a}ndez-Aguayo},
  {Koyama}, {Li}, {Mauland}, {Monaco}, {Moretti}, {Mota}, {Partmann},
  {Parimbelli}, {Potter}, {Schneider}, {Schulz}, {Smith}, {Springel}, {Stadel},
  {Tram}, {Viel}, {Villaescusa-Navarro}, {Winther}, {Wright}, {Zennaro},
  {Aghanim}, {Amendola}, {Auricchio}, {Bonino}, {Branchini}, {Brescia},
  {Camera}, {Capobianco}, {Cardone}, {Carretero}, {Castander}, {Castellano},
  {Cavuoti}, {Cimatti}, {Cledassou}, {Congedo}, {Conversi}, {Copin}, {Da
  Silva}, {Degaudenzi}, {Douspis}, {Dubath}, {Duncan}, {Dupac}, {Dusini},
  {Farrens}, {Ferriol}, {Fosalba}, {Frailis}, {Franceschi}, {Galeotta},
  {Garilli}, {Gillard}, {Gillis}, {Grazian}, {Haugan}, {Holmes}, {Hornstrup},
  {Jahnke}, {Kermiche}, {Kiessling}, {Kilbinger}, {Kitching}, {Kunz},
  {Kurki-Suonio}, {Lilje}, {Lloro}, {Mansutti}, {Marggraf}, {Marulli},
  {Massey}, {Medinaceli}, {Meneghetti}, {Meylan}, {Moresco}, {Moscardini},
  {Munari}, {Niemi}, {Padilla}, {Paltani}, {Pasian}, {Pedersen}, {Percival},
  {Pettorino}, {Polenta}, {Poncet}, {Popa}, {Raison}, {Rebolo}, {Renzi},
  {Rhodes}, {Riccio}, {Romelli}, {Roncarelli}, {Saglia}, {Sapone}, {Sartoris},
  {Schneider}, {Schrabback}, {Secroun}, {Seidel}, {Sirignano}, {Sirri},
  {Stanco}, {Starck}, {Tallada-Cresp{\'\i}}, {Taylor}, {Tereno},
  {Toledo-Moreo}, {Torradeflot}, {Tutusaus}, {Valenziano}, {Vassallo}, {Wang},
  {Weller}, {Zacchei}, {Zamorani}, {Zoubian}, {Fabbian}, \&
  {Scottez}}]{Euclid:2022qde}
{Adamek}, J., {Angulo}, R.~E., {Arnold}, C., {et~al.} 2023, \jcap, 06, 035

\bibitem[{Adamek {et~al.}(2016)Adamek, Daverio, Durrer, \&
  Kunz}]{Adamek:2015eda}
Adamek, J., Daverio, D., Durrer, R., \& Kunz, M. 2016, Nature Phys., 12, 346

\bibitem[{{Adamek} {et~al.}(2016){Adamek}, {Daverio}, {Durrer}, \&
  {Kunz}}]{Adamek:2016zes}
{Adamek}, J., {Daverio}, D., {Durrer}, R., \& {Kunz}, M. 2016, \jcap, 07, 053

\bibitem[{{Amendola}(2000)}]{Amendola:1999er}
{Amendola}, L. 2000, \prd, 62, 043511

\bibitem[{{Amendola}(2004)}]{Amendola:2003wa}
{Amendola}, L. 2004, \prd, 69, 103524

\bibitem[{{Amendola} {et~al.}(2018){Amendola}, {Appleby}, {Avgoustidis},
  {Bacon}, {Baker}, {Baldi}, {Bartolo}, {Blanchard}, {Bonvin}, {Borgani},
  {Branchini}, {Burrage}, \& {et al.}}]{Amendola:2016saw}
{Amendola}, L., {Appleby}, S., {Avgoustidis}, A., {et~al.} 2018, Living Rev.\
  Rel., 21, 2

\bibitem[{{Amendola} {et~al.}(2008){Amendola}, {Baldi}, \&
  {Wetterich}}]{Amendola:2007yx}
{Amendola}, L., {Baldi}, M., \& {Wetterich}, C. 2008, \prd, 78, 023015

\bibitem[{{Armendariz-Picon} {et~al.}(2000){Armendariz-Picon}, {Mukhanov}, \&
  {Steinhardt}}]{Armendariz-Picon:2000nqq}
{Armendariz-Picon}, C., {Mukhanov}, V., \& {Steinhardt}, P.~J. 2000, \prl, 85,
  4438

\bibitem[{{Armendariz-Picon} {et~al.}(2001){Armendariz-Picon}, {Mukhanov}, \&
  {Steinhardt}}]{Armendariz-Picon:2000ulo}
{Armendariz-Picon}, C., {Mukhanov}, V., \& {Steinhardt}, P.~J. 2001, \prd, 63,
  103510

\bibitem[{{Arnold} {et~al.}(2019){Arnold}, {Leo}, \& {Li}}]{Arnold:2019vpg}
{Arnold}, C., {Leo}, M., \& {Li}, B. 2019, Nature Astron., 3, 945

\bibitem[{Arnold {et~al.}(2022)Arnold, Li, Giblin, Harnois-D\'eraps, \&
  Cai}]{Arnold:2021xtm}
Arnold, C., Li, B., Giblin, B., Harnois-D\'eraps, J., \& Cai, Y.-C. 2022,
  \mnras, 515, 4161

\bibitem[{{Ayaita} {et~al.}(2016){Ayaita}, {Baldi}, {F{\"u}hrer}, {Puchwein},
  \& {Wetterich}}]{Ayaita:2014una}
{Ayaita}, Y., {Baldi}, M., {F{\"u}hrer}, F., {Puchwein}, E., \& {Wetterich}, C.
  2016, \prd, 93, 063511

\bibitem[{Baldi(2011)}]{Baldi:2010vv}
Baldi, M. 2011, \mnras, 411, 1077

\bibitem[{Baldi(2012{\natexlab{a}})}]{Baldi:2011qi}
Baldi, M. 2012{\natexlab{a}}, \mnras, 422, 1028

\bibitem[{Baldi(2012{\natexlab{b}})}]{Baldi:2011mt}
Baldi, M. 2012{\natexlab{b}}, ASP Conf. Ser., 453, 155

\bibitem[{{Baldi}(2023)}]{Baldi:2022uwb}
{Baldi}, M. 2023, \mnras, 521, 613

\bibitem[{{Baldi} {et~al.}(2010){Baldi}, {Pettorino}, {Robbers}, \&
  {Springel}}]{Baldi:2008ay}
{Baldi}, M., {Pettorino}, V., {Robbers}, G., \& {Springel}, V. 2010, \mnras,
  403, 1684

\bibitem[{{Baldi} \& {Simpson}(2015)}]{Baldi:2014ica}
{Baldi}, M. \& {Simpson}, F. 2015, \mnras, 449, 2239

\bibitem[{{Baldi} \& {Simpson}(2017)}]{Baldi:2016zom}
{Baldi}, M. \& {Simpson}, F. 2017, \mnras, 465, 653

\bibitem[{{Baldi} {et~al.}(2014){Baldi}, {Villaescusa-Navarro}, {Viel},
  {Puchwein}, {Springel}, \& {Moscardini}}]{Baldi:2013iza}
{Baldi}, M., {Villaescusa-Navarro}, F., {Viel}, M., {et~al.} 2014, \mnras, 440,
  75

\bibitem[{{Barreira} {et~al.}(2015){Barreira}, {Bose}, \&
  {Li}}]{Barreira:2015xvp}
{Barreira}, A., {Bose}, S., \& {Li}, B. 2015, \jcap, 12, 059

\bibitem[{{Barreira} {et~al.}(2013){Barreira}, {Li}, {Hellwing}, {Baugh}, \&
  {Pascoli}}]{Barreira:2013eea}
{Barreira}, A., {Li}, B., {Hellwing}, W.~A., {Baugh}, C.~M., \& {Pascoli}, S.
  2013, \jcap, 10, 027

\bibitem[{{Barreira} {et~al.}(2014){Barreira}, {Li}, {Hellwing}, {Baugh}, \&
  {Pascoli}}]{Barreira:2014kra}
{Barreira}, A., {Li}, B., {Hellwing}, W.~A., {Baugh}, C.~M., \& {Pascoli}, S.
  2014, \jcap, 09, 031

\bibitem[{{Becker} {et~al.}(2020){Becker}, {Arnold}, {Li}, \&
  {Heisenberg}}]{Becker:2020azq}
{Becker}, C., {Arnold}, C., {Li}, B., \& {Heisenberg}, L. 2020, \jcap, 10, 055

\bibitem[{{Bose} {et~al.}(2015){Bose}, {Hellwing}, \& {Li}}]{Bose:2014zba}
{Bose}, S., {Hellwing}, W.~A., \& {Li}, B. 2015, \jcap, 02, 034

\bibitem[{{Bose} {et~al.}(2017){Bose}, {Li}, {Barreira}, {He}, {Hellwing},
  {Koyama}, {Llinares}, \& {Zhao}}]{Bose:2016wms}
{Bose}, S., {Li}, B., {Barreira}, A., {et~al.} 2017, \jcap, 02, 050

\bibitem[{{Brando} {et~al.}(2022){Brando}, {Fiorini}, {Koyama}, \&
  {Winther}}]{Brando:2022gvg}
{Brando}, G., {Fiorini}, B., {Koyama}, K., \& {Winther}, H.~A. 2022, \jcap, 09,
  051

\bibitem[{{Brando} {et~al.}(2023){Brando}, {Koyama}, \&
  {Winther}}]{Brando:2023fzu}
{Brando}, G., {Koyama}, K., \& {Winther}, H.~A. 2023, \jcap, 06, 045

\bibitem[{Brandt(1977)}]{Brandt_1977}
Brandt, A. 1977, Mathematics of Computation, 31, 333

\bibitem[{{Brax} {et~al.}(2012){Brax}, {Davis}, {Li}, {Winther}, \&
  {Zhao}}]{2012JCAP...10..002B}
{Brax}, P., {Davis}, A.-C., {Li}, B., {Winther}, H.~A., \& {Zhao}, G.-B. 2012,
  \jcap, 10, 002

\bibitem[{{Brax} {et~al.}(2013){Brax}, {Davis}, {Li}, {Winther}, \&
  {Zhao}}]{Brax:2013mua}
{Brax}, P., {Davis}, A.-C., {Li}, B., {Winther}, H.~A., \& {Zhao}, G.-B. 2013,
  \jcap, 04, 029

\bibitem[{{Chan} \& {Scoccimarro}(2009)}]{Chan:2009ew}
{Chan}, K.~C. \& {Scoccimarro}, R. 2009, \prd, 80, 104005

\bibitem[{{Cheung} {et~al.}(2008){Cheung}, {Fitzpatrick}, {Kaplan}, {Senatore},
  \& {Creminelli}}]{Cheung:2007st}
{Cheung}, C., {Fitzpatrick}, A.~L., {Kaplan}, J., {Senatore}, L., \&
  {Creminelli}, P. 2008, JHEP, 3, 014

\bibitem[{{Chevallier} \& {Polarski}(2001)}]{Chevallier:2000qy}
{Chevallier}, M. \& {Polarski}, D. 2001, Int.\ J.\ Mod.\ Phys.\ D, 10, 213

\bibitem[{{Clifton} {et~al.}(2012){Clifton}, {Ferreira}, {Padilla}, \&
  {Skordis}}]{Clifton:2011jh}
{Clifton}, T., {Ferreira}, P.~G., {Padilla}, A., \& {Skordis}, C. 2012,
  \physrep, 513, 1

\bibitem[{Damour {et~al.}(1990)Damour, Gibbons, \& Gundlach}]{Damour:1990tw}
Damour, T., Gibbons, G.~W., \& Gundlach, C. 1990, \prl, 64, 123

\bibitem[{{DESI Collaboration: Aghamousa} {et~al.}(2016){DESI Collaboration:
  Aghamousa}, {Aguilar}, {Ahlen}, {Alam}, {Allen}, {Allende Prieto}, {Annis},
  {Bailey}, {Balland}, {Ballester}, {Baltay}, {Beaufore}, {Bebek}, {Beers},
  {Bell}, {Bernal}, {Besuner}, {Beutler}, {Blake}, {Bleuler}, {Blomqvist},
  {Blum}, {Bolton}, {Briceno}, {Brooks}, {Brownstein}, {Buckley-Geer},
  {Burden}, {Burtin}, {Busca}, {Cahn}, {Cai}, {Cardiel-Sas}, {Carlberg},
  {Carton}, {Casas}, {Castander}, {Cervantes-Cota}, {Claybaugh}, {Close},
  {Coker}, {Cole}, {Comparat}, {Cooper}, {Cousinou}, {Crocce}, {Cuby},
  {Cunningham}, {Davis}, {Dawson}, {de la Macorra}, {De Vicente}, {Delubac},
  {Derwent}, {Dey}, {Dhungana}, {Ding}, {Doel}, {Duan}, {Ealet}, {Edelstein},
  {Eftekharzadeh}, {Eisenstein}, {Elliott}, {Escoffier}, {Evatt}, {Fagrelius},
  {Fan}, {Fanning}, {Farahi}, {Farihi}, {Favole}, {Feng}, {Fernandez},
  {Findlay}, {Finkbeiner}, {Fitzpatrick}, {Flaugher}, {Flender}, {Font-Ribera},
  {Forero-Romero}, {Fosalba}, {Frenk}, {Fumagalli}, {Gaensicke}, {Gallo},
  {Garcia-Bellido}, {Gaztanaga}, {Pietro Gentile Fusillo}, {Gerard},
  {Gershkovich}, {Giannantonio}, {Gillet}, {Gonzalez-de-Rivera},
  {Gonzalez-Perez}, {Gott}, {Graur}, {Gutierrez}, {Guy}, {Habib}, {Heetderks},
  {Heetderks}, {Heitmann}, {Hellwing}, {Herrera}, {Ho}, {Holland}, {Honscheid},
  {Huff}, {Hutchinson}, {Huterer}, {Hwang}, {Illa Laguna}, {Ishikawa},
  {Jacobs}, {Jeffrey}, {Jelinsky}, {Jennings}, {Jiang}, {Jimenez}, {Johnson},
  {Joyce}, {Jullo}, {Juneau}, {Kama}, {Karcher}, {Karkar}, {Kehoe}, {Kennamer},
  {Kent}, {Kilbinger}, {Kim}, {Kirkby}, {Kisner}, {Kitanidis}, {Kneib},
  {Koposov}, {Kovacs}, {Koyama}, {Kremin}, {Kron}, {Kronig}, {Kueter-Young},
  {Lacey}, {Lafever}, {Lahav}, {Lambert}, {Lampton}, {Landriau}, {Lang},
  {Lauer}, {Le Goff}, {Le Guillou}, {Le Van Suu}, {Lee}, {Lee}, {Leitner},
  {Lesser}, {Levi}, {L'Huillier}, {Li}, {Liang}, {Lin}, {Linder}, {Loebman},
  {Luki{\'c}}, {Ma}, {MacCrann}, {Magneville}, {Makarem}, {Manera}, {Manser},
  {Marshall}, {Martini}, {Massey}, {Matheson}, {McCauley}, {McDonald},
  {McGreer}, {Meisner}, {Metcalfe}, {Miller}, {Miquel}, {Moustakas}, {Myers},
  {Naik}, {Newman}, {Nichol}, {Nicola}, {Nicolati da Costa}, {Nie}, {Niz},
  {Norberg}, {Nord}, {Norman}, {Nugent}, {O'Brien}, {Oh}, {Olsen}, {Padilla},
  {Padmanabhan}, {Padmanabhan}, {Palanque-Delabrouille}, {Palmese},
  {Pappalardo}, {P{\^a}ris}, {Park}, {Patej}, {Peacock}, {Peiris}, {Peng},
  {Percival}, {Perruchot}, {Pieri}, {Pogge}, {Pollack}, {Poppett}, {Prada},
  {Prakash}, {Probst}, {Rabinowitz}, {Raichoor}, {Ree}, {Refregier}, {Regal},
  {Reid}, {Reil}, {Rezaie}, {Rockosi}, {Roe}, {Ronayette}, {Roodman}, {Ross},
  {Ross}, {Rossi}, {Rozo}, {Ruhlmann-Kleider}, {Rykoff}, {Sabiu}, {Samushia},
  {Sanchez}, {Sanchez}, {Schlegel}, {Schneider}, {Schubnell}, {Secroun},
  {Seljak}, {Seo}, {Serrano}, {Shafieloo}, {Shan}, {Sharples}, {Sholl},
  {Shourt}, {Silber}, {Silva}, {Sirk}, {Slosar}, {Smith}, {Smoot}, {Som},
  {Song}, {Sprayberry}, {Staten}, {Stefanik}, {Tarle}, {Sien Tie}, {Tinker},
  {Tojeiro}, {Valdes}, {Valenzuela}, {Valluri}, {Vargas-Magana}, {Verde},
  {Walker}, {Wang}, {Wang}, {Weaver}, {Weaverdyck}, {Wechsler}, {Weinberg},
  {White}, {Yang}, {Yeche}, {Zhang}, {Zhao}, {Zheng}, {Zhou}, {Zhou}, {Zhu},
  {Zou}, \& {Zu}}]{DESI:2016fyo}
{DESI Collaboration: Aghamousa}, A., {Aguilar}, J., {Ahlen}, S., {et~al.} 2016,
  arXiv:1611.00036

\bibitem[{{Di Valentino} {et~al.}(2021{\natexlab{a}}){Di Valentino},
  {Anchordoqui}, {Akarsu}, {Ali-Haimoud}, {Amendola}, {Arendse}, {Asgari},
  {Ballardini}, {Basilakos}, {Battistelli}, {Benetti}, {Birrer}, {Bouchet},
  {Bruni}, {Calabrese}, {Camarena}, {Capozziello}, {Chen}, {Chluba},
  {Chudaykin}, {Colg{\'a}in}, {Cyr-Racine}, {de Bernardis}, {de Cruz
  P{\'e}rez}, {Delabrouille}, {Dunkley}, {Escamilla-Rivera}, {Fert{\'e}},
  {Finelli}, {Freedman}, {Frusciante}, {Giusarma}, {G{\'o}mez-Valent}, {Guy},
  {Handley}, {Harrison}, {Hart}, {Heavens}, {Hildebrandt}, {Holz}, {Huterer},
  {Ivanov}, {Joudaki}, {Kamionkowski}, {Karwal}, {Knox}, {Kumar}, {Lamagna},
  {Lesgourgues}, {Lucca}, {Marra}, {Masi}, {Matarrese}, {Mazumdar},
  {Melchiorri}, {Mena}, {Mersini-Houghton}, {Miranda}, {Moreno-Pulido}, {Mota},
  {Muir}, {Mukherjee}, {Niedermann}, {Notari}, {Nunes}, {Pace},
  {Paliathanasis}, {Palmese}, {Pan}, {Paoletti}, {Pettorino}, {Piacentini},
  {Poulin}, {Raveri}, {Riess}, {Salzano}, {Saridakis}, {Sen}, {Shafieloo},
  {Shajib}, {Silk}, {Silvestri}, {Sloth}, {Smith}, {Sol{\`a} Peracaula}, {van
  de Bruck}, {Verde}, {Visinelli}, {Wandelt}, {Wang}, {Wang}, {Yadav}, \&
  {Yang}}]{DiValentino:2020zio}
{Di Valentino}, E., {Anchordoqui}, L.~A., {Akarsu}, {\"O}., {et~al.}
  2021{\natexlab{a}}, Astropart. Phys., 131, 102605

\bibitem[{{Di Valentino} {et~al.}(2021{\natexlab{b}}){Di Valentino},
  {Anchordoqui}, {Akarsu}, {Ali-Haimoud}, {Amendola}, {Arendse}, {Asgari},
  {Ballardini}, {Basilakos}, {Battistelli}, {Benetti}, {Birrer}, {Bouchet},
  {Bruni}, {Calabrese}, {Camarena}, {Capozziello}, {Chen}, {Chluba},
  {Chudaykin}, {Colg{\'a}in}, {Cyr-Racine}, {de Bernardis}, {de Cruz
  P{\'e}rez}, {Delabrouille}, {Dunkley}, {Escamilla-Rivera}, {Fert{\'e}},
  {Finelli}, {Freedman}, {Frusciante}, {Giusarma}, {G{\'o}mez-Valent},
  {Handley}, {Harrison}, {Hart}, {Heavens}, {Hildebrandt}, {Holz}, {Huterer},
  {Ivanov}, {Joudaki}, {Kamionkowski}, {Karwal}, {Knox}, {Kumar}, {Lamagna},
  {Lesgourgues}, {Lucca}, {Marra}, {Masi}, {Matarrese}, {Mazumdar},
  {Melchiorri}, {Mena}, {Mersini-Houghton}, {Miranda}, {Moreno-Pulido}, {Mota},
  {Muir}, {Mukherjee}, {Niedermann}, {Notari}, {Nunes}, {Pace},
  {Paliathanasis}, {Palmese}, {Pan}, {Paoletti}, {Pettorino}, {Piacentini},
  {Poulin}, {Raveri}, {Riess}, {Salzano}, {Saridakis}, {Sen}, {Shafieloo},
  {Shajib}, {Silk}, {Silvestri}, {Sloth}, {Smith}, {Sol{\`a} Peracaula}, {van
  de Bruck}, {Verde}, {Visinelli}, {Wandelt}, {Wang}, {Wang}, {Yadav}, \&
  {Yang}}]{DiValentino:2020vvd}
{Di Valentino}, E., {Anchordoqui}, L.~A., {Akarsu}, {\"O}., {et~al.}
  2021{\natexlab{b}}, Astropart. Phys., 131, 102604

\bibitem[{{Dvali} {et~al.}(2000){Dvali}, {Gabadadze}, \&
  {Porrati}}]{Dvali:2000hr}
{Dvali}, G., {Gabadadze}, G., \& {Porrati}, M. 2000, Phys.\ Lett.\ B, 485, 208

\bibitem[{Elbers {et~al.}(2021)Elbers, Frenk, Jenkins, Li, \&
  Pascoli}]{Elbers:2020lbn}
Elbers, W., Frenk, C.~S., Jenkins, A., Li, B., \& Pascoli, S. 2021, \mnras,
  507, 2614

\bibitem[{{Euclid Collaboration: Adam} {et~al.}(2019){Euclid Collaboration:
  Adam}, {Vannier}, {Maurogordato}, {Biviano}, {Adami}, {Ascaso}, {Bellagamba},
  {Benoist}, {Cappi}, {D{\'\i}az-S{\'a}nchez}, {Durret}, {Farrens}, {Gonzalez},
  {Iovino}, {Licitra}, {Maturi}, {Mei}, {Merson}, {Munari}, {Pell{\'o}},
  {Ricci}, {Rocci}, {Roncarelli}, {Sarron}, {Amoura}, {Andreon}, {Apostolakos},
  {Arnaud}, {Bardelli}, {Bartlett}, {Baugh}, {Borgani}, {Brodwin}, {Castander},
  {Castignani}, {Cucciati}, {De Lucia}, {Dubath}, {Fosalba}, {Giocoli},
  {Hoekstra}, {Mamon}, {Melin}, {Moscardini}, {Paltani}, {Radovich},
  {Sartoris}, {Schultheis}, {Sereno}, {Weller}, {Burigana}, {Carvalho},
  {Corcione}, {Kurki-Suonio}, {Lilje}, {Sirri}, {Toledo-Moreo}, \&
  {Zamorani}}]{Euclid:2019bue}
{Euclid Collaboration: Adam}, R., {Vannier}, M., {Maurogordato}, S., {et~al.}
  2019, \aap, 627, A23

\bibitem[{{Euclid Collaboration: Ajani} {et~al.}(2023){Euclid Collaboration:
  Ajani}, {Baldi}, {Barthelemy}, {Boyle}, {Burger}, {Cardone}, {Cheng},
  {Codis}, {Giocoli}, {Harnois-D{\'e}raps}, {Heydenreich}, {Kansal},
  {Kilbinger}, {Linke}, {Llinares}, {Martinet}, {Parroni}, {Peel}, {Pires},
  {Porth}, {Tereno}, {Uhlemann}, {Vicinanza}, {Vinciguerra}, {Aghanim},
  {Auricchio}, {Bonino}, {Branchini}, {Brescia}, {Brinchmann}, {Camera},
  {Capobianco}, {Carbone}, {Carretero}, {Castander}, {Castellano}, {Cavuoti},
  {Cimatti}, {Cledassou}, {Congedo}, {Conselice}, {Conversi}, {Corcione},
  {Courbin}, {Cropper}, {Da Silva}, {Degaudenzi}, {Di Giorgio}, {Dinis},
  {Douspis}, {Dubath}, {Dupac}, {Farrens}, {Ferriol}, {Fosalba}, {Frailis},
  {Franceschi}, {Galeotta}, {Garilli}, {Gillis}, {Grazian}, {Grupp},
  {Hoekstra}, {Holmes}, {Hornstrup}, {Hudelot}, {Jahnke}, {Jhabvala},
  {K{\"u}mmel}, {Kitching}, {Kunz}, {Kurki-Suonio}, {Lilje}, {Lloro},
  {Maiorano}, {Mansutti}, {Marggraf}, {Markovic}, {Marulli}, {Massey}, {Mei},
  {Mellier}, {Meneghetti}, {Moresco}, {Moscardini}, {Niemi}, {Nightingale},
  {Nutma}, {Padilla}, {Paltani}, {Pedersen}, {Pettorino}, {Polenta}, {Poncet},
  {Popa}, {Raison}, {Renzi}, {Rhodes}, {Riccio}, {Romelli}, {Roncarelli},
  {Rossetti}, {Saglia}, {Sapone}, {Sartoris}, {Schneider}, {Schrabback},
  {Secroun}, {Seidel}, {Serrano}, {Sirignano}, {Stanco}, {Starck},
  {Tallada-Cresp{\'\i}}, {Taylor}, {Toledo-Moreo}, {Torradeflot}, {Tutusaus},
  {Valentijn}, {Valenziano}, {Vassallo}, {Wang}, {Weller}, {Zamorani},
  {Zoubian}, {Andreon}, {Bardelli}, {Boucaud}, {Bozzo}, {Colodro-Conde}, {Di
  Ferdinando}, {Fabbian}, {Farina}, {Graci{\'a}-Carpio}, {Keih{\"a}nen},
  {Lindholm}, {Maino}, {Mauri}, {Neissner}, {Schirmer}, {Scottez}, {Zucca},
  {Akrami}, {Baccigalupi}, {Balaguera-Antol{\'\i}nez}, {Ballardini},
  {Bernardeau}, {Biviano}, {Blanchard}, {Borgani}, {Borlaff}, {Burigana},
  {Cabanac}, {Cappi}, {Carvalho}, {Casas}, {Castignani}, {Castro}, {Chambers},
  {Cooray}, {Coupon}, {Courtois}, {Davini}, {de la Torre}, {De Lucia},
  {Desprez}, {Dole}, {Escartin}, {Escoffier}, {Ferrero}, {Finelli}, {Ganga},
  {Garcia-Bellido}, {George}, {Giacomini}, {Gozaliasl}, {Hildebrandt}, {Jimenez
  Mu{\~n}oz}, {Joachimi}, {Kajava}, {Kirkpatrick}, {Legrand}, {Loureiro},
  {Magliocchetti}, {Maoli}, {Marcin}, {Martinelli}, {Martins}, {Matthew},
  {Maurin}, {Metcalf}, {Monaco}, {Morgante}, {Nadathur}, {Nucita}, {Popa},
  {Potter}, {Pourtsidou}, {P{\"o}ntinen}, {Reimberg}, {S{\'a}nchez}, {Sakr},
  {Schneider}, {Sefusatti}, {Sereno}, {Shulevski}, {Spurio Mancini},
  {Steinwagner}, {Teyssier}, {Valiviita}, {Veropalumbo}, {Viel}, \&
  {Zinchenko}}]{Euclid:2023uha}
{Euclid Collaboration: Ajani}, V., {Baldi}, M., {Barthelemy}, A., {et~al.}
  2023, \aap, 675, A120

\bibitem[{{Euclid Collaboration: Bretonni{\`e}re} {et~al.}(2022){Euclid
  Collaboration: Bretonni{\`e}re}, {Huertas-Company}, {Boucaud}, {Lanusse},
  {Jullo}, {Merlin}, {Tuccillo}, {Castellano}, {Brinchmann}, {Conselice},
  {Dole}, {Cabanac}, {Courtois}, {Castander}, {Duc}, {Fosalba}, {Guinet},
  {Kruk}, {Kuchner}, {Serrano}, {Soubrie}, {Tramacere}, {Wang}, {Amara},
  {Auricchio}, {Bender}, {Bodendorf}, {Bonino}, {Branchini}, {Brau-Nogue},
  {Brescia}, {Capobianco}, {Carbone}, {Carretero}, {Cavuoti}, {Cimatti},
  {Cledassou}, {Congedo}, {Conversi}, {Copin}, {Corcione}, {Costille},
  {Cropper}, {Da Silva}, {Degaudenzi}, {Douspis}, {Dubath}, {Duncan}, {Dupac},
  {Dusini}, {Farrens}, {Ferriol}, {Frailis}, {Franceschi}, {Fumana}, {Garilli},
  {Gillard}, {Gillis}, {Giocoli}, {Grazian}, {Grupp}, {Haugan}, {Holmes},
  {Hormuth}, {Hudelot}, {Jahnke}, {Kermiche}, {Kiessling}, {Kilbinger},
  {Kitching}, {Kohley}, {K{\"u}mmel}, {Kunz}, {Kurki-Suonio}, {Ligori},
  {Lilje}, {Lloro}, {Maiorano}, {Mansutti}, {Marggraf}, {Markovic}, {Marulli},
  {Massey}, {Maurogordato}, {Melchior}, {Meneghetti}, {Meylan}, {Moresco},
  {Morin}, {Moscardini}, {Munari}, {Nakajima}, {Niemi}, {Padilla}, {Paltani},
  {Pasian}, {Pedersen}, {Pettorino}, {Pires}, {Poncet}, {Popa}, {Pozzetti},
  {Raison}, {Rebolo}, {Rhodes}, {Roncarelli}, {Rossetti}, {Saglia},
  {Schneider}, {Secroun}, {Seidel}, {Sirignano}, {Sirri}, {Stanco}, {Starck},
  {Tallada-Cresp{\'\i}}, {Taylor}, {Tereno}, {Toledo-Moreo}, {Torradeflot},
  {Valentijn}, {Valenziano}, {Wang}, {Welikala}, {Weller}, {Zamorani},
  {Zoubian}, {Baldi}, {Bardelli}, {Camera}, {Farinelli}, {Medinaceli}, {Mei},
  {Polenta}, {Romelli}, {Tenti}, {Vassallo}, {Zacchei}, {Zucca}, {Baccigalupi},
  {Balaguera-Antol{\'\i}nez}, {Biviano}, {Borgani}, {Bozzo}, {Burigana},
  {Cappi}, {Carvalho}, {Casas}, {Castignani}, {Colodro-Conde}, {Coupon}, {de la
  Torre}, {Fabricius}, {Farina}, {Ferreira}, {Flose-Reimberg}, {Fotopoulou},
  {Galeotta}, {Ganga}, {Garcia-Bellido}, {Gaztanaga}, {Gozaliasl}, {Hook},
  {Joachimi}, {Kansal}, {Kashlinsky}, {Keihanen}, {Kirkpatrick}, {Lindholm},
  {Mainetti}, {Maino}, {Maoli}, {Martinelli}, {Martinet}, {McCracken},
  {Metcalf}, {Morgante}, {Morisset}, {Nightingale}, {Nucita}, {Patrizii},
  {Potter}, {Renzi}, {Riccio}, {S{\'a}nchez}, {Sapone}, {Schirmer},
  {Schultheis}, {Scottez}, {Sefusatti}, {Teyssier}, {Tutusaus}, {Valiviita},
  {Viel}, {Whittaker}, \& {Knapen}}]{2022A&A...657A..90E}
{Euclid Collaboration: Bretonni{\`e}re}, H., {Huertas-Company}, M., {Boucaud},
  A., {et~al.} 2022, \aap, 657, A90

\bibitem[{{Euclid Collaboration: Bretonni{\`e}re} {et~al.}(2023){Euclid
  Collaboration: Bretonni{\`e}re}, {Kuchner}, {Huertas-Company}, {Merlin},
  {Castellano}, {Tuccillo}, {Buitrago}, {Conselice}, {Boucaud},
  {H{\"a}u{\ss}ler}, {K{\"u}mmel}, {Hartley}, {Alvarez Ayllon}, {Bertin},
  {Ferrari}, {Ferreira}, {Gavazzi}, {Hern{\'a}ndez-Lang}, {Lucatelli},
  {Robotham}, {Schefer}, {Wang}, {Cabanac}, {Dom{\'\i}nguez S{\'a}nchez},
  {Duc}, {Fotopoulou}, {Kruk}, {La Marca}, {Margalef-Bentabol}, {Marleau},
  {Tortora}, {Aghanim}, {Amara}, {Auricchio}, {Azzollini}, {Baldi}, {Bender},
  {Bodendorf}, {Branchini}, {Brescia}, {Brinchmann}, {Camera}, {Capobianco},
  {Carbone}, {Carretero}, {Castander}, {Cavuoti}, {Cimatti}, {Cledassou},
  {Congedo}, {Conversi}, {Copin}, {Corcione}, {Courbin}, {Cropper}, {Da Silva},
  {Degaudenzi}, {Dinis}, {Dubath}, {Duncan}, {Dupac}, {Dusini}, {Farrens},
  {Ferriol}, {Frailis}, {Franceschi}, {Fumana}, {Galeotta}, {Garilli},
  {Gillis}, {Giocoli}, {Grazian}, {Grupp}, {Haugan}, {Hoekstra}, {Holmes},
  {Hormuth}, {Hornstrup}, {Hudelot}, {Jahnke}, {Kermiche}, {Kiessling},
  {Kohley}, {Kunz}, {Kurki-Suonio}, {Ligori}, {Lilje}, {Lloro}, {Mansutti},
  {Marggraf}, {Markovic}, {Marulli}, {Massey}, {McCracken}, {Medinaceli},
  {Melchior}, {Meneghetti}, {Meylan}, {Moresco}, {Moscardini}, {Munari},
  {Niemi}, {Padilla}, {Paltani}, {Pasian}, {Pedersen}, {Percival}, {Pettorino},
  {Polenta}, {Poncet}, {Pozzetti}, {Raison}, {Rebolo}, {Renzi}, {Rhodes},
  {Riccio}, {Romelli}, {Rosset}, {Rossetti}, {Saglia}, {Sapone}, {Sartoris},
  {Schneider}, {Secroun}, {Seidel}, {Sirignano}, {Sirri}, {Skottfelt},
  {Starck}, {Tallada-Cresp{\'\i}}, {Taylor}, {Tereno}, {Toledo-Moreo},
  {Tutusaus}, {Valentijn}, {Valenziano}, {Vassallo}, {Wang}, {Weller},
  {Zamorani}, {Zoubian}, {Andreon}, {Bardelli}, {Colodro-Conde}, {Di
  Ferdinando}, {Graci{\'a}-Carpio}, {Lindholm}, {Mauri}, {Mei}, {Scottez},
  {Zucca}, {Baccigalupi}, {Ballardini}, {Bernardeau}, {Biviano}, {Borgani},
  {Borlaff}, {Burigana}, {Cappi}, {Carvalho}, {Casas}, {Castignani}, {Cooray},
  {Coupon}, {Courtois}, {Davini}, {De Lucia}, {Desprez}, {Escartin},
  {Escoffier}, {Fabricius}, {Farina}, {Fontana}, {Ganga}, {Garcia-Bellido},
  {George}, {Gozaliasl}, {Hildebrandt}, {Hook}, {Ilbert}, {Ili{\'c}},
  {Joachimi}, {Kansal}, {Keihanen}, {Kirkpatrick}, {Loureiro}, {Macias-Perez},
  {Magliocchetti}, {Maoli}, {Marcin}, {Martinelli}, {Martinet}, {Maturi},
  {Monaco}, {Morgante}, {Nadathur}, {Nucita}, {Patrizii}, {Popa}, {Porciani},
  {Potter}, {Pourtsidou}, {P{\"o}ntinen}, {Reimberg}, {S{\'a}nchez}, {Sakr},
  {Schirmer}, {Sefusatti}, {Sereno}, {Stadel}, {Teyssier}, {Valiviita}, {van
  Mierlo}, {Veropalumbo}, {Viel}, {Weaver}, \& {Scott}}]{2023A&A...671A.102E}
{Euclid Collaboration: Bretonni{\`e}re}, H., {Kuchner}, U., {Huertas-Company},
  M., {et~al.} 2023, \aap, 671, A102

\bibitem[{{Euclid Collaboration: Castander} {et~al.}(2024){Euclid
  Collaboration: Castander}, {Fosalba}, {Stadel}, {et~al.}}]{Euclid:2024few}
{Euclid Collaboration: Castander}, F.~J., {Fosalba}, P., {Stadel}, J., {et~al.}
  2024 [\eprint[arXiv]{2405.13495}]

\bibitem[{{Euclid Collaboration: Castro} {et~al.}(2023){Euclid Collaboration:
  Castro}, {Fumagalli}, {Angulo}, {Bocquet}, {Borgani}, {Carbone}, {Dakin},
  {Dolag}, {Giocoli}, {Monaco}, {Ragagnin}, {Saro}, {Sefusatti}, {Costanzi},
  {Le Brun}, {Corasaniti}, {Amara}, {Amendola}, {Baldi}, {Bender}, {Bodendorf},
  {Branchini}, {Brescia}, {Camera}, {Capobianco}, {Carretero}, {Castellano},
  {Cavuoti}, {Cimatti}, {Cledassou}, {Congedo}, {Conversi}, {Copin},
  {Corcione}, {Courbin}, {Da Silva}, {Degaudenzi}, {Douspis}, {Dubath},
  {Duncan}, {Dupac}, {Farrens}, {Ferriol}, {Fosalba}, {Frailis}, {Franceschi},
  {Galeotta}, {Garilli}, {Gillis}, {Grazian}, {Grupp}, {Haugan}, {Hormuth},
  {Hornstrup}, {Hudelot}, {Jahnke}, {Kermiche}, {Kitching}, {Kunz},
  {Kurki-Suonio}, {Lilje}, {Lloro}, {Mansutti}, {Marggraf}, {Marulli},
  {Meneghetti}, {Merlin}, {Meylan}, {Moresco}, {Moscardini}, {Munari}, {Niemi},
  {Padilla}, {Paltani}, {Pasian}, {Pedersen}, {Pettorino}, {Pires}, {Polenta},
  {Poncet}, {Popa}, {Pozzetti}, {Raison}, {Rebolo}, {Renzi}, {Rhodes},
  {Riccio}, {Romelli}, {Saglia}, {Sapone}, {Sartoris}, {Schneider}, {Seidel},
  {Sirri}, {Stanco}, {Tallada Cresp{\'\i}}, {Taylor}, {Toledo-Moreo},
  {Torradeflot}, {Tutusaus}, {Valentijn}, {Valenziano}, {Vassallo}, {Wang},
  {Weller}, {Zacchei}, {Zamorani}, {Andreon}, {Bardelli}, {Bozzo},
  {Colodro-Conde}, {Di Ferdinando}, {Farina}, {Graci{\'a}-Carpio}, {Lindholm},
  {Neissner}, {Scottez}, {Tenti}, {Zucca}, {Baccigalupi},
  {Balaguera-Antol{\'\i}nez}, {Ballardini}, {Bernardeau}, {Biviano},
  {Blanchard}, {Borlaff}, {Burigana}, {Cabanac}, {Cappi}, {Carvalho}, {Casas},
  {Castignani}, {Cooray}, {Coupon}, {Courtois}, {Davini}, {De Lucia},
  {Desprez}, {Dole}, {Escartin}, {Escoffier}, {Finelli}, {Ganga},
  {Garcia-Bellido}, {George}, {Gozaliasl}, {Hildebrandt}, {Hook}, {Ili{\'c}},
  {Kansal}, {Keihanen}, {Kirkpatrick}, {Loureiro}, {Macias-Perez},
  {Magliocchetti}, {Maoli}, {Marcin}, {Martinelli}, {Martinet}, {Matthew},
  {Maturi}, {Metcalf}, {Morgante}, {Nadathur}, {Nucita}, {Patrizii}, {Peel},
  {Popa}, {Porciani}, {Potter}, {Pourtsidou}, {P{\"o}ntinen}, {S{\'a}nchez},
  {Sakr}, {Schirmer}, {Sereno}, {Spurio Mancini}, {Teyssier}, {Valiviita},
  {Veropalumbo}, \& {Viel}}]{Euclid:2022dbc}
{Euclid Collaboration: Castro}, T., {Fumagalli}, A., {Angulo}, R.~E., {et~al.}
  2023, \aap, 671, A100

\bibitem[{{Euclid Collaboration: Desprez} {et~al.}(2020){Euclid Collaboration:
  Desprez}, {Paltani}, {Coupon}, {Almosallam}, {Alvarez-Ayllon}, {Amaro},
  {Brescia}, {Brodwin}, {Cavuoti}, {De Vicente-Albendea}, {Fotopoulou},
  {Hatfield}, {Hartley}, {Ilbert}, {Jarvis}, {Longo}, {Rau}, {Saha}, {Speagle},
  {Tramacere}, {Castellano}, {Dubath}, {Galametz}, {Kuemmel}, {Laigle},
  {Merlin}, {Mohr}, {Pilo}, {Salvato}, {Andreon}, {Auricchio}, {Baccigalupi},
  {Balaguera-Antol{\'\i}nez}, {Baldi}, {Bardelli}, {Bender}, {Biviano},
  {Bodendorf}, {Bonino}, {Bozzo}, {Branchini}, {Brinchmann}, {Burigana},
  {Cabanac}, {Camera}, {Capobianco}, {Cappi}, {Carbone}, {Carretero},
  {Carvalho}, {Casas}, {Casas}, {Castander}, {Castignani}, {Cimatti},
  {Cledassou}, {Colodro-Conde}, {Congedo}, {Conselice}, {Conversi}, {Copin},
  {Corcione}, {Courtois}, {Cuby}, {Da Silva}, {de la Torre}, {Degaudenzi}, {Di
  Ferdinando}, {Douspis}, {Duncan}, {Dupac}, {Ealet}, {Fabbian}, {Fabricius},
  {Farrens}, {Ferreira}, {Finelli}, {Fosalba}, {Fourmanoit}, {Frailis},
  {Franceschi}, {Fumana}, {Galeotta}, {Garilli}, {Gillard}, {Gillis},
  {Giocoli}, {Gozaliasl}, {Graci{\'a}-Carpio}, {Grupp}, {Guzzo}, {Hailey},
  {Haugan}, {Holmes}, {Hormuth}, {Humphrey}, {Jahnke}, {Keihanen}, {Kermiche},
  {Kilbinger}, {Kirkpatrick}, {Kitching}, {Kohley}, {Kubik}, {Kunz},
  {Kurki-Suonio}, {Ligori}, {Lilje}, {Lloro}, {Maino}, {Maiorano}, {Marggraf},
  {Markovic}, {Martinet}, {Marulli}, {Massey}, {Maturi}, {Mauri},
  {Maurogordato}, {Medinaceli}, {Mei}, {Meneghetti}, {Metcalf}, {Meylan},
  {Moresco}, {Moscardini}, {Munari}, {Niemi}, {Padilla}, {Pasian}, {Patrizii},
  {Pettorino}, {Pires}, {Polenta}, {Poncet}, {Popa}, {Potter}, {Pozzetti},
  {Raison}, {Renzi}, {Rhodes}, {Riccio}, {Rossetti}, {Saglia}, {Sapone},
  {Schneider}, {Scottez}, {Secroun}, {Serrano}, {Sirignano}, {Sirri}, {Stanco},
  {Stern}, {Sureau}, {Tallada Cresp{\'\i}}, {Tavagnacco}, {Taylor}, {Tenti},
  {Tereno}, {Toledo-Moreo}, {Torradeflot}, {Valenziano}, {Valiviita},
  {Vassallo}, {Viel}, {Wang}, {Welikala}, {Whittaker}, {Zacchei}, {Zamorani},
  {Zoubian}, \& {Zucca}}]{Euclid:2020gbk}
{Euclid Collaboration: Desprez}, G., {Paltani}, S., {Coupon}, J., {et~al.}
  2020, \aap, 644, A31

\bibitem[{{Euclid Collaboration: Ilbert} {et~al.}(2021){Euclid Collaboration:
  Ilbert}, {de la Torre}, {Martinet}, {Wright}, {Paltani}, {Laigle},
  {Davidzon}, {Jullo}, {Hildebrandt}, {Masters}, {Amara}, {Conselice},
  {Andreon}, {Auricchio}, {Azzollini}, {Baccigalupi},
  {Balaguera-Antol{\'\i}nez}, {Baldi}, {Balestra}, {Bardelli}, {Bender},
  {Biviano}, {Bodendorf}, {Bonino}, {Borgani}, {Boucaud}, {Bozzo}, {Branchini},
  {Brescia}, {Burigana}, {Cabanac}, {Camera}, {Capobianco}, {Cappi}, {Carbone},
  {Carretero}, {Carvalho}, {Casas}, {Castander}, {Castellano}, {Castignani},
  {Cavuoti}, {Cimatti}, {Cledassou}, {Colodro-Conde}, {Congedo}, {Conversi},
  {Copin}, {Corcione}, {Costille}, {Coupon}, {Courtois}, {Cropper}, {Cuby}, {Da
  Silva}, {Degaudenzi}, {Di Ferdinando}, {Dubath}, {Duncan}, {Dupac}, {Dusini},
  {Ealet}, {Fabricius}, {Farrens}, {Ferreira}, {Finelli}, {Fosalba},
  {Fotopoulou}, {Franceschi}, {Franzetti}, {Galeotta}, {Garilli}, {Gillard},
  {Gillis}, {Giocoli}, {Gozaliasl}, {Graci{\'a}-Carpio}, {Grupp}, {Guzzo},
  {Haugan}, {Holmes}, {Hormuth}, {Jahnke}, {Keihanen}, {Kermiche}, {Kiessling},
  {Kirkpatrick}, {Kunz}, {Kurki-Suonio}, {Ligori}, {Lilje}, {Lloro}, {Maino},
  {Maiorano}, {Marggraf}, {Markovic}, {Marulli}, {Massey}, {Maturi}, {Mauri},
  {Maurogordato}, {McCracken}, {Medinaceli}, {Mei}, {Metcalf}, {Moresco},
  {Morin}, {Moscardini}, {Munari}, {Nakajima}, {Neissner}, {Niemi},
  {Nightingale}, {Padilla}, {Pasian}, {Patrizii}, {Pedersen}, {Pello},
  {Pettorino}, {Pires}, {Polenta}, {Poncet}, {Popa}, {Potter}, {Pozzetti},
  {Raison}, {Renzi}, {Rhodes}, {Riccio}, {Romelli}, {Roncarelli}, {Rossetti},
  {Saglia}, {S{\'a}nchez}, {Sapone}, {Schneider}, {Schrabback}, {Scottez},
  {Secroun}, {Seidel}, {Serrano}, {Sirignano}, {Sirri}, {Stanco}, {Sureau},
  {Tallada Cresp{\'a}}, {Tenti}, {Teplitz}, {Tereno}, {Toledo-Moreo},
  {Torradeflot}, {Tramacere}, {Valentijn}, {Valenziano}, {Valiviita},
  {Vassallo}, {Wang}, {Welikala}, {Weller}, {Whittaker}, {Zacchei}, {Zamorani},
  {Zoubian}, \& {Zucca}}]{Euclid:2021upd}
{Euclid Collaboration: Ilbert}, O., {de la Torre}, S., {Martinet}, N., {et~al.}
  2021, \aap, 647, A117

\bibitem[{{Euclid Collaboration: Mellier} {et~al.}(2024)}]{Euclid:2024yrr}
{Euclid Collaboration: Mellier}, Y. {et~al.} 2024 [\eprint[arXiv]{2405.13491}]

\bibitem[{{Euclid Collaboration: Merlin} {et~al.}(2023){Euclid Collaboration:
  Merlin}, {Castellano}, {Bretonni{\`e}re}, {Huertas-Company}, {Kuchner},
  {Tuccillo}, {Buitrago}, {Peterson}, {Conselice}, {Caro}, {Dimauro}, {Nemani},
  {Fontana}, {K{\"u}mmel}, {H{\"a}u{\ss}ler}, {Hartley}, {Alvarez Ayllon},
  {Bertin}, {Dubath}, {Ferrari}, {Ferreira}, {Gavazzi}, {Hern{\'a}ndez-Lang},
  {Lucatelli}, {Robotham}, {Schefer}, {Tortora}, {Aghanim}, {Amara},
  {Amendola}, {Auricchio}, {Baldi}, {Bender}, {Bodendorf}, {Branchini},
  {Brescia}, {Camera}, {Capobianco}, {Carbone}, {Carretero}, {Castander},
  {Cavuoti}, {Cimatti}, {Cledassou}, {Congedo}, {Conversi}, {Copin},
  {Corcione}, {Courbin}, {Cropper}, {Da Silva}, {Degaudenzi}, {Dinis},
  {Douspis}, {Dubath}, {Duncan}, {Dupac}, {Dusini}, {Farrens}, {Ferriol},
  {Frailis}, {Franceschi}, {Franzetti}, {Galeotta}, {Garilli}, {Gillis},
  {Giocoli}, {Grazian}, {Grupp}, {Haugan}, {Hoekstra}, {Holmes}, {Hormuth},
  {Hornstrup}, {Hudelot}, {Jahnke}, {Kermiche}, {Kiessling}, {Kitching},
  {Kohley}, {Kunz}, {Kurki-Suonio}, {Ligori}, {Lilje}, {Lloro}, {Mansutti},
  {Marggraf}, {Markovic}, {Marulli}, {Massey}, {McCracken}, {Medinaceli},
  {Melchior}, {Meneghetti}, {Meylan}, {Moresco}, {Moscardini}, {Munari},
  {Niemi}, {Padilla}, {Paltani}, {Pasian}, {Pedersen}, {Percival}, {Polenta},
  {Poncet}, {Popa}, {Pozzetti}, {Raison}, {Rebolo}, {Renzi}, {Rhodes},
  {Riccio}, {Romelli}, {Rossetti}, {Saglia}, {Sapone}, {Sartoris}, {Schneider},
  {Secroun}, {Seidel}, {Sirignano}, {Sirri}, {Skottfelt}, {Starck},
  {Tallada-Cresp{\'\i}}, {Taylor}, {Tereno}, {Toledo-Moreo}, {Tutusaus},
  {Valenziano}, {Vassallo}, {Wang}, {Weller}, {Zacchei}, {Zamorani}, {Zoubian},
  {Andreon}, {Bardelli}, {Boucaud}, {Colodro-Conde}, {Di Ferdinando},
  {Graci{\'a}-Carpio}, {Lindholm}, {Mauri}, {Mei}, {Neissner}, {Scottez},
  {Tramacere}, {Zucca}, {Baccigalupi}, {Balaguera-Antol{\'\i}nez},
  {Ballardini}, {Bernardeau}, {Biviano}, {Borgani}, {Borlaff}, {Burigana},
  {Cabanac}, {Cappi}, {Carvalho}, {Casas}, {Castignani}, {Cooray}, {Coupon},
  {Courtois}, {Cucciati}, {Davini}, {De Lucia}, {Desprez}, {Escartin},
  {Escoffier}, {Farina}, {Ganga}, {Garcia-Bellido}, {George}, {Gozaliasl},
  {Hildebrandt}, {Hook}, {Ilbert}, {Ili{\'c}}, {Joachimi}, {Kansal},
  {Keihanen}, {Kirkpatrick}, {Loureiro}, {Macias-Perez}, {Magliocchetti},
  {Mainetti}, {Maoli}, {Marcin}, {Martinelli}, {Martinet}, {Matthew}, {Maturi},
  {Metcalf}, {Monaco}, {Morgante}, {Nadathur}, {Nucita}, {Patrizii}, {Popa},
  {Porciani}, {Potter}, {Pourtsidou}, {P{\"o}ntinen}, {Reimberg},
  {S{\'a}nchez}, {Sakr}, {Schirmer}, {Sereno}, {Stadel}, {Teyssier}, {Valieri},
  {Valiviita}, {van Mierlo}, {Veropalumbo}, {Viel}, {Weaver}, \&
  {Scott}}]{2023A&A...671A.101E}
{Euclid Collaboration: Merlin}, E., {Castellano}, M., {Bretonni{\`e}re}, H.,
  {et~al.} 2023, \aap, 671, A101

\bibitem[{{Euclid Collaboration: Scaramella} {et~al.}(2022){Euclid
  Collaboration: Scaramella}, {Amiaux}, {Mellier}, {Burigana}, {Carvalho},
  {Cuillandre}, {Da Silva}, {Derosa}, {Dinis}, {Maiorano}, {Maris}, {Tereno},
  {Laureijs}, {Boenke}, {Buenadicha}, {Dupac}, {Gaspar Venancio},
  {G{\'o}mez-{\'A}lvarez}, {Hoar}, {Lorenzo Alvarez}, {Racca},
  {Saavedra-Criado}, {Schwartz}, {Vavrek}, {Schirmer}, {Aussel}, {Azzollini},
  {Cardone}, {Cropper}, {Ealet}, {Garilli}, {Gillard}, {Granett}, {Guzzo},
  {Hoekstra}, {Jahnke}, {Kitching}, {Maciaszek}, {Meneghetti}, {Miller},
  {Nakajima}, {Niemi}, {Pasian}, {Percival}, {Pottinger}, {Sauvage},
  {Scodeggio}, {Wachter}, {Zacchei}, {Aghanim}, {Amara}, {Auphan}, {Auricchio},
  {Awan}, {Balestra}, {Bender}, {Bodendorf}, {Bonino}, {Branchini},
  {Brau-Nogue}, {Brescia}, {Candini}, {Capobianco}, {Carbone}, {Carlberg},
  {Carretero}, {Casas}, {Castander}, {Castellano}, {Cavuoti}, {Cimatti},
  {Cledassou}, {Congedo}, {Conselice}, {Conversi}, {Copin}, {Corcione},
  {Costille}, {Courbin}, {Degaudenzi}, {Douspis}, {Dubath}, {Duncan}, {Dusini},
  {Farrens}, {Ferriol}, {Fosalba}, {Fourmanoit}, {Frailis}, {Franceschi},
  {Franzetti}, {Fumana}, {Gillis}, {Giocoli}, {Grazian}, {Grupp}, {Haugan},
  {Holmes}, {Hormuth}, {Hudelot}, {Kermiche}, {Kiessling}, {Kilbinger},
  {Kohley}, {Kubik}, {K{\"u}mmel}, {Kunz}, {Kurki-Suonio}, {Lahav}, {Ligori},
  {Lilje}, {Lloro}, {Mansutti}, {Marggraf}, {Markovic}, {Marulli}, {Massey},
  {Maurogordato}, {Melchior}, {Merlin}, {Meylan}, {Mohr}, {Moresco}, {Morin},
  {Moscardini}, {Munari}, {Nichol}, {Padilla}, {Paltani}, {Peacock},
  {Pedersen}, {Pettorino}, {Pires}, {Poncet}, {Popa}, {Pozzetti}, {Raison},
  {Rebolo}, {Rhodes}, {Rix}, {Roncarelli}, {Rossetti}, {Saglia}, {Schneider},
  {Schrabback}, {Secroun}, {Seidel}, {Serrano}, {Sirignano}, {Sirri},
  {Skottfelt}, {Stanco}, {Starck}, {Tallada-Cresp{\'\i}}, {Tavagnacco},
  {Taylor}, {Teplitz}, {Toledo-Moreo}, {Torradeflot}, {Trifoglio}, {Valentijn},
  {Valenziano}, {Verdoes Kleijn}, {Wang}, {Welikala}, {Weller}, {Wetzstein},
  {Zamorani}, {Zoubian}, {Andreon}, {Baldi}, {Bardelli}, {Boucaud}, {Camera},
  {Di Ferdinando}, {Fabbian}, {Farinelli}, {Galeotta}, {Graci{\'a}-Carpio},
  {Maino}, {Medinaceli}, {Mei}, {Neissner}, {Polenta}, {Renzi}, {Romelli},
  {Rosset}, {Sureau}, {Tenti}, {Vassallo}, {Zucca}, {Baccigalupi},
  {Balaguera-Antol{\'\i}nez}, {Battaglia}, {Biviano}, {Borgani}, {Bozzo},
  {Cabanac}, {Cappi}, {Casas}, {Castignani}, {Colodro-Conde}, {Coupon},
  {Courtois}, {Cuby}, {de la Torre}, {Desai}, {Dole}, {Fabricius}, {Farina},
  {Ferreira}, {Finelli}, {Flose-Reimberg}, {Fotopoulou}, {Ganga}, {Gozaliasl},
  {Hook}, {Keihanen}, {Kirkpatrick}, {Liebing}, {Lindholm}, {Mainetti},
  {Martinelli}, {Martinet}, {Maturi}, {McCracken}, {Metcalf}, {Morgante},
  {Nightingale}, {Nucita}, {Patrizii}, {Potter}, {Riccio}, {S{\'a}nchez},
  {Sapone}, {Schewtschenko}, {Schultheis}, {Scottez}, {Teyssier}, {Tutusaus},
  {Valiviita}, {Viel}, {Vriend}, \& {Whittaker}}]{Euclid:2021icp}
{Euclid Collaboration: Scaramella}, R., {Amiaux}, J., {Mellier}, Y., {et~al.}
  2022, \aap, 662, A112

\bibitem[{{Fiorini} {et~al.}(2023){Fiorini}, {Koyama}, \&
  {Baker}}]{Fiorini:2023fjl}
{Fiorini}, B., {Koyama}, K., \& {Baker}, T. 2023, \jcap, 12, 045

\bibitem[{{Giocoli} {et~al.}(2018){Giocoli}, {Baldi}, \&
  {Moscardini}}]{Giocoli:2018gqh}
{Giocoli}, C., {Baldi}, M., \& {Moscardini}, L. 2018, \mnras, 481, 2813

\bibitem[{{Gleyzes} {et~al.}(2014){Gleyzes}, {Langlois}, \&
  {Vernizzi}}]{Gleyzes:2014rba}
{Gleyzes}, J., {Langlois}, D., \& {Vernizzi}, F. 2014, Int.\ J.\ Mod.\ Phys.\
  D, 23, 1443010

\bibitem[{Gordon {et~al.}(2024)Gordon, de~Aguiar, Rebou\c{c}as, Brando,
  Falciano, Miranda, Koyama, \& Winther}]{Gordon:2024jaj}
Gordon, J., de~Aguiar, B.~F., Rebou\c{c}as, J.~a., {et~al.} 2024
  [\eprint[arXiv]{2404.12344}]

\bibitem[{Gronke {et~al.}(2014)Gronke, Llinares, \& Mota}]{Gronke:2013mea}
Gronke, M.~B., Llinares, C., \& Mota, D.~F. 2014, \aap, 562, A9

\bibitem[{Guillet \& Teyssier(2011)}]{GUILLET}
Guillet, T. \& Teyssier, R. 2011, J. Comp. Phys., 230, 4756

\bibitem[{Hagala {et~al.}(2016)Hagala, Llinares, \& Mota}]{Hagala:2015paa}
Hagala, R., Llinares, C., \& Mota, D.~F. 2016, \aap, 585, A37

\bibitem[{Hagala {et~al.}(2017)Hagala, Llinares, \& Mota}]{Hagala:2016fks}
Hagala, R., Llinares, C., \& Mota, D.~F. 2017, \prl, 118, 101301

\bibitem[{Hammami {et~al.}(2015)Hammami, Llinares, Mota, \&
  Winther}]{Hammami:2015iwa}
Hammami, A., Llinares, C., Mota, D.~F., \& Winther, H.~A. 2015, \mnras, 449,
  3635

\bibitem[{Harnois-D\'eraps {et~al.}(2023)Harnois-D\'eraps, Hernandez-Aguayo,
  Cuesta-Lazaro, Arnold, Li, Davies, \& Cai}]{Harnois-Deraps:2022bie}
Harnois-D\'eraps, J., Hernandez-Aguayo, C., Cuesta-Lazaro, C., {et~al.} 2023,
  \mnras, 525, 6336

\bibitem[{{Hassani} {et~al.}(2019){Hassani}, {Adamek}, {Kunz}, \&
  {Vernizzi}}]{Hassani:2019lmy}
{Hassani}, F., {Adamek}, J., {Kunz}, M., \& {Vernizzi}, F. 2019, \jcap, 12, 011

\bibitem[{{Hassani} {et~al.}(2020){Hassani}, {L'Huillier}, {Shafieloo}, {Kunz},
  \& {Adamek}}]{Hassani:2019wed}
{Hassani}, F., {L'Huillier}, B., {Shafieloo}, A., {Kunz}, M., \& {Adamek}, J.
  2020, \jcap, 04, 039

\bibitem[{{Hern{\'a}ndez-Aguayo} {et~al.}(2021){Hern{\'a}ndez-Aguayo},
  {Arnold}, {Li}, \& {Baugh}}]{Hernandez-Aguayo:2020kgq}
{Hern{\'a}ndez-Aguayo}, C., {Arnold}, C., {Li}, B., \& {Baugh}, C.~M. 2021,
  \mnras, 503, 3867

\bibitem[{{Hu} \& {Sawicki}(2007)}]{Hu:2007nk}
{Hu}, W. \& {Sawicki}, I. 2007, \prd, 76, 064004

\bibitem[{{Ivezi{\'c}} {et~al.}(2019){Ivezi{\'c}}, {Kahn}, {Tyson}, {Abel},
  {Acosta}, {Allsman}, {Alonso}, {AlSayyad}, {Anderson}, {Andrew}, {Angel},
  {Angeli}, {Ansari}, {Antilogus}, {Araujo}, {Armstrong}, {Arndt}, {Astier},
  {Aubourg}, {Auza}, {Axelrod}, {Bard}, {Barr}, {Barrau}, {Bartlett}, {Bauer},
  {Bauman}, {Baumont}, {Bechtol}, {Bechtol}, {Becker}, {Becla}, {Beldica},
  {Bellavia}, {Bianco}, {Biswas}, {Blanc}, {Blazek}, {Blandford}, {Bloom},
  {Bogart}, {Bond}, {Booth}, {Borgland}, {Borne}, {Bosch}, {Boutigny},
  {Brackett}, {Bradshaw}, {Brandt}, {Brown}, {Bullock}, {Burchat}, {Burke},
  {Cagnoli}, {Calabrese}, {Callahan}, {Callen}, {Carlin}, {Carlson},
  {Chandrasekharan}, {Charles-Emerson}, {Chesley}, {Cheu}, {Chiang}, {Chiang},
  {Chirino}, {Chow}, {Ciardi}, {Claver}, {Cohen-Tanugi}, {Cockrum}, {Coles},
  {Connolly}, {Cook}, {Cooray}, {Covey}, {Cribbs}, {Cui}, {Cutri}, {Daly},
  {Daniel}, {Daruich}, {Daubard}, {Daues}, {Dawson}, {Delgado}, {Dellapenna},
  {de Peyster}, {de Val-Borro}, {Digel}, {Doherty}, {Dubois},
  {Dubois-Felsmann}, {Durech}, {Economou}, {Eifler}, {Eracleous}, {Emmons},
  {Fausti Neto}, {Ferguson}, {Figueroa}, {Fisher-Levine}, {Focke}, {Foss},
  {Frank}, {Freemon}, {Gangler}, {Gawiser}, {Geary}, {Gee}, {Geha}, {Gessner},
  {Gibson}, {Gilmore}, {Glanzman}, {Glick}, {Goldina}, {Goldstein}, {Goodenow},
  {Graham}, {Gressler}, {Gris}, {Guy}, {Guyonnet}, {Haller}, {Harris},
  {Hascall}, {Haupt}, {Hernandez}, {Herrmann}, {Hileman}, {Hoblitt}, {Hodgson},
  {Hogan}, {Howard}, {Huang}, {Huffer}, {Ingraham}, {Innes}, {Jacoby}, {Jain},
  {Jammes}, {Jee}, {Jenness}, {Jernigan}, {Jevremovi{\'c}}, {Johns}, {Johnson},
  {Johnson}, {Jones}, {Juramy-Gilles}, {Juri{\'c}}, {Kalirai}, {Kallivayalil},
  {Kalmbach}, {Kantor}, {Karst}, {Kasliwal}, {Kelly}, {Kessler}, {Kinnison},
  {Kirkby}, {Knox}, {Kotov}, {Krabbendam}, {Krughoff}, {Kub{\'a}nek},
  {Kuczewski}, {Kulkarni}, {Ku}, {Kurita}, {Lage}, {Lambert}, {Lange},
  {Langton}, {Le Guillou}, {Levine}, {Liang}, {Lim}, {Lintott}, {Long},
  {Lopez}, {Lotz}, {Lupton}, {Lust}, {MacArthur}, {Mahabal}, {Mandelbaum},
  {Markiewicz}, {Marsh}, {Marshall}, {Marshall}, {May}, {McKercher}, {McQueen},
  {Meyers}, {Migliore}, {Miller}, {Mills}, {Miraval}, {Moeyens}, {Moolekamp},
  {Monet}, {Moniez}, {Monkewitz}, {Montgomery}, {Morrison}, {Mueller},
  {Muller}, {Mu{\~n}oz Arancibia}, {Neill}, {Newbry}, {Nief}, {Nomerotski},
  {Nordby}, {O'Connor}, {Oliver}, {Olivier}, {Olsen}, {O'Mullane}, {Ortiz},
  {Osier}, {Owen}, {Pain}, {Palecek}, {Parejko}, {Parsons}, {Pease},
  {Peterson}, {Peterson}, {Petravick}, {Libby Petrick}, {Petry},
  {Pierfederici}, {Pietrowicz}, {Pike}, {Pinto}, {Plante}, {Plate}, {Plutchak},
  {Price}, {Prouza}, {Radeka}, {Rajagopal}, {Rasmussen}, {Regnault}, {Reil},
  {Reiss}, {Reuter}, {Ridgway}, {Riot}, {Ritz}, {Robinson}, {Roby}, {Roodman},
  {Rosing}, {Roucelle}, {Rumore}, {Russo}, {Saha}, {Sassolas}, {Schalk},
  {Schellart}, {Schindler}, {Schmidt}, {Schneider}, {Schneider}, {Schoening},
  {Schumacher}, {Schwamb}, {Sebag}, {Selvy}, {Sembroski}, {Seppala}, {Serio},
  {Serrano}, {Shaw}, {Shipsey}, {Sick}, {Silvestri}, {Slater}, {Smith},
  {Smith}, {Sobhani}, {Soldahl}, {Storrie-Lombardi}, {Stover}, {Strauss},
  {Street}, {Stubbs}, {Sullivan}, {Sweeney}, {Swinbank}, {Szalay}, {Takacs},
  {Tether}, {Thaler}, {Thayer}, {Thomas}, {Thornton}, {Thukral}, {Tice},
  {Trilling}, {Turri}, {Van Berg}, {Vanden Berk}, {Vetter}, {Virieux},
  {Vucina}, {Wahl}, {Walkowicz}, {Walsh}, {Walter}, {Wang}, {Wang}, {Warner},
  {Wiecha}, {Willman}, {Winters}, {Wittman}, {Wolff}, {Wood-Vasey}, {Wu},
  {Xin}, {Yoachim}, \& {Zhan}}]{LSST:2008ijt}
{Ivezi{\'c}}, {\v{Z}}., {Kahn}, S.~M., {Tyson}, J.~A., {et~al.} 2019, \apj,
  873, 111

\bibitem[{{Joyce} {et~al.}(2016){Joyce}, {Lombriser}, \&
  {Schmidt}}]{Joyce:2016vqv}
{Joyce}, A., {Lombriser}, L., \& {Schmidt}, F. 2016, Ann.\ Rev.\ Nucl.\ Part.\
  Sci., 66, 95

\bibitem[{{Khoury} \& {Wyman}(2009)}]{Khoury:2009tk}
{Khoury}, J. \& {Wyman}, M. 2009, \prd, 80, 064023

\bibitem[{{Laureijs} {et~al.}(2011){Laureijs}, {Amiaux}, {Arduini},
  {Augu{\`e}res}, {Brinchmann}, {Cole}, {Cropper}, {Dabin}, {Duvet}, {Ealet},
  {Garilli}, {Gondoin}, {Guzzo}, {Hoar}, {Hoekstra}, {Holmes}, {Kitching},
  {Maciaszek}, {Mellier}, {Pasian}, {Percival}, {Rhodes}, {Saavedra Criado},
  {Sauvage}, {Scaramella}, {Valenziano}, {Warren}, {Bender}, {Castander},
  {Cimatti}, {Le F{\`e}vre}, {Kurki-Suonio}, {Levi}, {Lilje}, {Meylan},
  {Nichol}, {Pedersen}, {Popa}, {Rebolo Lopez}, {Rix}, {Rottgering},
  {Zeilinger}, {Grupp}, {Hudelot}, {Massey}, {Meneghetti}, {Miller}, {Paltani},
  {Paulin-Henriksson}, {Pires}, {Saxton}, {Schrabback}, {Seidel}, {Walsh},
  {Aghanim}, {Amendola}, {Bartlett}, {Baccigalupi}, {Beaulieu}, {Benabed},
  {Cuby}, {Elbaz}, {Fosalba}, {Gavazzi}, {Helmi}, {Hook}, {Irwin}, {Kneib},
  {Kunz}, {Mannucci}, {Moscardini}, {Tao}, {Teyssier}, {Weller}, {Zamorani},
  {Zapatero Osorio}, {Boulade}, {Foumond}, {Di Giorgio}, {Guttridge}, {James},
  {Kemp}, {Martignac}, {Spencer}, {Walton}, {Bl{\"u}mchen}, {Bonoli},
  {Bortoletto}, {Cerna}, {Corcione}, {Fabron}, {Jahnke}, {Ligori}, {Madrid},
  {Martin}, {Morgante}, {Pamplona}, {Prieto}, {Riva}, {Toledo}, {Trifoglio},
  {Zerbi}, {Abdalla}, {Douspis}, {Grenet}, {Borgani}, {Bouwens}, {Courbin},
  {Delouis}, {Dubath}, {Fontana}, {Frailis}, {Grazian}, {Koppenh{\"o}fer},
  {Mansutti}, {Melchior}, {Mignoli}, {Mohr}, {Neissner}, {Noddle}, {Poncet},
  {Scodeggio}, {Serrano}, {Shane}, {Starck}, {Surace}, {Taylor},
  {Verdoes-Kleijn}, {Vuerli}, {Williams}, {Zacchei}, {Altieri}, {Escudero
  Sanz}, {Kohley}, {Oosterbroek}, {Astier}, {Bacon}, {Bardelli}, {Baugh},
  {Bellagamba}, {Benoist}, {Bianchi}, {Biviano}, {Branchini}, {Carbone},
  {Cardone}, {Clements}, {Colombi}, {Conselice}, {Cresci}, {Deacon}, {Dunlop},
  {Fedeli}, {Fontanot}, {Franzetti}, {Giocoli}, {Garcia-Bellido}, {Gow},
  {Heavens}, {Hewett}, {Heymans}, {Holland}, {Huang}, {Ilbert}, {Joachimi},
  {Jennins}, {Kerins}, {Kiessling}, {Kirk}, {Kotak}, {Krause}, {Lahav}, {van
  Leeuwen}, {Lesgourgues}, {Lombardi}, {Magliocchetti}, {Maguire}, {Majerotto},
  {Maoli}, {Marulli}, {Maurogordato}, {McCracken}, {McLure}, {Melchiorri},
  {Merson}, {Moresco}, {Nonino}, {Norberg}, {Peacock}, {Pello}, {Penny},
  {Pettorino}, {Di Porto}, {Pozzetti}, {Quercellini}, {Radovich}, {Rassat},
  {Roche}, {Ronayette}, {Rossetti}, {Sartoris}, {Schneider}, {Semboloni},
  {Serjeant}, {Simpson}, {Skordis}, {Smadja}, {Smartt}, {Spano}, {Spiro},
  {Sullivan}, {Tilquin}, {Trotta}, {Verde}, {Wang}, {Williger}, {Zhao},
  {Zoubian}, \& {Zucca}}]{2011arXiv1110.3193L}
{Laureijs}, R., {Amiaux}, J., {Arduini}, S., {et~al.} 2011, arXiv:1110.3193

\bibitem[{{Lesgourgues}(2011)}]{Lesgourgues:2011re}
{Lesgourgues}, J. 2011, arXiv:1104.2932

\bibitem[{{Lewis} {et~al.}(2000){Lewis}, {Challinor}, \&
  {Lasenby}}]{Lewis:1999bs}
{Lewis}, A., {Challinor}, A., \& {Lasenby}, A. 2000, \apj, 538, 473

\bibitem[{Li(2018)}]{Li_book}
Li, B. 2018, Simulating Large-Scale Structure for Models of Cosmic
  Acceleration, 2514-3433 (IOP Publishing)

\bibitem[{{Li} {et~al.}(2013{\natexlab{a}}){Li}, {Barreira}, {Baugh},
  {Hellwing}, {Koyama}, {Pascoli}, \& {Zhao}}]{Li:2013tda}
{Li}, B., {Barreira}, A., {Baugh}, C.~M., {et~al.} 2013{\natexlab{a}}, \jcap,
  11, 012

\bibitem[{{Li} \& {Barrow}(2011)}]{Li:2010re}
{Li}, B. \& {Barrow}, J.~D. 2011, \prd, 83, 024007

\bibitem[{{Li} {et~al.}(2013{\natexlab{b}}){Li}, {Zhao}, \&
  {Koyama}}]{Li:2013nua}
{Li}, B., {Zhao}, G.-B., \& {Koyama}, K. 2013{\natexlab{b}}, \jcap, 05, 023

\bibitem[{{Li} {et~al.}(2012){Li}, {Zhao}, {Teyssier}, \& {Koyama}}]{Li:2011vk}
{Li}, B., {Zhao}, G.-B., {Teyssier}, R., \& {Koyama}, K. 2012, \jcap, 01, 051

\bibitem[{{Linder}(2003)}]{Linder:2002et}
{Linder}, E.~V. 2003, \prl, 90, 091301

\bibitem[{{Llinares}(2018)}]{Llinares_2018}
{Llinares}, C. 2018, Int.\ J.\ Mod.\ Phys.\ D, 27, 1848003

\bibitem[{Llinares {et~al.}(2020)Llinares, Hagala, \& Mota}]{Llinares:2019rbe}
Llinares, C., Hagala, R., \& Mota, D.~F. 2020, \mnras, 491, 1868

\bibitem[{Llinares \& Mota(2013)}]{Llinares:2013qbh}
Llinares, C. \& Mota, D. 2013, \prl, 110, 161101

\bibitem[{{Llinares} \& {Mota}(2014)}]{Llinares:2013jua}
{Llinares}, C. \& {Mota}, D.~F. 2014, \prd, 89, 084023

\bibitem[{Llinares {et~al.}(2014)Llinares, Mota, \& Winther}]{Llinares:2013jza}
Llinares, C., Mota, D.~F., \& Winther, H.~A. 2014, \aap, 562, A78

\bibitem[{{Lombriser}(2016)}]{Lombriser:2016zfz}
{Lombriser}, L. 2016, \jcap, 11, 039

\bibitem[{{Martinelli} {et~al.}(2021){Martinelli}, {Martins}, {Nesseris},
  {Tutusaus}, {Blanchard}, {Camera}, {Carbone}, {Casas}, {Pettorino}, {Sakr},
  {Yankelevich}, {Sapone}, {Amara}, {Auricchio}, {Bodendorf}, {Bonino},
  {Branchini}, {Capobianco}, {Carretero}, {Castellano}, {Cavuoti}, {Cimatti},
  {Cledassou}, {Corcione}, {Costille}, {Degaudenzi}, {Douspis}, {Dubath},
  {Dusini}, {Ealet}, {Ferriol}, {Frailis}, {Franceschi}, {Garilli}, {Giocoli},
  {Grazian}, {Grupp}, {Haugan}, {Holmes}, {Hormuth}, {Jahnke}, {Kiessling},
  {K{\"u}mmel}, {Kunz}, {Kurki-Suonio}, {Ligori}, {Lilje}, {Lloro}, {Mansutti},
  {Marggraf}, {Markovic}, {Massey}, {Meneghetti}, {Meylan}, {Moscardini},
  {Niemi}, {Padilla}, {Paltani}, {Pasian}, {Pedersen}, {Pires}, {Poncet},
  {Popa}, {Raison}, {Rebolo}, {Rhodes}, {Roncarelli}, {Rossetti}, {Saglia},
  {Secroun}, {Seidel}, {Serrano}, {Sirignano}, {Sirri}, {Starck}, {Tavagnacco},
  {Taylor}, {Tereno}, {Toledo-Moreo}, {Valenziano}, {Wang}, {Zamorani},
  {Zoubian}, {Baldi}, {Brescia}, {Congedo}, {Conversi}, {Copin}, {Fabbian},
  {Farinelli}, {Medinaceli}, {Mei}, {Polenta}, {Romelli}, \&
  {Vassallo}}]{Euclid:2021cfn}
{Martinelli}, M., {Martins}, C.~J.~A.~P., {Nesseris}, S., {et~al.} 2021, \aap,
  654, A148

\bibitem[{Mauland {et~al.}(2024)Mauland, Winther, \& Ruan}]{Mauland:2023pjt}
Mauland, R., Winther, H.~A., \& Ruan, C.-Z. 2024, \aap, 685, A156

\bibitem[{Mitchell {et~al.}(2022)Mitchell, Arnold, \& Li}]{Mitchell:2021ter}
Mitchell, M.~A., Arnold, C., \& Li, B. 2022, \mnras, 514, 3349

\bibitem[{Mota {et~al.}(2008)Mota, Pettorino, Robbers, \&
  Wetterich}]{Mota:2008nj}
Mota, D.~F., Pettorino, V., Robbers, G., \& Wetterich, C. 2008, Phys. Lett. B,
  663, 160

\bibitem[{{Nesseris} {et~al.}(2022){Nesseris}, {Sapone}, {Martinelli},
  {Camarena}, {Marra}, {Sakr}, {Garcia-Bellido}, {Martins}, {Clarkson}, {Da
  Silva}, {Fleury}, {Lombriser}, {Mimoso}, {Casas}, {Pettorino}, {Tutusaus},
  {Amara}, {Auricchio}, {Bodendorf}, {Bonino}, {Branchini}, {Brescia},
  {Capobianco}, {Carbone}, {Carretero}, {Castellano}, {Cavuoti}, {Cimatti},
  {Cledassou}, {Congedo}, {Conversi}, {Copin}, {Corcione}, {Courbin},
  {Cropper}, {Degaudenzi}, {Douspis}, {Dubath}, {Duncan}, {Dupac}, {Dusini},
  {Ealet}, {Farrens}, {Fosalba}, {Frailis}, {Franceschi}, {Fumana}, {Garilli},
  {Gillis}, {Giocoli}, {Grazian}, {Grupp}, {Haugan}, {Holmes}, {Hormuth},
  {Jahnke}, {Kermiche}, {Kiessling}, {Kitching}, {K{\"u}mmel}, {Kunz},
  {Kurki-Suonio}, {Ligori}, {Lilje}, {Lloro}, {Mansutti}, {Marggraf},
  {Markovic}, {Marulli}, {Massey}, {Meneghetti}, {Merlin}, {Meylan}, {Moresco},
  {Moscardini}, {Munari}, {Niemi}, {Padilla}, {Paltani}, {Pasian}, {Pedersen},
  {Percival}, {Poncet}, {Popa}, {Racca}, {Raison}, {Rhodes}, {Roncarelli},
  {Saglia}, {Sartoris}, {Schneider}, {Secroun}, {Seidel}, {Serrano},
  {Sirignano}, {Sirri}, {Stanco}, {Starck}, {Tallada-Cresp{\'\i}}, {Taylor},
  {Tereno}, {Toledo-Moreo}, {Torradeflot}, {Valentijn}, {Valenziano}, {Wang},
  {Welikala}, {Zamorani}, {Zoubian}, {Andreon}, {Baldi}, {Camera},
  {Medinaceli}, {Mei}, \& {Renzi}}]{Euclid:2021frk}
{Nesseris}, S., {Sapone}, D., {Martinelli}, M., {et~al.} 2022, \aap, 660, A67

\bibitem[{{Oyaizu}(2008)}]{Oyaizu:2008sr}
{Oyaizu}, H. 2008, \prd, 78, 123523

\bibitem[{Palma \& Candlish(2023)}]{Palma:2023ggq}
Palma, D. \& Candlish, G.~N. 2023, \mnras, 526, 1904

\bibitem[{{Pettorino} \& {Baccigalupi}(2008)}]{Pettorino:2008ez}
{Pettorino}, V. \& {Baccigalupi}, C. 2008, \prd, 77, 103003

\bibitem[{Pillepich {et~al.}(2018)}]{Pillepich:2017jle}
Pillepich, A. {et~al.} 2018, \mnras, 473, 4077

\bibitem[{Potter {et~al.}(2017)Potter, Stadel, \& Teyssier}]{Potter:2016ttn}
Potter, D., Stadel, J., \& Teyssier, R. 2017, Comput. Astrophys. Cosmol., 4, 2

\bibitem[{Pourtsidou {et~al.}(2013)Pourtsidou, Skordis, \&
  Copeland}]{Pourtsidou:2013nha}
Pourtsidou, A., Skordis, C., \& Copeland, E.~J. 2013, \prd, 88, 083505

\bibitem[{{Puchwein} {et~al.}(2013){Puchwein}, {Baldi}, \&
  {Springel}}]{Puchwein:2013lza}
{Puchwein}, E., {Baldi}, M., \& {Springel}, V. 2013, \mnras, 436, 348

\bibitem[{Ramachandra {et~al.}(2021)Ramachandra, Valogiannis, Ishak, \&
  Heitmann}]{Ramachandra:2020lue}
Ramachandra, N., Valogiannis, G., Ishak, M., \& Heitmann, K. 2021, \prd, 103,
  123525

\bibitem[{{Ruan} {et~al.}(2024){Ruan}, {Cuesta-Lazaro}, {Eggemeier}, {Li},
  {Baugh}, {Arnold}, {Bose}, {Hern{\'a}ndez-Aguayo}, {Zarrouk}, \&
  {Davies}}]{Ruan:2023mgq}
{Ruan}, C.-Z., {Cuesta-Lazaro}, C., {Eggemeier}, A., {et~al.} 2024, \mnras,
  527, 2490

\bibitem[{{Ruan} {et~al.}(2022){Ruan}, {Hern{\'a}ndez-Aguayo}, {Li}, {Arnold},
  {Baugh}, {Klypin}, \& {Prada}}]{Ruan:2021wup}
{Ruan}, C.-Z., {Hern{\'a}ndez-Aguayo}, C., {Li}, B., {et~al.} 2022, \jcap, 05,
  018

\bibitem[{S\'aez-Casares {et~al.}(2023)S\'aez-Casares, Rasera, \&
  Li}]{Saez-Casares:2023olw}
S\'aez-Casares, I.~n., Rasera, Y., \& Li, B. 2023 [\eprint[arXiv]{2303.08899}]

\bibitem[{{Sawicki} \& {Bellini}(2015)}]{Sawicki:2015zya}
{Sawicki}, I. \& {Bellini}, E. 2015, \prd, 92, 084061

\bibitem[{Schmidt(2009{\natexlab{a}})}]{Schmidt:2009sv}
Schmidt, F. 2009{\natexlab{a}}, \prd, 80, 123003

\bibitem[{Schmidt(2009{\natexlab{b}})}]{Schmidt:2009sg}
Schmidt, F. 2009{\natexlab{b}}, \prd, 80, 043001

\bibitem[{Schneider {et~al.}(2016)Schneider, Teyssier, Potter, Stadel, Onions,
  Reed, Smith, Springel, Pearce, \& Scoccimarro}]{Schneider:2015yka}
Schneider, A., Teyssier, R., Potter, D., {et~al.} 2016, \jcap, 04, 047

\bibitem[{{Simpson}(2010)}]{Simpson:2010vh}
{Simpson}, F. 2010, \prd, 82, 083505

\bibitem[{{Skordis} {et~al.}(2015){Skordis}, {Pourtsidou}, \&
  {Copeland}}]{Skordis:2015yra}
{Skordis}, C., {Pourtsidou}, A., \& {Copeland}, E.~J. 2015, \prd, 91, 083537

\bibitem[{{Spergel} {et~al.}(2015){Spergel}, {Gehrels}, {Baltay}, {Bennett},
  {Breckinridge}, {Donahue}, {Dressler}, {Gaudi}, {Greene}, {Guyon}, {Hirata},
  {Kalirai}, {Kasdin}, {Macintosh}, {Moos}, {Perlmutter}, {Postman},
  {Rauscher}, {Rhodes}, {Wang}, {Weinberg}, {Benford}, {Hudson}, {Jeong},
  {Mellier}, {Traub}, {Yamada}, {Capak}, {Colbert}, {Masters}, {Penny},
  {Savransky}, {Stern}, {Zimmerman}, {Barry}, {Bartusek}, {Carpenter}, {Cheng},
  {Content}, {Dekens}, {Demers}, {Grady}, {Jackson}, {Kuan}, {Kruk}, {Melton},
  {Nemati}, {Parvin}, {Poberezhskiy}, {Peddie}, {Ruffa}, {Wallace}, {Whipple},
  {Wollack}, \& {Zhao}}]{2015arXiv150303757S}
{Spergel}, D., {Gehrels}, N., {Baltay}, C., {et~al.} 2015, arXiv:1503.03757

\bibitem[{{Springel}(2005)}]{Springel:2005mi}
{Springel}, V. 2005, \mnras, 364, 1105

\bibitem[{{Springel}(2010)}]{Springel:2009aa}
{Springel}, V. 2010, \mnras, 401, 791

\bibitem[{Tassev {et~al.}(2013)Tassev, Zaldarriaga, \&
  Eisenstein}]{Tassev:2013pn}
Tassev, S., Zaldarriaga, M., \& Eisenstein, D. 2013, JCAP, 06, 036

\bibitem[{{Teyssier}(2002)}]{Teyssier:2001cp}
{Teyssier}, R. 2002, \aap, 385, 337

\bibitem[{Tsujikawa(2010)}]{Tsujikawa:2010zza}
Tsujikawa, S. 2010, Lect. Notes Phys., 800, 99

\bibitem[{Vainshtein(1972)}]{Vainshtein:1972sx}
Vainshtein, A.~I. 1972, Phys.\ Lett.\ B, 39, 393

\bibitem[{{Weinberger} {et~al.}(2017){Weinberger}, {Springel}, {Hernquist},
  {Pillepich}, {Marinacci}, {Pakmor}, {Nelson}, {Genel}, {Vogelsberger},
  {Naiman}, \& {Torrey}}]{Weinberger:2017MNRAS}
{Weinberger}, R., {Springel}, V., {Hernquist}, L., {et~al.} 2017, \mnras, 465,
  3291

\bibitem[{{Weinberger} {et~al.}(2020){Weinberger}, {Springel}, \&
  {Pakmor}}]{Weinberger:2020dt}
{Weinberger}, R., {Springel}, V., \& {Pakmor}, R. 2020, \apjs, 248, 32

\bibitem[{Wesseling(2004)}]{wesseling2004}
Wesseling, P. 2004, An Introduction to Multigrid Methods (Philadelphia: R.T.
  Edwards)

\bibitem[{Wetterich(1995)}]{Wetterich:1994bg}
Wetterich, C. 1995, \aap, 301, 321

\bibitem[{Wetterich(2004)}]{Wetterich:2004pv}
Wetterich, C. 2004, Phys.\ Lett.\ B, 594, 17

\bibitem[{{Will}(2014)}]{Will:2014kxa}
{Will}, C.~M. 2014, Living Rev.\ Rel., 17, 4

\bibitem[{{Winther} \& {Ferreira}(2015{\natexlab{a}})}]{Winther:2014cia}
{Winther}, H.~A. \& {Ferreira}, P.~G. 2015{\natexlab{a}}, \prd, 91, 123507

\bibitem[{{Winther} \& {Ferreira}(2015{\natexlab{b}})}]{Winther:2015pta}
{Winther}, H.~A. \& {Ferreira}, P.~G. 2015{\natexlab{b}}, \prd, 92, 064005

\bibitem[{Winther {et~al.}(2017)Winther, Koyama, Manera, Wright, \&
  Zhao}]{Winther:2017jof}
Winther, H.~A., Koyama, K., Manera, M., Wright, B.~S., \& Zhao, G.-B. 2017,
  \jcap, 08, 006

\bibitem[{{Winther} {et~al.}(2015){Winther}, {Schmidt}, {Barreira}, {Arnold},
  {Bose}, {Llinares}, {Baldi}, {Falck}, {Hellwing}, {Koyama}, {Li}, {Mota},
  {Puchwein}, {Smith}, \& {Zhao}}]{Winther:2015wla}
{Winther}, H.~A., {Schmidt}, F., {Barreira}, A., {et~al.} 2015, \mnras, 454,
  4208

\bibitem[{Wright {et~al.}(2017)Wright, Winther, \& Koyama}]{Wright:2017dkw}
Wright, B.~S., Winther, H.~A., \& Koyama, K. 2017, \jcap, 10, 054

\bibitem[{{Zumalac{\'a}rregui} {et~al.}(2017){Zumalac{\'a}rregui}, {Bellini},
  {Sawicki}, {Lesgourgues}, \& {Ferreira}}]{Zumalacarregui:2016pph}
{Zumalac{\'a}rregui}, M., {Bellini}, E., {Sawicki}, I., {Lesgourgues}, J., \&
  {Ferreira}, P.~G. 2017, \jcap, 08, 019

\end{thebibliography}

\appendix
\section{Computational cost}

\begin{figure}
    \includegraphics[width=\columnwidth]{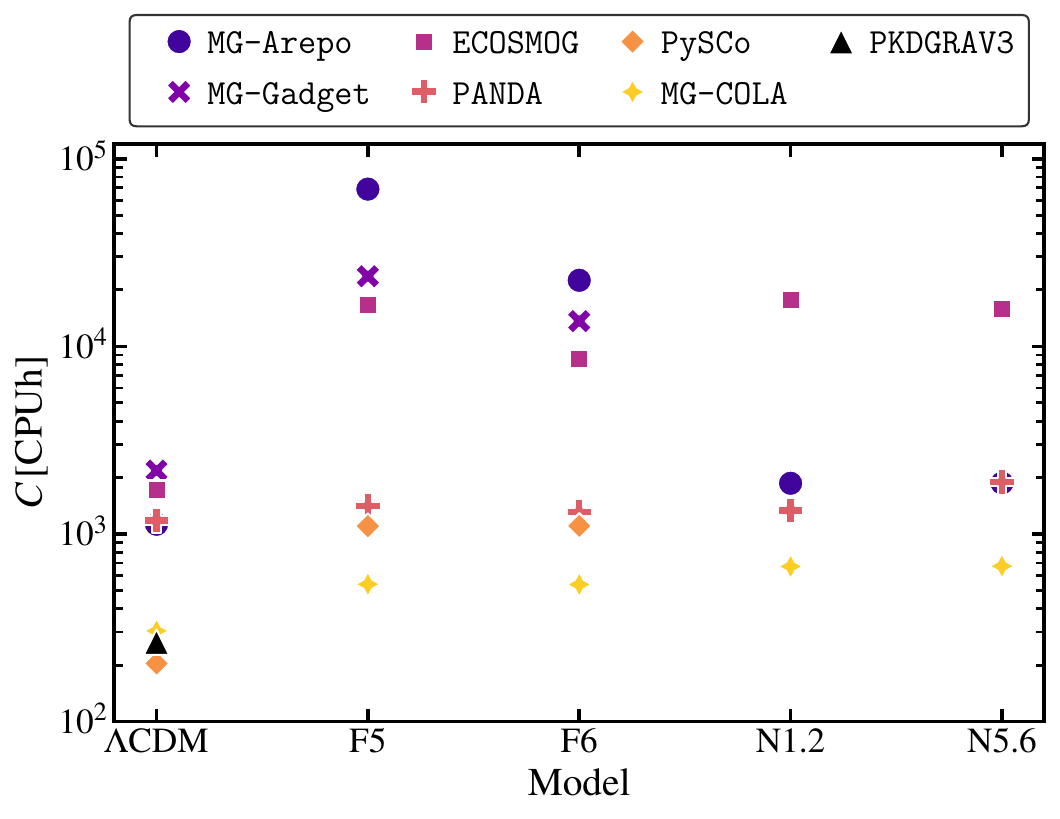}
    \caption{Comparison of the computational costs of simulations run under different gravity models (labels on the $x$-axis) and with different codes (markers and colours as described in the legend).}
    \label{fig:CompCost}
\end{figure}

One of the obstacles in producing accurate predictions in the nonlinear regime of structure formation is presented by the computational cost of running large {\it N}-body simulations. This issue becomes even more pronounced in the case of MG 
simulations, due to the additional computational cost associated with the solution of the Klein--Gordon equation for the scalar field describing the additional degree of freedom in these theories.
To provide some insights into the trade-offs between accuracy and time-to-solution, 
we attempt here a comparative analysis of the computational cost of the simulations run for this paper. Given that these simulations were run on different machines and with different parallelisation settings, we cannot conduct a precise assessment of their computational cost. Instead, we limit this analysis to an order-of-magnitude comparison of the simulations run with the various codes and models discussed in this paper.

For this exercise, we use the information on the wall-clock time, $T_{\rm real}$, and the number of cores, $N_{\rm cores}$, as recorded in the log files of the simulation runs. This information was available for all simulations except for the ones run with \texttt{ISIS}. We estimate the computational cost of the simulations, $C$, as the product of the wall-clock time and the number of cores:
\begin{equation}
    C \equiv N_{\rm cores} \; T_{\rm real} \, .
\end{equation}
The estimates of the computational cost for the simulations are compared in Fig.\,\ref{fig:CompCost}.
We can see that the cost of $\Lambda$CDM simulations of Tree-PM and AMR codes is $C \sim 10^3 \, {\rm CPUh}$, while for \texttt{MG-COLA} and \texttt{PySCo} the cost is about one order of magnitude lower at $C \sim 10^2 \, {\rm CPUh}$. For $f(R)$ gravity models instead, the computational cost increases significantly (approximately by a factor of ten) for the codes that solve the full Klein--Gordon equation of the scalar field, namely \texttt{MG-Arepo}, \texttt{MG-Gadget}, \texttt{ECOSMOG}, and \texttt{PySCo}, while the overhead is smaller for the codes that adopt screening approximations, namely \texttt{PANDA} and \texttt{MG-COLA}. Finally, in nDGP gravity, only \texttt{ECOSMOG} has a significant overhead compared to $\Lambda$CDM, while \texttt{MG-Arepo} and the approximate codes, \texttt{PANDA} and \texttt{MG-COLA}, have just a small overhead.

The performance of multigrid codes depends on the convergence criterion chosen. For \pysco, we chose an extremely conservative approach (see Sect.\,\ref{sec:pysco}) with a very low tolerance threshold, resulting in more V-cycles and almost double the CPU time needed to solve the linear Poisson equation compared to a more standard setup. Regarding the nonlinear solver, we use two F-cycles instead of a single one (which should in principle be enough, but it is not the goal of the present paper to provide a convergence study), therefore roughly doubling the CPU time needed for the $f(R)$ gravity models.

We stress that a thorough assessment of the efficiency of the codes is beyond the scope of this paper and would have required a much more methodical effort including (but not limited to)
\begin{itemize}
    \item running the simulations in a controlled environment,
    \item conducting convergence tests for the various hyper-parameters,
    \item testing the scaling performance of each code.
\end{itemize}

In fact, when focusing only on predictions of the amplification factors, it is possible to achieve a similar level of accuracy with lower force, mass or time resolution, since resolution effects mostly cancel out when taking ratios of quantities affected by the same inaccuracies \citep{Brando:2022gvg}.
This has been shown to be the case for \texttt{MG-COLA} simulations in \cite{Fiorini:2023fjl}, where the use of a lower resolution allowed accurate predictions of power spectrum boosts in nDGP gravity with a theoretical gain of $\sim 300$ with respect to the computational cost of the COLA simulations presented here. 

Such large speed-ups have paved the way for creating emulators for the nonlinear amplification of the power spectrum in models beyond $\Lambda$CDM in a cost-effective way, i.e.\ without the need for supercomputers. This has already been done for some of the models we consider in this paper \citep[see e.g.][]{Ramachandra:2020lue,Mauland:2023pjt,Fiorini:2023fjl}. For instance, \citet{Fiorini:2023fjl} found that an emulator for the nDGP model can be constructed with as little as a few thousand CPUh worth of computational time. Likewise, \citet{Mauland:2023pjt} who presented a generic pipeline for using COLA to create such emulators, used $f(R)$ gravity as an example and found similar numbers for the required computational time. \citet{Gordon:2024jaj} described a simulation setup that can also be used for emulating the full power spectrum (up to a reasonable high wavenumber $k\sim 1\,h\,\mathrm{Mpc}^{-1}$), requiring around $\sim 100$ CPUh per simulation on a modern CPU. Emulators have also been constructed for beyond-$\Lambda$CDM models using high-resolution direct simulations in the same way as has been done for $\Lambda$CDM. This approach generally gives more accurate emulators than those created with approximate methods, but this comes at a much higher cost. For example, both \citet{Saez-Casares:2023olw} and \citet{Arnold:2021xtm} have each presented a high-fidelity emulator for the $f(R)$ model considered in this paper, but at a higher cost of about 3.5-4 million CPUh.

\end{document}